# Quantum Chemistry Model of Surface Reactions and Kinetic Model of Diamond Growth: Effects of $CH_3$ Radicals and $C_2H_2$ Molecules at Low-Temperatures CVD

Yu. Barsukov[1], I.D. Kaganovich[1], M. Mokrov[1], and A. Khrabry[2]

[1] *Princeton Plasma Physics Laboratory, Princeton University, Princeton, NJ USA, 08540*

[2] *Princeton University, Princeton, NJ, USA, 08544*

**Abstract**

The objective of this study is to explore conditions that facilitate a significant reduction in substrate temperature during diamond growth. The typical temperature for this process is around 1200K; we aim to reduce it to a much lower level. To achieve this, we need to understand processes that limit the diamond growth at low temperatures. Therefore, we developed a detailed chemical kinetic model to analyze diamond growth on the (100) surface. This model accounts for variations in substrate temperature and gas composition. Using an ab initio quantum chemistry, we calculated the reaction rates of all major gas phase reactants with the diamond surface, totaling 91 elemental surface reactions. Consistent with previous studies, the model identifies that $CH_3$ is a major precursor of diamond growth, and the contribution from $C_2H_2$ to the growth is significantly smaller. However, $C_2H_2$ can also contribute to forming a $sp^2$-phase instead of a $sp^3$-phase, and this process becomes dominant below a critical temperature. As a result, $C_2H_2$ flux inhibits diamond growth at low temperatures. To quantify this deleterious process, we developed a new mechanism for $sp^2$-phase nucleation on the (100) surface. Similar to the so-called HACA mechanism for soot formation it involves hydrogen abstraction and $C_2H_2$ addition. Consequently, optimal low-temperature CVD growth could be realized in a reactor designed to maximize the $CH_3$ radical production, while minimizing the generation of $C_2H_2$ and other sp and $sp^2$ hydrocarbons.

## 1. Introduction

Diamond is a unique material with outstanding mechanical, optical, and thermal properties. Since the 1980s, the well-established method of diamond growth through chemical vapor deposition (CVD) has been the subject of extensive research [1–5]. Hot-filament (HF) and microwave (MW) reactors are commonly used for diamond chemical vapor deposition (CVD) [6]. A typical feed gas for this process consists of hydrocarbons, with methane being the most commonly used, diluted in hydrogen [7]. One major drawback of modern diamond CVD is the requirement for a very high substrate temperature, around 1200K [8–20]. At this temperature, the growth rate curve has a sharp peak, leading to a notable sensitivity of film uniformity to variations in substrate temperature [21]. The high substrate temperature limits the applications of modern CVD processes [22,23]. Therefore, developing a method of low-temperature diamond CVD (with a substrate temperature of 800K and below) is the subject of great interest [24–27].

Several experiments were conducted aimed at measuring diamond growth rate as a function of temperature. Mono- [28], poly- [29], and nanocrystalline [30–36] diamonds have been grown using hot-filament [28,30,37–52], hot graphite plate [53], microwave plasma-assisted [15,24,49–58], laser-induced [64–66], and flame reactors [19,67–75]. The measured growth rates are a strong function of the substrate temperature, with a maximum at around 1200K. The variation of the growth rate with the substrate temperature is dependent on the reactor type and operating conditions [76]. In conventional CVD with $CH_4/H_2$ feed gases, the growth rate decreases sharply as temperature decreases below 1200K. Introducing oxygen-contained gases such as $O_2$ [77,78], CO [57] or $CO_2$ [25] into the $CH_4/H_2$ mixtures can sustain diamond growth even at temperatures of 700-800K. Kawato and Kondo [79] demonstrated that oxygen reduces acetylene concentration, thereby suppressing the deposition of graphitic and amorphous carbon. This leads to an increase in the rate of high-quality diamond deposition rate in the hot-filament reactor. However, the fraction of $O_2$ must not exceed the $CH_4$ fraction; otherwise, no deposition occurs [80]. Recently, Malakoutian et al. [81] developed a technique that enables diamond growth at a temperature of 700K. They showed that the $CH_4/H_2$ mixture yields a soot-like $sp^2$ carbon structure characterized by a low $sp^3$ content. The addition of a small amount of $O_2$ to the original mixture significantly improves the film quality, resulting in crystalline diamond deposition, as shown in Fig. 1. Further increasing the $O_2$ concentration continues to improve the film quality but reduces the deposition rate.

To quantify the contribution of $C_2H_2$ into $sp^2$-phase, we developed a comprehensive model that incorporates both $sp^3$-phase diamond growth and $sp^2$-phase nucleation. This model integrates the most critical mechanisms identified in previous studies along with an introduced here novel mechanism for $sp^2$-phase nucleation that was not previously considered. The model elucidates why the conventional $CH_4/H_2$ mixture, typically used for diamond growth at around 1200K, leads to $sp^2$-phase production at lower temperatures, as reported by Malakoutian et al. Drawing on the results obtained, we discuss the role of oxygen in enhancing diamond growth in Section 4.

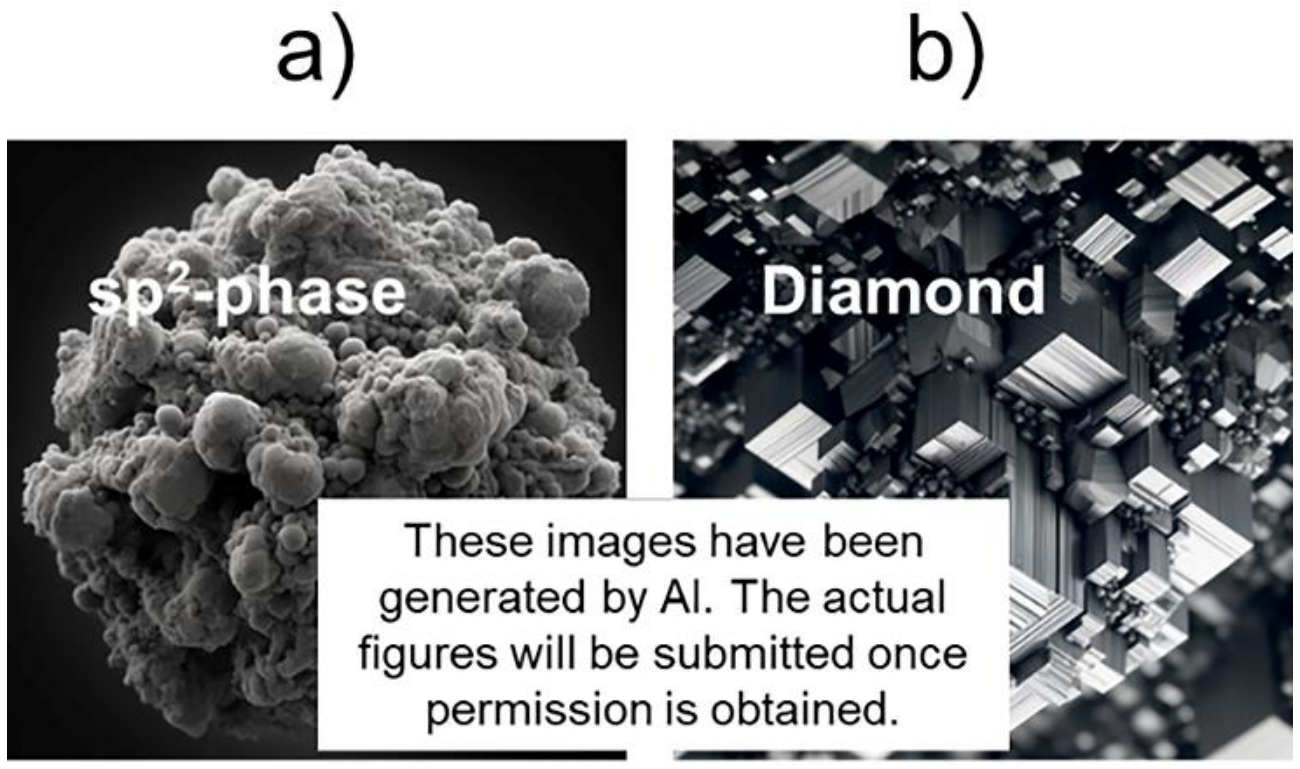

**Figure 1.** Cross-section view of carbon films deposited at 700K using $CH_4/H_2$ (a) and $CH_4/H_2/O_2$ (b) mixtures. Material from Malakoutian et al., MRS Advances 9, 7 (2024). Published by MRS Springer [81].

It is well-established that diamond growth from carbon precursors occurs on reactive surface sites. During CVD, most of the surface is covered (passivated) by hydrogen. Gas-phase atomic hydrogen abstracts hydrogen from the surface (H atom recombination), activates the surface and produces the reactive sites with dangling bonds. The heterogenous probability of H atom loss on the polycrystalline diamond surface was indirectly measured by Krasnoperov et al. [82], and by Harris and Weiner [83]. Krasnoperov et al. measured the densities of H and $H_2$ with and without a diamond substrate using a mass-spectrometer. Harris and Weiner measured the emissivity of diamond, which is a function of the surface heating and relates to H atom recombination on the surface.

Two major carbon reactive species are produced in typical CVD reactors: $CH_3$ and $C_2H_2$ [3,39,84–86]. There is wide consensus that the $CH_3$ radical is the main precursor of diamond growth. Using quantum chemistry calculations Frenklach et al. [4,87–89] theoretically predicted that $C_2H_2$ can insert into a diamond crystal network on the (100) and (111) surfaces and can act as a growth precursor. Several experiments were conducted to identify the contributions of $CH_3$ and $C_2H_2$ to the diamond growth. In the experiments with $C^{13}$ and $C^{12}$ isotope labeling [90–93], it was concluded that the $CH_3$ radical contributes predominantly to the diamond growth, while the contribution from $C_2H_2$ is minor. Similar results were obtained by Yarbrough and co-authors [94] for hot-filament diamond CVD.

Martin and co-workers [3,95–97] conducted experiments in a flow-tube reactor to identify roles of $CH_3$ and $C_2H_2$ to the diamond growth. In addition to the flow tube, they also used a remote plasma source to dissociate $H_2$ gas. The dissociated $H_2$ gas was mixed with either $CH_4$ or $C_2H_2$ in the reaction chamber. The substrate was heated up to 1100K. The gas pressure of 3 Torr was chosen to prevent the conversion of $CH_4$ into $C_2H_2$. The modeling of gas chemistry was performed in Ref. [3] and confirmed that conversion of $CH_4$ into $C_2H_2$ was indeed prevented under the experimental conditions. When the feed gas was $CH_4$, H-atoms generate $CH_3$ radicals from $CH_4$ in the mixing area, while the production of $C_2H_2$ from $CH_4$ is negligible at such low pressure. In the other case, when the feed gas was $C_2H_2$, H-atoms do not react fast with $C_2H_2$, and the reactive mixture remains predominantly $C_2H_2$. This enabled Martin et al. to clearly distinguish between the contributions of $CH_3$ and $C_2H_2$ into diamond growth. The authors concluded that $CH_3$ radical is much more effective for growing diamonds and produces better quality films than $C_2H_2$, because $C_2H_2$ is responsible for both diamond and non-diamond ($sp^2$-phase) carbon formation [3].

The fact that $C_2H_2$ is a precursor of $sp^2$-carbon growth is well-established. The well-known HACA mechanism (Hydrogen Abstraction Carbon Addition) [98–102] of $sp^2$-carbon growth was developed by Frenklach *et al.* for soot. Frenklach and Wang [103] also considered the $sp^2$-phase formation during diamond CVD. They assumed that the $sp^2$-phase nucleation proceeds through $C_6H_6$ adsorption and predicted that the $sp^2$-phase covers the diamond surface at low temperatures. Coltrin and Dandy [104] proposed a reaction kinetic model of the $sp^2$-phase and diamond growth involving $CH_3$, $C_2H_2$, and C. This model does not elaborate on elementary reaction steps but describes graphite phase growth by $C_2H_2$ using a qualitative kinetic model for most probable reaction pathways with effective reaction rates (not based on quantum chemistry calculations). A detailed mechanism of the $sp^2$-phase nucleation on a diamond surface due to $C_2H_2$ addition has been unexplored so far. In this paper, we report a new mechanism for the $sp^2$-phase nucleation on the (100) diamond surface, consisting of steps similar to those in the HACA mechanism.

As was mentioned above, to describe the $sp^2$-phase nucleation, we utilized a new chemical reaction mechanism developed in this study. For the contribution of $C_2H_2$ in diamond growth, we adopted the Skokov-Weiner-Frenklach mechanism [105]. This mechanism assumes that the surface undergoes reconstruction and consists of carbon dimers [106,107]. Acetylene inserts into the C-C bond of the surface dimers resulting in the reconstruction being repeated followed by growth. The insertion of $C_2H_2$ involves three steps: isomerization of $C_2H_2$ adsorbate into $CCH_2$ (vinylidene), dimer opening, and vinylidene bridging. This mechanism of $C_2H_2$ insertion bears similarities to the $CH_3$ insertion into diamond. The insertion of $CH_3$ proceeds through H abstraction from $CH_3$ adsorbate, resulting in $CH_2$ (methylene) formation, followed by dimer opening and methylene bridging steps [88,108–112], as reported by Garrison et al. [108] and by Huang and Frenklach [112].

Battaile et al. [113] reported about the importance of considering of the high temperature surface cleaning from carbon adsorbates. Namely, they showed that taking into account the $CH_3$ thermal desorption and the "etching" of $CH_2$ adsorbate by H-atom results in the growth rate reduction at high temperatures [114]. Moreover, including these reactions into the kinetic Monte Carlo modeling yields better agreement of modeling results with the measured surface roughness on the (111) facet. Including these reactions into the chemical model is not necessary to describe low-temperature CVD, but they are essential for understanding of the growth rate peak as they describe the growth rate decrease with temperature for substrate temperatures higher than 1200K . Therefore, we also incorporated these reactions leading to surface cleaning into the chemical model as detailed in Appendix D.

The paper is organized as follows. Section 2 provides computational details, including surface site nomenclature, quantum chemistry methods, integrated model. Section 3 summarizes theory of diamond growth and the gas phase chemistry in the hot filament reactors. The discussion of the results and conclusions are given in Sections 4 and 5, respectively.

## 2. Methods

First, we identify all possible states of the diamond surface as a full set of distinct surface sites and set of all possible surface reactions with major gas-phase reactants, including $H_2$, H, $CH_4$, $CH_3$, $C_2H_4$, and $C_2H_2$. For calculations of the surface reactions, we use the $C_9H_{14}$ cluster depicted in Fig. 2 to represent the $HC_dH$ site on the pristine (100) surface, where $C_d$ denotes diamond structure. Figure 3 shows a set of 21 surface sites. The proposed kinetic model consists of 91 surface reactions listed in Table 1. These reactions involve transformations of surface sites both with and without direct interaction with gas-phase reactants. The sites 0-13 have been considered previously and represent intermediate structures on the surface during diamond growth. In this study, we have developed a mechanism for the formation of the $sp^2$-site (20). As we will show later, this $sp^2$-site (20) plays a key role during low-temperature CVD. To describe the formation of this site, intermediates sites 14-19 were added to consideration.

We employ the notation X-$C_d$-Y (m) for the sites, where "$C_d$" represents the carbon dimer on the reconstructed (100) surface, "X-" and "Y-" are the substitutes on the $C_d$ dimer, "m" is a serial number of the site in our calculation of the surface reactions as shown in Fig. 3. $\theta_m$ denotes the population fraction of the "m" surface site or surface density of these sites. For example, $CH_3$-$C_dH$ (3) represents a surface site 3 with $CH_3$ and H substitutes, and $\theta_3$ is the fraction of this particular site.

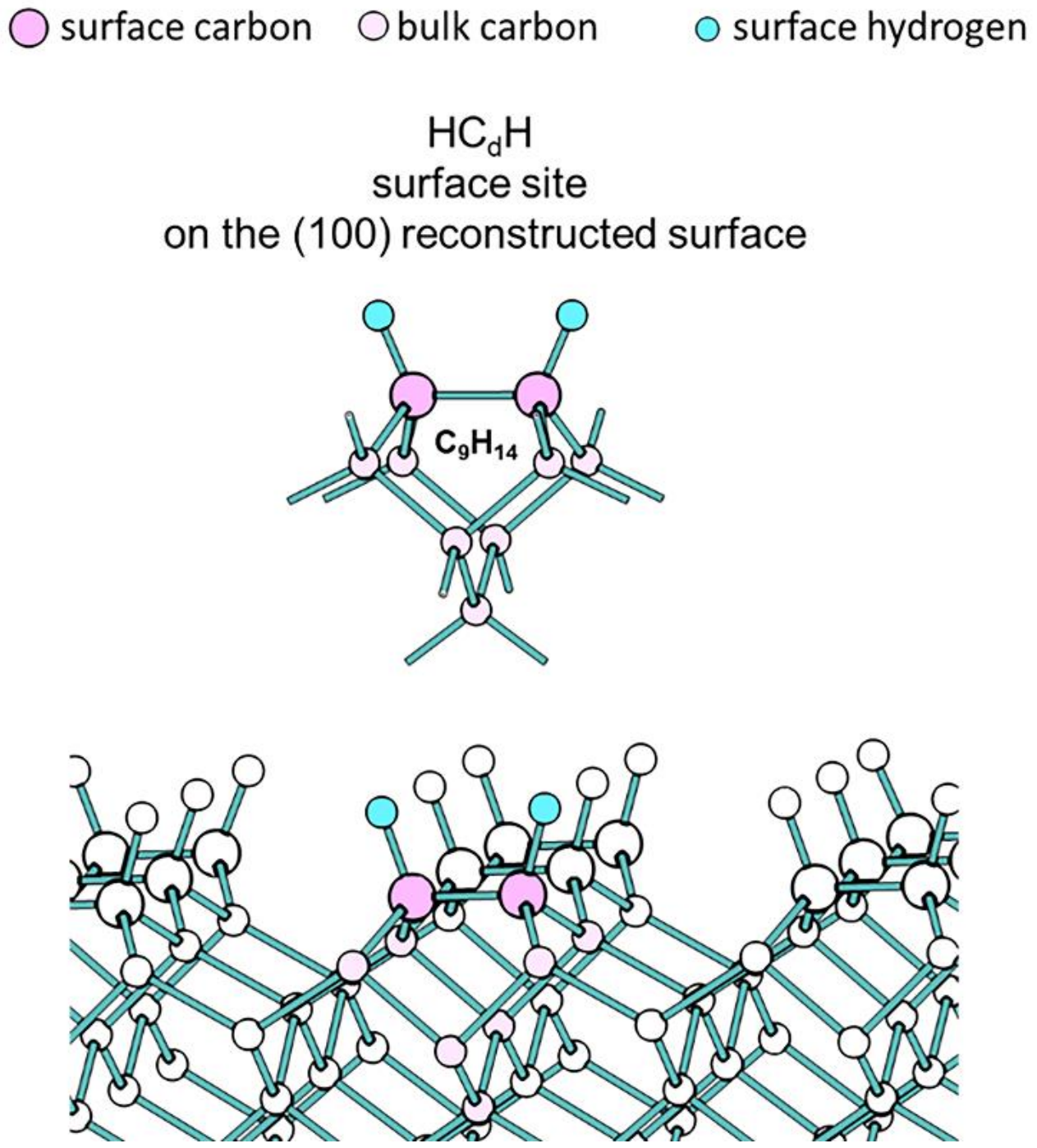

**Figure 2.** Reconstructed surface of diamond (bottom) and corresponding $C_9H_{14}$ cluster (top), mimicking the $HC_dH$ dimer on the reconstructed surface.

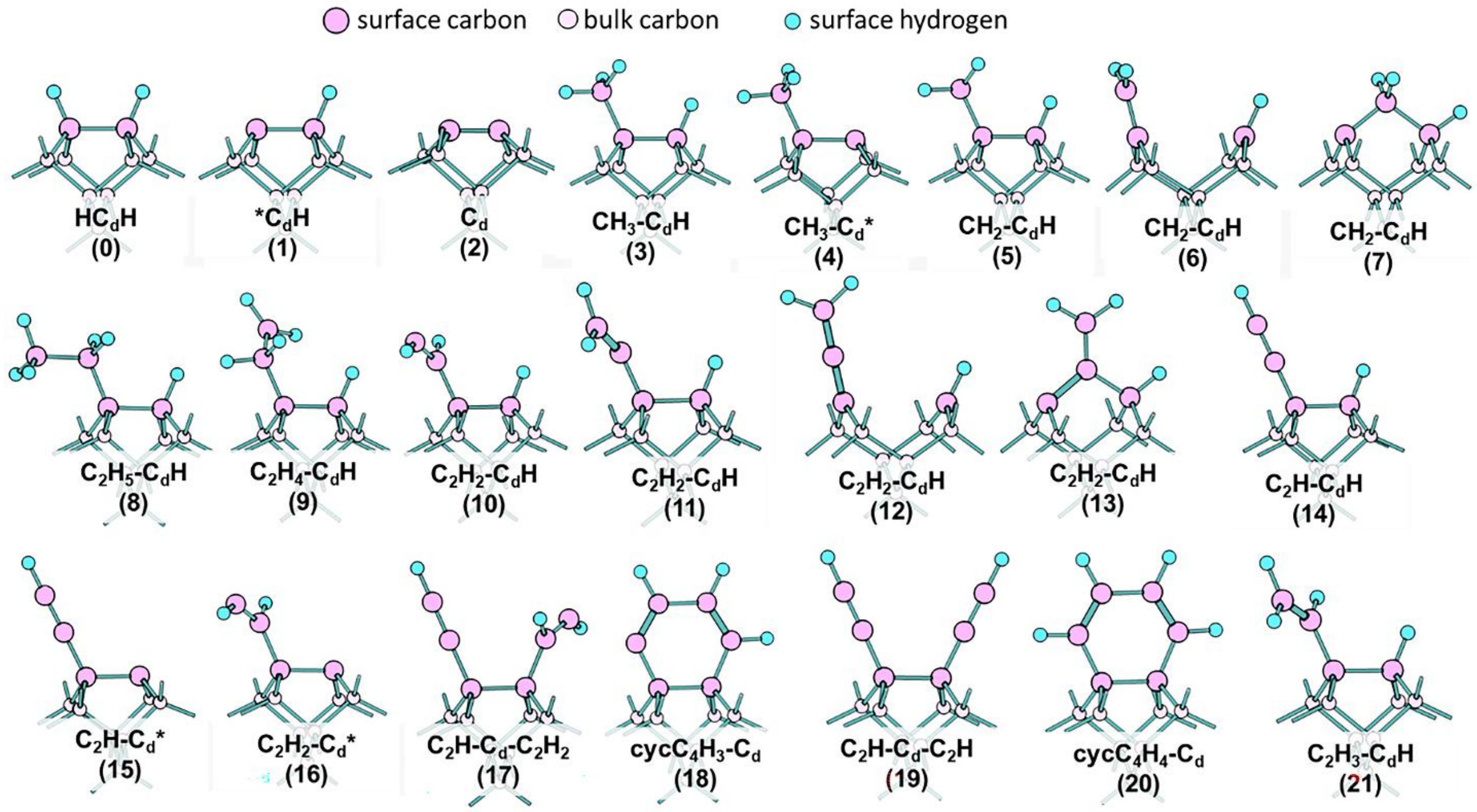

**Figure 3.** X-C$_d$-Y (m) surface sites on the diamond surface considered in the modeling.

The rate constants of the elementary surface reactions were calculated within the framework of transition state theory. The WB97X-D DFT (density functional theory) functional, along with 6-31+G(d) basis set for C-atoms and 6-311+G(d,p) for H-atoms, were chosen. The WB97X-D functional includes the dispersion and long-range correction terms, which are crucial for modeling hydrogen and π bonds. Appendix A contains the validation data for the WB97X-D functional. The calculations were performed using Gaussian 16 software [115].

We assume a rate constant to be well approximated in the Arrhenius form for first-order surface reactions, describing transformation of surface sites without direct interaction with gas-phase species:

$$k(T) = A \exp\left(-\frac{E_a}{RT}\right), \tag{1}$$

where A is the pre-exponential factor, expressed in s$^{-1}$, $E_a$ is the activation energy. For second-order reactions, which describe the transformation of surface sites through reactions with gas-phase reactants, we used dimensionless probabilities:

$$\gamma(T) = \gamma' \exp\left(-\frac{E_a}{RT}\right). \tag{2}$$

Here, $\gamma'$ were calculated using the transition state theory [116]:

$$A = \frac{k_B T}{h} \frac{Z_{vib}^{TS}(T)}{Z_{vib}^{surf}(T)}, \gamma' = \frac{\sqrt{2\pi m k_B T}}{h} p_S \frac{Z_{vib}^{TS}(T)}{Z_{vib}^{surf}(T) Z_{tot}^{gas}(T)},$$

where $k_B$ is the Boltzmann constant, $h$ is Planck's constant, $Z_{vib}^{surf}(T)$ is the vibrational partition function of a surface (or clusters which model the surface), $Z_{vib}^{TS}(T)$ is the vibrational partition function of the transition state, $Z_{tot}^{gas}(T)$ is the total partition function of gas-phase reactant, $p_S$ is surface site density, $m$ is mass of gas phase reactant.

The calculated $A$ and $\gamma'$ were fitted to pre-exponential factors of the modified Arrhenius equation:

$$A = A_0 \left(\frac{T}{298.15}\right)^n \tag{3a}$$

and

$$\gamma' = \gamma_0 \left(\frac{T}{298.15}\right)^n . \tag{3b}$$

For the barrierless reactions the transition state theory is not applicable, in this case the ab initio molecular dynamics (AIMD) with periodic boundary conditions was used, as detailed in Appendix D. The simulation was performed using VASP software [117–120]. Specifically, we explored the reaction of atomic hydrogen association with $CH_2$ absorbate that occurs through vibrational excitation. For this reaction, long-range interaction can be neglected, and, consequently, we can utilize the less time-consuming PBE0 DFT functional.

For integrated modeling of the surface and gas-phase chemistry for the diamond growth we added a reduced model for gas-phase chemistry. The main purpose of the model is to quickly evaluate the radical fluxes impinging on the diamond substrate. For this paper we only focused on hot filament CVD reactor and microwave reactor will be considered in future studies. Below we give a brief overview of our gas-phase chemistry model the details of which are provided in Appendix H. To calculate the concentrations of the gas-phase diamond precursors H, $CH_4$, $CH_3$, and $C_2H_2$ for the typical conditions of the hot-filament reactors, we solve reaction-diffusion equations for the active chemical species in $CH_4/H_2$ mixture. The chemically activated mixture is limited to the most important 12 species which include H, $CH_4$, $CH_3$, $CH_2$(S) (excited specie), $CH_2$, $C_2H_4$, $C_2H_3$, $C_2H_5$, $C_2H_2$, $C_2H_6$, $C_2H$, and $H_2$. The reactions between them and the corresponding rate constants are taken from previous thorough studies [121–123]. The total flux of each species is the sum of the diffusion and thermal diffusion fluxes as in [121]. Because in the diamond CVD process molecular hydrogen is predominantly background gas and other gas species including methane, $CH_4$, the products of its decomposition, and H atoms are highly diluted in $H_2$, it can be assumed that all these species diffuse in $H_2$ independently of each other. Their diffusion coefficients and thermal diffusion ratios were estimated as in Ref. [124] using the Chapman-Enskog kinetic theory for a binary gas mixture. Given the total pressure in the reactor p, we solve the reaction-diffusion equations along the radial coordinate r for all species except for $H_2$. The molar concentration of the molecular hydrogen, $c_{H2}(r)$, is determined from the pressure balance, i.e., from the condition $p =$ $c(r)RT(r) = const.$ Here $c(r)$ is the molar concentration of the total mixture, $T(r)$ is the known gas temperature distribution and R is the universal gas constant. The temperature distribution $T(r)$ in the reactor is approximated by the analytical formula [121] giving the temperature distribution around one hot filament. The equations for the species are discretized by the finite volume method on equidistant mesh and solved numerically up to the steady state.

# 3. Theory

## 3.1 Chemical modeling in the volume of a hot-filament reactor

Previous experimental and theoretical [121–123] studies of the gas chemistry in the hot filament reactors show that $CH_4/H_2$ mixtures with a few percent of $CH_4$ do not attain local thermodynamic equilibrium at the conditions typical for these reactors. This is primarily because the chemical sources and sinks of H atoms, which activate the gas mixture, are separated in space. The H atoms are formed by $H_2$ dissociation at the filament surface, and then they diffuse to the substrate as well as recombine in the reactor volume and the substrate surface. Since the gas near the tungsten filament is heated up to temperatures not higher than 2500 K, the amount of H atoms produced by the volume dissociation of $H_2$ molecules is small. Another factor that prevents the mixture from attaining the chemical equilibrium is that the “hot” ($T > 1700$ K) and “cold” regions of a CVD reactor are only several mm apart. The rates of production of chemical species are strong functions of the gas temperature and the created chemical species diffuse between “hot” and “cold” zones of the reactor. Therefore, the steady-state spatial profiles of H, $CH_3$, $C_2H_2$ and $CH_4$ and the other products of $CH_4/H_2$ mixture decomposition cannot be found accurately without accounting for both their chemical production and diffusive transport.

The one of the most important pyrolysis processes - conversion of $CH_4$ into $C_2H_2$ is a complex process occurring through several stages and intermediate products as shown in Figure 4. We analyzed the reaction set and determined that only 19 elementary chemical reactions between the following species H, $CH_4$, $CH_3$, $CH_2(S)$, $CH_2$, $C_2H_4$, $C_2H_3$, $C_2H_5$, $C_2H_2$, $C_2H_6$, $C_2H$, and $H_2$, with forward and reverse reactions needed to be considered. A total list of the reactions is given in Table H1 of Appendix H. These reactions include H-atom abstraction reactions leading to formation of $CH_3$ from $CH_4$ and $CH_2$ from $CH_3$. Subsequent reactions between $CH_3$ and $CH_2$ lead to the formation of $C_2H_4$ which in its turn converts to $C_2H_2$ via two H-atom abstractions and formation of $C_2H_3$ as an intermediate product [5].

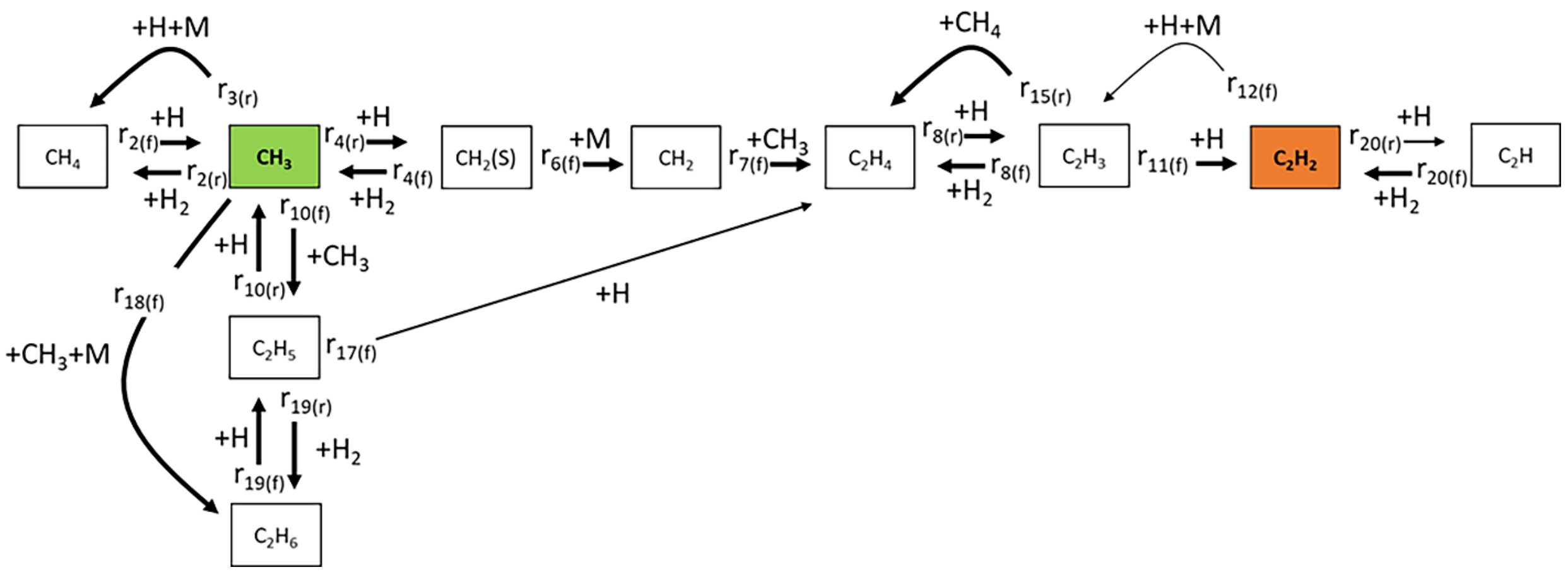

**Figure 4.** Reaction mechanism showing key reactions involved in the decomposition of methane within a $CH_4/H_2$ mixture, $r_i$ is reaction rates of gas-phase reactions, i is the number in Table H1.

## 3.2 Full and reduced surface reaction mechanisms

We developed a reaction kinetic model that incorporates crucial surface processes described previously:

- Reactive surface sites formation (dehydrogenation/hydrogenation) [4,76,87,125]
- Insertion of $CH_3$ radical into the diamond network, we call this mechanism as “dimer-opening-methylene-bridging” mechanism [108,112]
- Insertion of $C_2H_2$ molecule into the diamond network, we call this mechanism as “dimer-opening- vinylene-bridging” mechanism [105]
- Desorption of hydrocarbon chains from the surface by the β-scission mechanism [5]
- High-temperature adsorbate desorption [113,114]

Figures 5b and 5c show the insertion of $CH_3$ and $C_2H_2$ into the diamond network, respectively. $CH_3$ radical adsorbs on the reactive sites (transformation of site 1 to 3 in reaction $s_{15}$, as illustrated in Fig. 5b). The loss of the dangling bond leads to surface deactivation. Therefore, another hydrogen abstraction from $CH_3$ adsorbate (transformation site 3 to site 5 in the reaction $s_{25}$), is necessary to regenerate the dangling bond and reactivate the surface site and produce $CH_2$ adsorbate (methylene). The activated $CH_2$ adsorbate is then inserted into the diamond network. The insertion of $CH_2$ adsorbate, which is called here the “dimer-opening-methylene-bridging” mechanism, involves the following steps [88,108,109,112]: dimer opening (transformation sites 5 to 6 in the reaction $s_{43}$) and methylene bridging (transformation sites 6 to 7 in the reaction $s_{45}$). The list of all surface reactions "$s_i$" can be found in Table 1, where “i” is the serial number of the reaction in Table 1.

Dangling bond remains after the adsorption of $C_2H_2$ molecule on the reactive site (transformation site 1 to site 10 in the reaction $s_{48}$ shown in Fig. 5c); therefore, additional hydrogen abstraction from the adsorbed $C_2H_2$ molecule is not necessary for activation. The $C_2H_2$ (vinylene) adsorbate is then inserted into the diamond network similar to $CH_2$ adsorbate. Analogous to the “dimer-opening-methylene-bridging” mechanism, Skokov et al. [105] developed a “dimer-opening-vinylene-bridging” mechanism for the insertion of $C_2H_2$ into the diamond network. The insertion involves the following steps: hydrogen migration (transformation site 10 to 11 in the reaction $s_{85}$), dimer opening (transformation site 11 to 12 in the reaction $s_{87}$), and $CH_2C$ (vinylene) bridging (transformation site 12 to 13 in the reaction $s_{89}$). These insertions ultimately contribute to diamond growth. The final reaction product of the insertions, site 7 and site 13 (see Fig.3), undergo reconstruction, resulting in the initial $HC_dH$ sites 0 being reproduced on the newly deposited layer. We assume rapid reconstruction; therefore, this step is not considered in this study.

Both the insertion of $CH_3$ and $C_2H_2$, and the subsequent growth, initiate at the $*C_dH$ (1) reactive sites. The contribution of these reactive sites to the diamond growth is crucial and, therefore, studied in great details here. The formation and removal of the reactive site involve dehydrogenation and hydrogenation, respectively, as illustrated in Fig. 5a. It is well established that atomic hydrogen generates the reactive site through hydrogen abstraction (transformation site 0 to 1 in the reaction $s_1$) and simultaneously removes them via hydrogen addition (transformation site 1 to 0 in the reaction $s_2$). In addition, we consider the removal of the reactive site via well-known dissociative chemisorption of $H_2$ molecule ($s_3$ reaction).

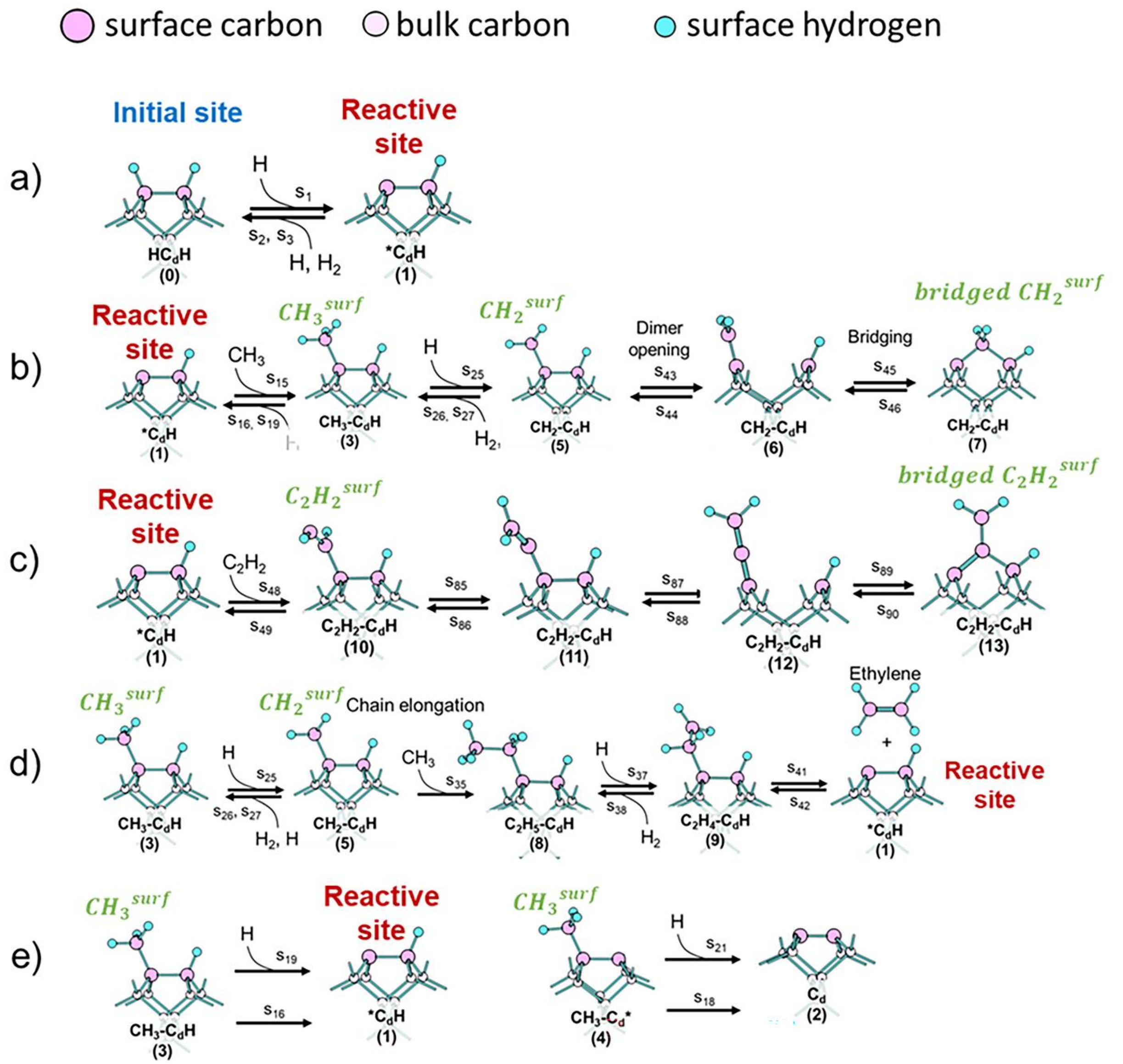

**Figure 5.** Reaction mechanisms describing the reactive site production (a), “dimer-opening-methylene-bridging” growth mechanism (b), “dimer-opening-vinylene-bridging” growth mechanism (c), desorption of hydrocarbon chains from the surface by the β-scission (d), thermal desorption and $CH_3$ adsorbate abstraction surface passivation (e). The surface reactions "$s_i$" are given in Table 1.

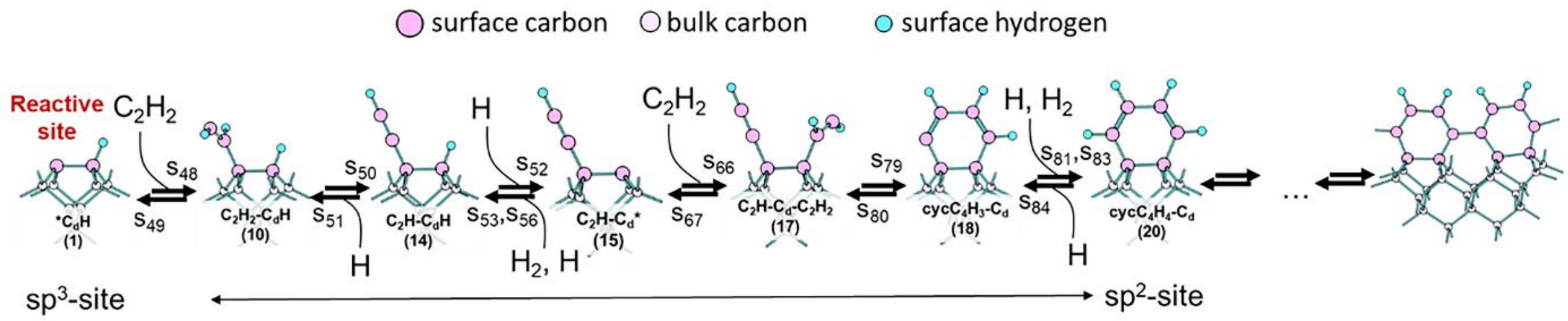

**Figure 6.** Reaction mechanism describing the $sp^2$-phase nucleation/decomposition on the carbon dimers of (100) reconstructed surface. The surface reactions "$s_i$" are given in Table 1.

The important mechanism of hydrocarbon chain desorption through β-scission (see Appendix F) was proposed by Butler and Oleynik in Refs. [5,126]. This mechanism prevents the elongation of hydrocarbon onto the diamond surface, as shown in Fig. 5d. The $s_{35}$ reaction, which involves the adsorption of $CH_3$ radical onto the dangling bond of $CH_2$ adsorbate, leading to the formation of a $C_2H_5$ adsorbate. Hydrogen abstraction activates the $C_2H_5$ chain (see transformation of site 8 to 9 in the reaction $s_{37}$ shown in Fig. 5d), facilitating the ethylene desorption via β-scission (see transformation of site 9 to 1 in the reaction $s_{41}$). As a result, the surface remains free of elongated hydrocarbon chains.

Battaile et al. [114] demonstrated that kinetic Monte Carlo modeling more accurately reproduces the experimental surface roughness and growth rates when reactions involving high-temperature surface cleaning of adsorbates are included. A mechanism of adsorbate removal considered in this study is shown in Fig. 5e.

*However, a detailed mechanism of $sp^2$-phase nucleation on the reactive sites has not been developed so far*. To this end, we developed a new mechanism, shown in Fig. 6 and detailed in Appendix G. Acetylene initiates the $sp^2$-phase nucleation. Two $C_2H_2$ molecules form a conjugated system (see site 18 in Fig.6), forming a $cycC_4H_4$-$C_d$ six-membered cycle, resembling a benzene ring (site 20, see Fig.6). Consequently, the $sp^2$-phase is represented by the stable site 20. This $sp^2$-site exhibits low reactivity, similar to benzene. Even if activated (by H), further growth on this site will contribute to the $sp^2$-phase growth, not diamond. Acetylene molecules convert the diamond reactive sites into the less reactive $sp^2$-sites, passivating the surface. *The $sp^2$-site nucleation is a reversible process and is highly dependent on the $C_2H_2$ density and surface temperature.* Low temperature and high $C_2H_2$ density shift the reaction equilibrium towards the $sp^2$-sites, making the $sp^2$-phase a dominant process in low-temperature CVD.

Finally, the full set of 91 chemical reactions discussed above is assembled and shown in Fig. 7. Including these 91 reactions is necessary to self-consistently describe all surface processes occurring in the diamond growth. In the figure, we have highlighted the most important processes: mechanisms of diamond growth involving $CH_3$ and $C_2H_2$ insertions and the $sp^2$-phase nucleation by $C_2H_2$.

As discussed in the Introduction, the contribution of $C_2H_2$ to diamond growth is typically small compared to the contribution of $CH_3$. In contrast, the $sp^2$-phase grows from $C_2H_2$, not $CH_3$. Neglecting the contribution of $C_2H_2$ to diamond growth simplifies the model and yields a reduced reaction set (mechanism) for the diamond growth. In this reduced mechanism, $C_2H_2$ contributes solely to the $sp^2$-phase nucleation, whereas $CH_3$ contributes exclusively to the diamond growth, as shown in Fig. 8. To reduce the number of processes in the reduced mechanism, we substitute multiple surface reactions by effective processes which are a result of many elementary surface reactions representing initial site transformation into the final site with an effective rate of the all intermediate elementary steps combined.

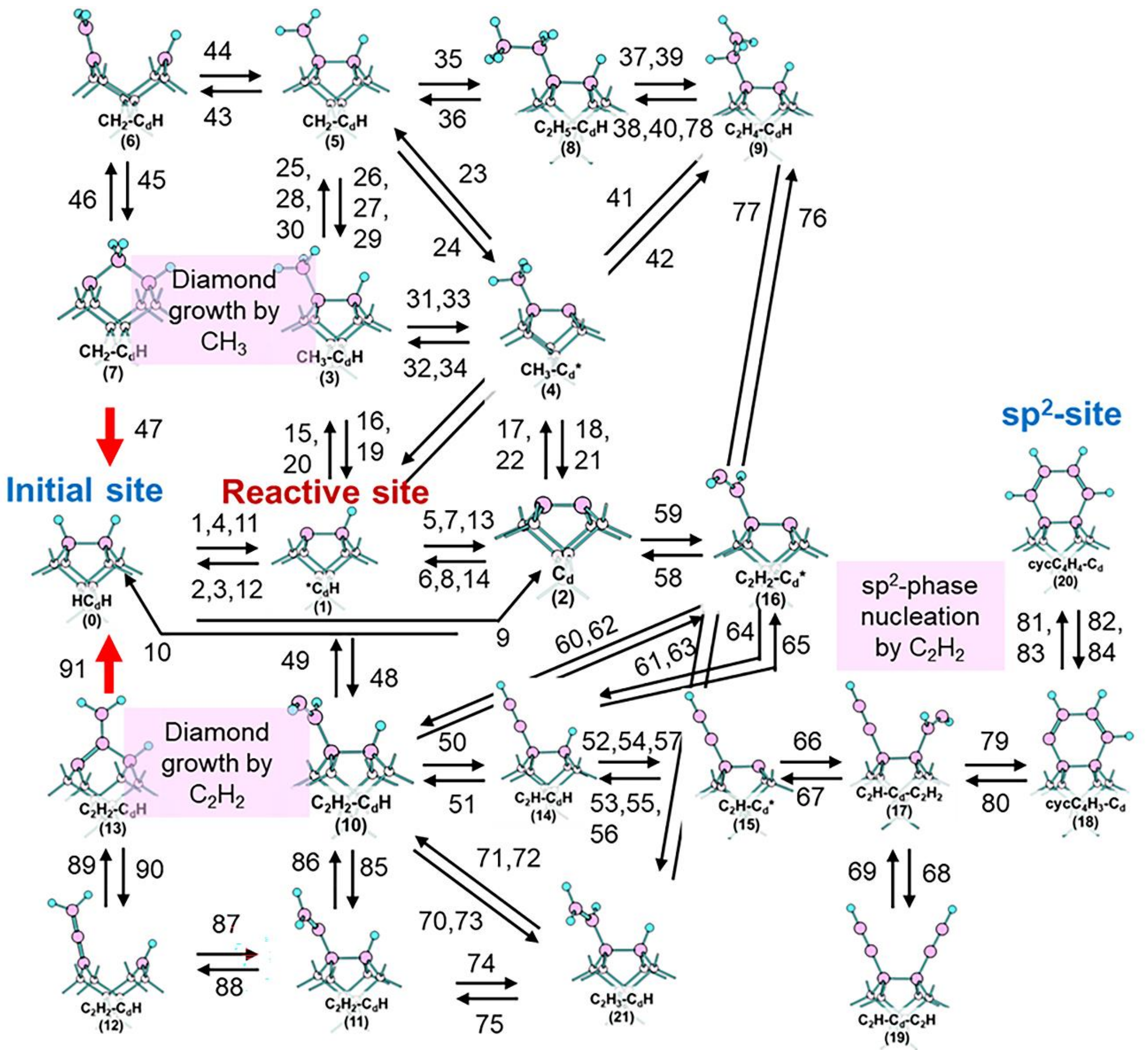

**Figure 7.** The full chemical reactions network of 21 surface sites consisting of the 91 surface reactions. The numbers adjacent to the arrows denote the serial numbers of the surface reactions in Table 1. The red arrows denote the reactions of growth reproducing the initial site 0.

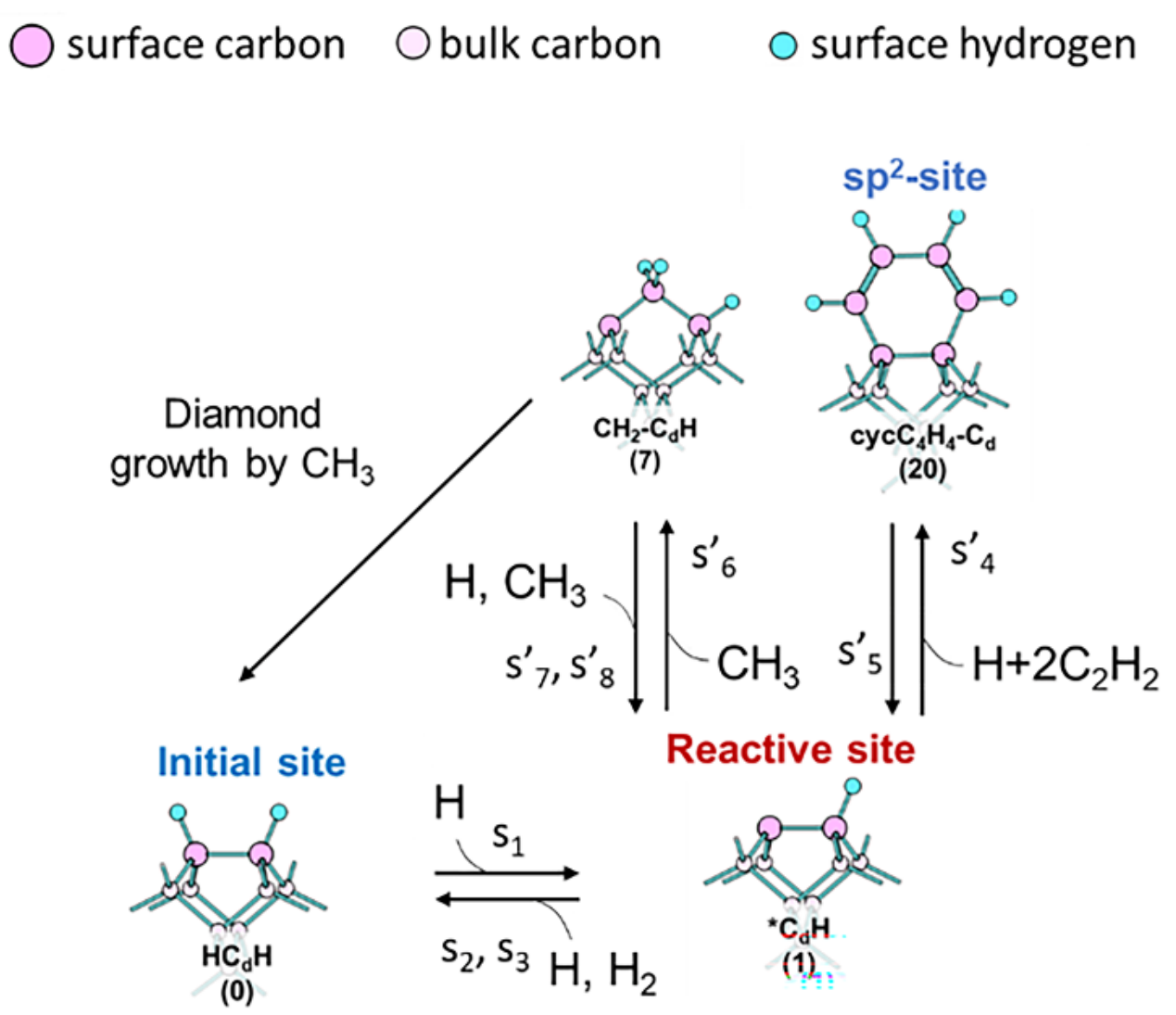

**Figure 8.** The reduced chemical reactions network composed of eight surface reactions. The mechanism includes elementary reactions (the same as in the full model) responsible for the formation and loss of reactive sites ($s_{1\text{-}3}$), effective reactions (marked by an apostrophe) leading to reactive site passivation by $C_2H_2$ ($s'_{4\text{-}5}$), and reversible transformation of site 1 to 7 ($s`_{6\text{-}8}$). Probabilities for the elementary reactions $s_{1\text{-}3}$ are listed in Table 1, the rate constant of the effective reactions $s`_{4\text{-}8}$ are calculated from the full reaction kinetic modeling as discussed in Appendix B and E.

## 3.3 Models for the reactive site fractions and growth rates

As shown earlier [121] the diamond growth rate is proportional to the reactive site fraction $\theta_1$. According to the developed mechanism in Section 3.2, acetylene can produce a benzene-like ring on the (100) diamond surface, closing the reactive sites $\theta_1$. Taking into account this effect of $C_2H_2$ closing of the reactive sites, we have derived the equation for the fraction $\theta_1$ (details are given in Appendix B), which reads

$$\theta_1 = \frac{\gamma_1}{\gamma_1 + 1 + \gamma_3 \dfrac{1-\alpha}{2\sqrt{2\alpha}} + \dfrac{2\alpha}{1+\alpha}\gamma_1 \dfrac{k'_4}{k'_5} n^2_{\mathrm{C2H2}} n_0}, \tag{4}$$

where $n_0$ is the number density of $H_2$ in the feedstock gases and

$$\alpha \equiv \frac{n_{\mathrm{H}}}{2n_{\mathrm{H}_2} + n_{\mathrm{H}}}, \tag{5}$$

$$\frac{k`_4}{k`_5} \approx 5.0 \times 10^{-67} \exp\left(\frac{4.5eV}{RT}\right), cm^9.$$

As detailed in Appendix D, the revised model shows that the growth rate is actually proportional to $\theta_7$, not $\theta_1$. The difference in the two models is not significant at temperatures below 1200K, where $\theta_1$ and $\theta_7$ are approximately equal. Above 1200K, the distinction between $\theta_1$ and $\theta_7$ becomes significant because $\theta_7$ and, correspondingly, the growth rate begins to decrease, while $\theta_1$continues to increase. Site $\theta_7$ consists of site $\theta_1$ with carbon adatom incorporated into its network. Above 1200K, the reverse reactions that lead to the extraction of the carbon adatom and their subsequent desorption become significant, resulting in a decrease in $\theta_7$. The fraction $\theta_7$ is given by

$$\theta_7 = \frac{k_6'/k_8'}{\frac{n_{\mathrm{H}}}{n_{\mathrm{CH3}}} k_7'/k_8' + 1} \theta_1 \tag{6}$$

where

$$k_6'/k_8' = 1.1,$$

$$k_7'/k_8' = 7.0 \times 10^{18} \exp\left(-\frac{5.1eV}{RT}\right).$$

As shown in Appendix E, the growth rate is given by

$$GR \approx \gamma_{CH3} \frac{v_{th}^{CH3}}{4} n_{CH3} \frac{k_6'/k_8'}{\frac{n_{\mathrm{H}}}{n_{\mathrm{CH3}}} k_7'/k_8' + 1} \theta_1 \frac{M}{p * N_A}, \tag{7}$$

where $v_{th}^{CH3}$ is the thermal velocity and $n_{CH3}$ is the number density of $CH_3$ radical; $M$ is the molar mass, $p$ is bulk density of diamond, $N_A$ is the Avogadro number. Substituting Eq. (4) for $\theta_1$, into Eq. (7) gives

$$GR \approx \gamma_{CH3} \frac{v_{th}^{CH3}}{4} n_{CH3} \frac{\frac{k_6'}{k_8'}}{\frac{\frac{n_{\mathrm{H}}}{n_{\mathrm{CH3}}} k_7'}{k_8'} + 1} \frac{\gamma_1}{\gamma_1 + 1 + \gamma_3 \frac{1-\alpha}{2\sqrt{2\alpha}} + \frac{2\alpha}{1+\alpha} \gamma_1 \frac{k'_4}{k'_5} n_{\mathrm{C2H2}}^2 n_0} \frac{M}{p * N_A} \tag{8}$$

where, as detailed in Appendix E,

$$\gamma_{CH3} = 0.85.$$

In Eq. 8, the deleterious effect of $C_2H_2$ on the diamond growth rate is accounted for by including the effective reactions s'$_4$ and s'$_5$ shown in Fig. 8.

## 4. Results and discussions

Before we discuss the surface chemistry model, let us summarize the important volumetric physical and chemical processes which govern $H_2$ dissociation degree and the concentrations of $CH_3$ and $C_2H_2$ in a hot-filament reactor, as follows from the gas phase chemistry simulations (see Appendix H).

The qualitative schematics of these processes is shown in Fig. 9. The chemical activation of the mixture occurs by the $H_2$ dissociation on the filament surface, where H atoms are produced. The H atoms diffuse to the substrate on which they recombine, whereas the $H_2$ dissociation and the H-atoms recombination in the gas volume do not contribute significantly to the balance of H-atoms (see Appendix H).

In a hot zone near the filament, the H atoms activate the $CH_4$ decomposition. The H abstraction reactions result in the formation of $CH_2$ radicals through the following chain $CH_4 \rightarrow CH_3 \rightarrow CH_2(S) \rightarrow CH_2$. The reactions between the $CH_2$ and $CH_3$ radicals lead to the formation of $C_2H_4$ and $C_2H_2$ via the following chemical reactions $C_2H_4 \rightarrow C_2H_3 \rightarrow C_2H_2$ [5]. The resulting $CH_3$ and $C_2H_2$ molecules diffuse from the "hot" to "colder" parts of the reactor, where $CH_3$ is converted back to $CH_4$ by the reactions $CH_3 + H_2 \rightarrow CH_4 + H$ or $H + CH_3 + H_2 \rightarrow CH_4 + H_2$. The $CH_4$ molecules diffuse in the opposite direction, from the substrate to the "hot" zone near the filament (see Appendix H). This qualitative description is similar to that obtained previously in [123] from the accurate 3D simulations and more comprehensive chemistry model.

However, because the actual geometry of the reactor is three-dimensional and whole system is interdependent (the fluxes of particles between the "hot" and "cold" zones affect the chemistry in these zones) , the full quantitative description of the reactor can be obtained only from the 3D simulations.

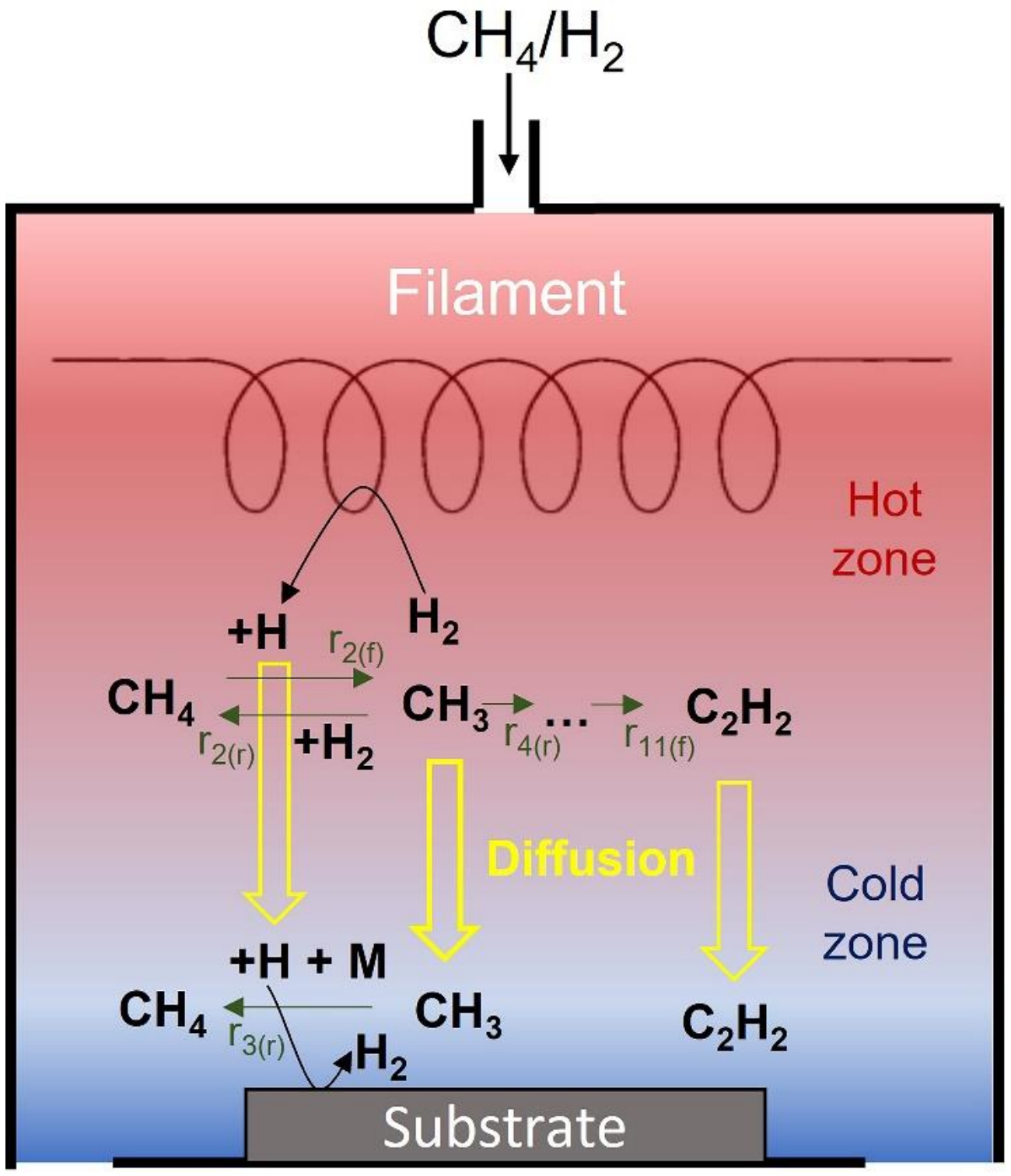

**Figure 9.** The schematic diagram of the hot-filament reactor illustrating the main chemical transformations and physical processes taking place in the reactor volume during the diamond growth.

In this paper to qualitatively but quickly describe the gas-phase chemistry, we introduced, an approximation of complex three dimensional processes, the characteristic residence time of particle movement due to diffusion in directions parallel to substrate (See Appendix H). This diffusion describes the flux of $CH_4$ from the periphery of the reactor (where its molar fraction is close to that in the initial mixture) to the reaction zones (where the chemical transformations take place) and the flux of H atoms and hydrocarbons from the reaction zones to the reactor periphery where their concentrations are zero. This approximation is akin of the so-called global model often used in plasma processing [127]. Full 3D modelling of these processes for the specific conditions of the diamond growth and the coupling of this model to the surface chemistry model are reserved for the follow-up future studies.

As our simulations show, the balance of $C_2H_2$ molecules near the substrate is determined by the their influx due to diffusion from the "hot" zone and the their outflux due to diffusion to the reactor periphery (see Appendix H). Because of the importance of the latter process, the $C_2H_2$ concentration depends on the choice of the residence time. The lower is the residence time, the lower is the $C_2H_2$ concentration near the substrate surface. This is in accord with findings of Refs. [121,122] where the 3D model for the actual reactor geometry yields a factor of 5 lower concentration of $C_2H_2$ molecules as compared with the 2D simulation results [121] due to the enhanced diffusive transport of $C_2H_2$ along the substrate surface in the 3D modeling [122]. The possibility to control the $C_2H_2$ concentration is very important for design of future hot-filament CVD reactors since the decrease in $C_2H_2$ concentration near the substrate is critical for obtaining high-quality diamond films, as the following discussion of the surface chemistry in this Section shows.

As discussed in Section 3.3, the diamond growth rate is proportional to the fraction of the reactive site, denoted as $\theta_1$. Therefore, understanding the critical factors affecting on the formation of these reactive sites is important. While this paper primarily focuses on growth on the (100) facet, it also presents modeling results for the formation of on the (110) and (111) facets induced by $H/H_2$ mixture (details are available in Appendices A and B). (It should be noted that exposure to hydrogen plasma roughens the (110) surface, resulting in the formation of pits faceted with the (111) orientation [128]. This effect of surface etching falls outside the scope of our research). As shown in Fig. 10 , the formation of the reactive sites is facet-selective and dependent on the $H_2$ dissociation degree, $\alpha$, on the diamond facets. The quantity of $\theta_1$ is increased in the following order of (111) > (110) > (100). The fraction $\theta_1$ increases with the hydrogen dissociation degree for all considered facets. Guillaume et al. [129] reported a similar modeling result for the (100) facet. As predicted by modeling [84,85], in both hot-filament and microwave reactors, the $H_2$ dissociation degree on the diamond surface varies from 0.001 to 0.005, within which $\theta_1$ is a linear function of $\alpha$. Consequently, the growth rate should also exhibit a linear relationship with the $H_2$ dissociation degree under typical CVD conditions.

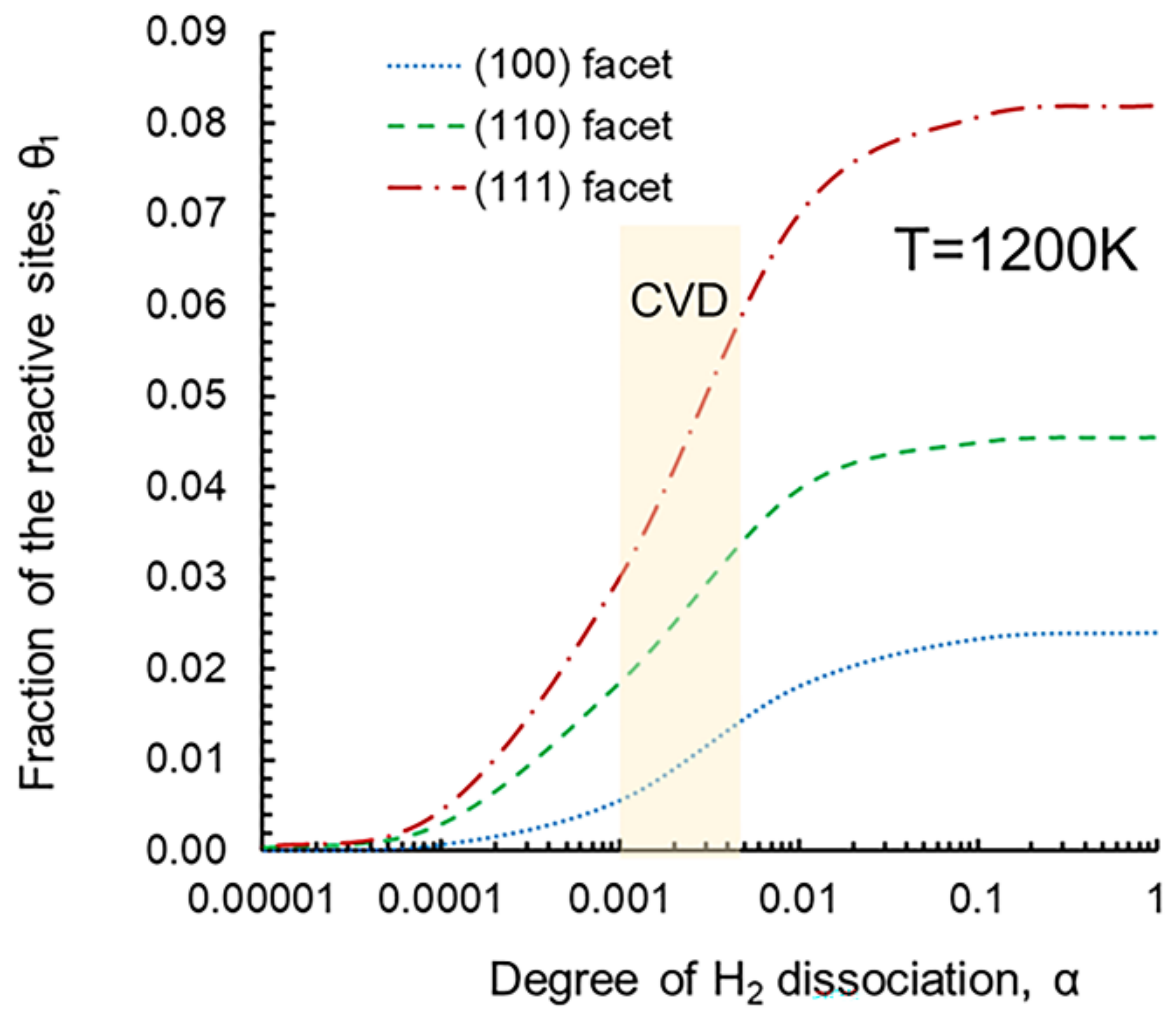

**Figure 10.** Fraction of the reactive site $\theta_1$ as a function of $H_2$ dissociation degree on the substrate level, calculated using Eq. (4) for different diamond facets neglecting the contribution of $C_2H_2$ in the diamond growth and forming the sp$^2$-phase on diamond. The yellow band indicates the typical values of $\alpha$ during conventional hot-filament and microwave CVD processes. The probabilities $\gamma_1$ and $\gamma_3$ are listed in Table 2.

**Table 2**. Calculated $\gamma_1$ and $\gamma_3$ reaction probabilities of the hydrogen abstraction and $H_2$ dissociation reactions on the hydrogenated (100), (110), and (111) diamond facets. The reaction rates are given in the form $\gamma_{1,3}(T) = \gamma_0 \left(\frac{T}{T_0}\right)^n \exp\left(-\frac{\text{Ea}}{RT}\right)$, $T_0$ is the room temperature.

| Surface orientation | | $\gamma_0$ | n | $E_a$, eV |
|---|---|---|---|---|
| 100 | $\gamma_1$ | 0.2987 | 0.5 | 0.33 |
| | $\gamma_3$ | 0.1729 | 0.0 | 0.30 |
| 110 | $\gamma_1$ | 0.2663 | 0.5 | 0.25 |
| | $\gamma_3$ | 0.1539 | 0.0 | 0.37 |
| 111 | $\gamma_1$ | 0.2606 | 0.55 | 0.19 |
| | $\gamma_3$ | 0.1415 | 0.0 | 0.34 |

To investigate the influence of $C_2H_2$ on diamond growth, we studied the reactions between $C_2H_2$ and the reactive sites on the (100) diamond surface, that lead to the sp$^2$-phase formation and reactive site deactivation (see Appendix G). Our modeling indicates the impact of $C_2H_2$ on the fraction $\theta_1$ is insignificant during conventional CVD processes, where substrate temperature is too high for the sp$^2$-phase nucleation. The sp$^2$-phase begins to nucleate at temperatures below the critical $T_{cr}$. As shown in Appendix B, the critical temperature is given by

$$T_{cr} \approx \frac{46418}{110 - \ln[P(Torr) n_{C2H2}^2 (cm^{-6}) \alpha]}. \quad (9)$$

The critical temperature depends on the $C_2H_2$ density because the nucleation is a reversible process. Fig. 11 shows that below $T_{cr}$, the reactive sites $\theta_1$ on the diamond surface rapidly disappears. HFCVD and MWCVD reactors typically produce $C_2H_2$ at concentrations exceeding $10^{14}$ $cm^{-3}$ [84,85]. Consequently, to facilitate diamond deposition in standard HFCVD and MWCVD reactors at temperatures of 800K and lower, it is crucial to significantly suppress $C_2H_2$ generation.

The described above effect of acetylene on the diamond growth can explain experimental observations for CVD at 700K, which is discussed in the Introduction and shown in Fig. 1. MWCVD reactors using the $CH_4/H_2$ mixture produce $C_2H_2$ at high levels and corresponding $T_{cr}$ exceeds 700K, below that temperature acetylene leads to the production of $sp^2$ soot-like carbon. This is also consistent with the results from Refs. [25,68] where it has been observed that as temperature decreases, diamonds containing graphite inclusions – referred to as ballas-type diamonds [68,130] – form instead of pure diamond crystals. The addition of oxygen into the gas feed suppresses $C_2H_2$ generation, which results in reduced $sp^2$-phase deposition and better quality diamonds produced [79]. However, oxygen also consumes $CH_3$ radicals, which are the primary precursors for diamond growth. Consequently, further increasing the oxygen concentration leads to a reduction in the diamond growth rate, eventually dropping to zero. It should be noted that oxygen not only suppresses $sp^2$-phase nucleation but also cleans the surface by removing the $sp^2$-phase, as reported in Refs. [131,132].

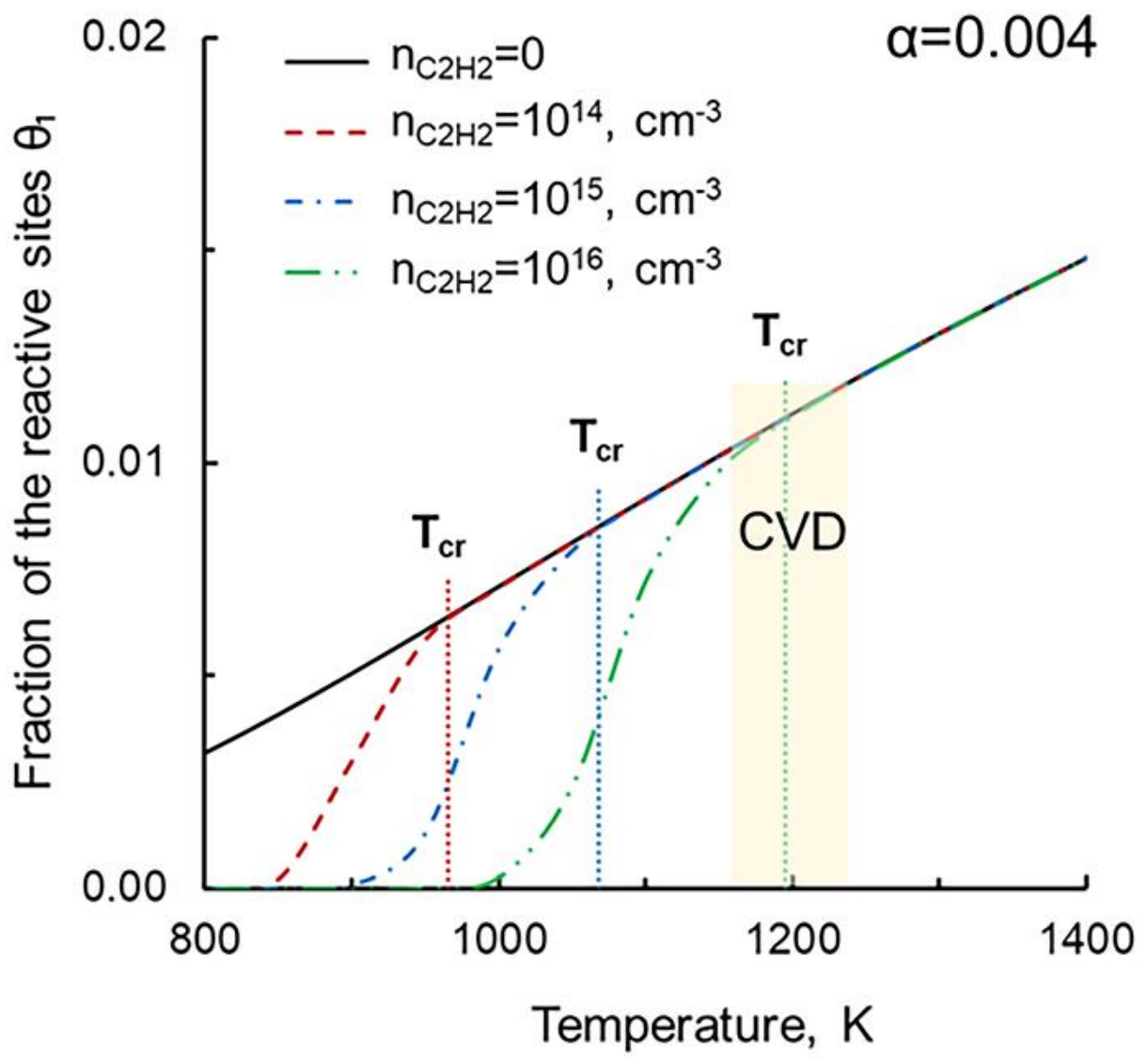

**Figure 11.** Fraction of the reactive site $\theta_1$ as a function of substrate level at different values of the $C_2H_2$ density, calculated using Eq. (4). The yellow band highlights the typical temperature values during conventional hot-filament and microwave CVD processes as calculated in Refs. [84,85].

The growth rate model, as given by Eq. (8), was validated by reproducing the experimental results reported by Kondoh et al. [38]. Growth rates as a function of substrate temperature were measured for three specific distances between the filament and the substrate (6, 8, and 10 mm). Appendix H summarizes the model that describes both gas transport and chemistry within the volume of the hot-filament reactor. Fig. 12 shows a good agreement between the modeling and experiments.

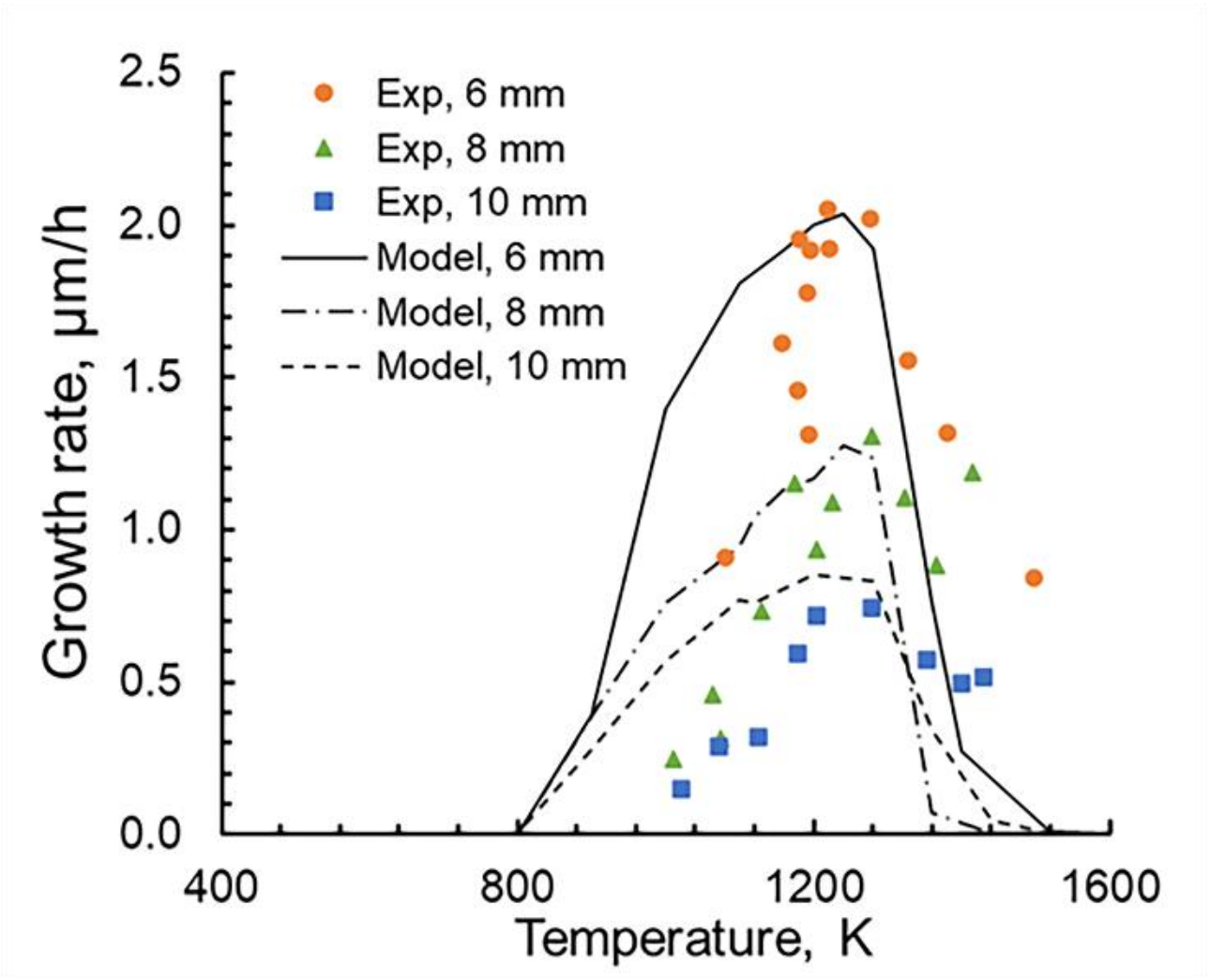

**Figure 12.** Comparison of calculated using Eq. (8) and measured [38] diamond growth rates. The gas-phase densities for Eq. (8) were calculated as described in Appendix H.

Fig. 13 presents the modeling results, which predict the surface composition in hot-filament and microwave reactors, and for a limiting case of an ideal reactor that produces only H and $CH_3$ without generating $C_2H_2$. The temperature range is divided into blue, green, and red zones. In the blue zone, $C_2H_2$ completely covers the diamond surface with the $sp^2$-phase, preventing any deposition. In the red zone, high-temperature adsorbate desorption prevents diamond growth (see Appendix D). The green zone, above the critical temperature, represents the temperature range for diamond growth. According to Eq. (9), the growth window narrows at higher $C_2H_2$ densities. Mankelevich's modeling indicates that the microwave reactor produces approximately 400 times more $C_2H_2$ than the hot-filament reactor [84,85] resulting in a narrower growth window for the former. For MWCVD, the critical temperature closely approaches the peak temperature; therefore, at temperatures near the growth rate peak, the diamond surface is partly covered by the $sp^2$-phase. It implies that for high-quality diamond CVD, the growth temperature should be set slightly above the peak value. Additionally, even in the absence of $C_2H_2$, the growth rate decreases as temperature decreases, limited by the rate of reactions that form reactive sites through the hydrogen abstraction mechanism, see Fig. 13c.

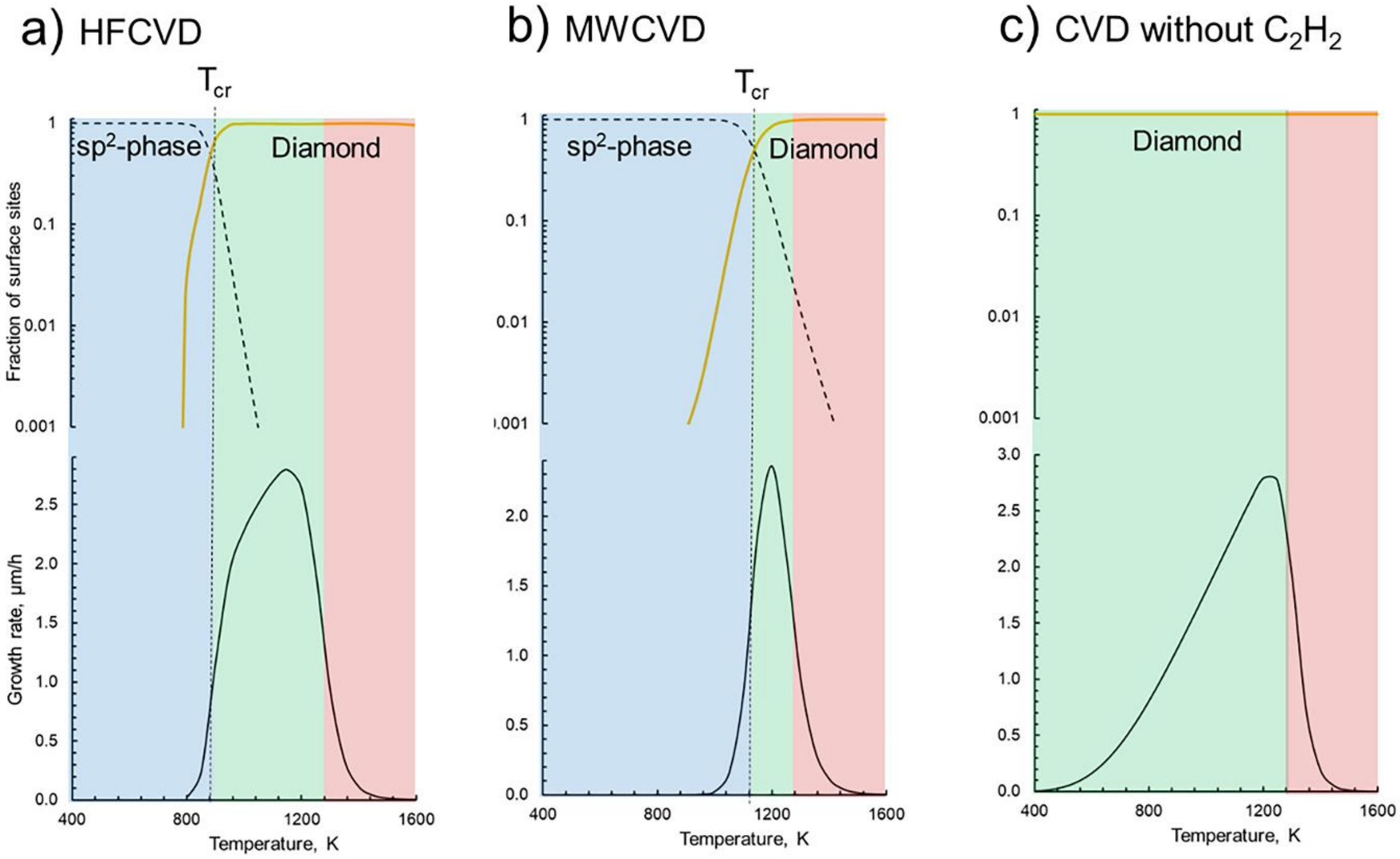

**Figure 13.** The calculated fractions of sp$^2$ ($\theta_{20}$) and diamond ($\theta_0$) surface sites, represented by a black dashed line and a yellow solid line, respectively, in the top section. The bottom section displays the calculated diamond growth rates for hot-filament chemical vapor deposition (HFCVD) (a), microwave-enhanced chemical vapor deposition (MWCVD) (b), and in a reactor that does not produce $C_2H_2$ at all. The densities of the gas-phase reactants were sourced from Refs. [84,85].

Consequently, the production of acetylene is highly undesirable in low-temperature diamond CVD processes. In standard CVD processes, the presence of acetylene is less critical. Eliminating $C_2H_2$ production is a necessary condition for successful low-temperature CVD. However, even in the absence of acetylene, the growth rate still declines as the temperature decreases.

## 5. Conclusion

We developed a detailed reaction kinetic model that predicts the composition of (100) diamond surfaces and growth rates across a broad temperature range. In agreement with previous models, at temperatures below 1200K, the growth rate is proportional to the fraction of reactive sites. To clarify the role of $C_2H_2$ in diamond CVD, a new mechanism of sp$^2$-phase nucleation on the (100) surface has been developed. While previous reports indicated that the contribution of $C_2H_2$ to diamond growth is minor at temperatures around 1200K, our findings reveal that below a critical temperature, $C_2H_2$ acts as a precursor for sp$^2$-phase growth, rather than diamond. Consequently, during standard CVD, the production of $C_2H_2$ is not critical. However, in low-temperature CVD, the generation of $C_2H_2$ is highly undesirable. Adding $O_2$ to the $CH_4/H_2$ reactive mixture probably transforms carbon species into CO, rather than $C_2H_2$. Removing $C_2H_2$ from the reactive mixture suppresses $sp^2$-phase growth, which, in turn, promotes diamond growth at low temperatures.

We derived analytical formula for the growth rate Eq. (8), as a function of degree of hydrogen dissociation, $\alpha$, and concentrations of methil , $n_{CH3}$, and acetylene, $n_{C2H2}$, and substrate temperature. Consistent with experimental data, the primary precursor of diamond growth is $CH_3$ radical. In contrast, acetylene passivates the (100) surface below the critical temperature. The critical temperature, given by Eq. (9), is a function of $\alpha$ and $n_{C2H2}$.

In addition, modeling demonstrates that the formation of reactive sites depends on crystal facets, thus offering new insights into the mechanisms underlying polycrystalline diamond growth. Furthermore, we found that during standard CVD, the fraction of these reactive sites correlates linearly with the degree of $H_2$ dissociation at the substrate level. This discovery highlights the critical impact of $H_2$ dissociation on the reproducibility of CVD processes.

# Acknowledgments

The research is supported by U.S. Department of Energy (DOE) under the “microelectronics co-design” research DOE national laboratory program. The authors gratefully acknowledge discussions with James E. Butler, Alastair Stacey, Nathalie de Leon, and Yevgeny Raitses. This research used computing resources on the Princeton University Adroit Cluster and Stellar Cluster.

**Table 1**. List of reactions of the comprehensive reaction kinetic model predicting surface composition during diamond CVD. The rate constants ($s^{-1}$) and dimensionless probabilities of surface reactions are presented in the modified Arrhenius forms: $k_i = A_0(\frac{T}{298.15})^n \exp(-\frac{E_a}{RT})$ and $\gamma_i = \gamma_0(\frac{T}{298.15})^n \exp(-\frac{E_a}{RT})$, respectively.

| $s_i$ | Reactions | $\gamma_0$ or $A_0$ | n | $E_a$, eV |
|---|---|---|---|---|
| Hydrogenation/dehydrogenation | | | | |
| 1 | $H + \theta_0 = H_2 + \theta_1$ | 0.2987 | 0.50 | 0.33 |
| 2 | $H + \theta_1 = \theta_0$ | 1 | 0 | 0 |
| 3 | $H_2 + \theta_1 = H + \theta_0$ | 0.1729 | 0.00 | 0.30 |
| 4 | $\theta_0 = H + \theta_1$ | $8.59*10^{13}$, 1/s | 2.30 | 4.40 |
| 5 | $H + \theta_1 = H_2 + \theta_2$ | 0.2939 | 0.55 | 0.16 |
| 6 | $H_2 + \theta_2 = H + \theta_1$ | 0.1817 | 0.00 | 1.11 |
| 7 | $\theta_1 = H + \theta_2$ | $8.59*10^{13}$, 1/s | 2.30 | 3.46 |
| 8 | $H + \theta_2 = \theta_1$ | 1 | 0 | 0 |
| 9 | $\theta_0 = H_2 + \theta_2$ | $6.76*10^{12}$, 1/s | 1.40 | 5.67 |
| 10 | $H_2 + \theta_2 = \theta_0$ | 0.1014 | -1.00 | 2.21 |
| 11 | $CH_3 + \theta_0 = CH_4 + \theta_1$ | 0.0022 | 1.50 | 0.41 |
| 12 | $CH_4 + \theta_1 = CH_3 + \theta_0$ | 0.0309 | 2.00 | 0.44 |
| 13 | $CH_3 + \theta_1 = CH_4 + \theta_2$ | 0.0021 | 1.50 | 0.21 |
| 14 | $CH_4 + \theta_2 = CH_3 + \theta_1$ | 0.0314 | 2.00 | 1.22 |
| $CH_3$ adsorption/desorption & $CH_2$ adsorbate formation | | | | |
| 15 | $CH_3 + \theta_1 = \theta_3$ | 0.0001 | 0 | 0 |
| 16 | $\theta_3 = CH_3 + \theta_1$ | $2.46*10^{12}$, 1/s | 1.00 | 4.07 |
| 17 | $CH_3 + \theta_2 = \theta_4$ | 0.0001 | 0 | 0 |
| 18 | $\theta_4 = CH_3 + \theta_2$ | $2.20*10^{12}$, 1/s | 1.00 | 3.08 |
| 19 | $H + \theta_3 = CH_4 + \theta_1$ | 0.2824 | 0.20 | 1.58 |
| 20 | $CH_4 + \theta_1 = H + \theta_3$ | 0.0027 | 1.00 | 1.95 |
| 21 | $H + \theta_4 = CH_4 + \theta_2$ | 0.2663 | 0.20 | 1.39 |
| 22 | $CH_4 + \theta_2 = H + \theta_4$ | 0.0028 | 1.30 | 2.75 |
| 23 | $\theta_4 = \theta_5$ | $4.20*10^{12}$, 1/s | 0.70 | 1.71 |
| 24 | $\theta_5 = \theta_4$ | $3.14*10^{12}$, 1/s | 0.35 | 1.87 |
| 25 | $H + \theta_3 = H_2 + \theta_5$ | 0.3021 | 0.40 | 0.37 |
| 26 | $H_2 + \theta_5 = H + \theta_3$ | 0.1250 | -0.80 | 0.49 |
| 27 | $H + \theta_5 = \theta_3$ | 0.1077 | -1.10 | 0 |
| 28 | $\theta_3 = H + \theta_5$ | $8.74*10^{12}$, 1/s | 1.00 | 4.26 |
| 29 | $CH_3 + \theta_3 = CH_4 + \theta_5$ | 0.0016 | 1.30 | 0.44 |
| 30 | $CH_4 + \theta_5 = CH_3 + \theta_3$ | 0.0158 | 1.30 | 0.62 |
| 31 | $H + \theta_3 = H_2 + \theta_4$ | 0.2987 | 0.50 | 0.33 |
| 32 | $H_2 + \theta_4 = H + \theta_3$ | 0.1729 | 0.00 | 0.30 |
| 33 | $\theta_3 = H + \theta_4$ | $8.59*10^{13}$, 1/s | 2.30 | 4.40 |
| 34 | $H + \theta_4 = \theta_3$ | 1 | 0 | 0 |
| Desorption of hydrocarbon chains by β-scission | | | | |
| 35 | $CH_3 + \theta_5 = \theta_8$ | $1.92*10^{-5}$ | 0.40 | 0 |
| 36 | $\theta_8 = CH_3 + \theta_5$ | $2.39*10^{12}$, 1/s | 1.00 | 3.77 |
| 37 | $H + \theta_8 = H_2 + \theta_9$ | 0.3021 | 0.40 | 0.37 |
| 38 | $H_2 + \theta_9 = H + \theta_8$ | 0.0293 | 0.40 | 0.49 |
| 39 | $CH_3 + \theta_8 = CH_4 + \theta_9$ | 0.0016 | 1.30 | 0.44 |
| 40 | $CH_4 + \theta_9 = CH_3 + \theta_8$ | 0.0158 | 1.30 | 0.62 |
| 41 | $\theta_9 = C_2H_4 + \theta_1$ | $3.11*10^{13}$, 1/s | 1.00 | 1.46 |
| 42 | $C_2H_4 + \theta_1 = \theta_9$ | 0.0014 | 1.75 | 0.05 |
| "Dimer-opening-methylene-bridging" | | | | |
| 43 | $\theta_5 = \theta_6$ | $4.33*10^{12}$, 1/s | 0.50 | 0.65 |
| 44 | $\theta_6 = \theta_5$ | $1.35*10^{12}$, 1/s | -0.10 | 0.42 |
| 45 | $\theta_6 = \theta_7$ | $1.07*10^{12}$, 1/s | -0.40 | 0.57 |
| 46 | $\theta_7 = \theta_6$ | $8.84*10^{12}$, 1/s | 1.30 | 1.28 |
| 47 | $CH_3 + \theta_7 = \theta_0$ | 0.0001 | 0 | 0 |
| $C_2H_2$ adsorption/desorption and their adsorbate hydrogenation/dehydrogenation | | | | |
| 48 | $C_2H_2 + \theta_1 = \theta_{10}$ | 0.0133 | 1.60 | 0.12 |
| 49 | $\theta_{10} = C_2H_2 + \theta_1$ | $7.82*10^{13}$, 1/s | 1.80 | 1.68 |
| 50 | $\theta_{10} = H + \theta_{14}$ | $6.20*10^{12}$, 1/s | 1.60 | 1.79 |
| 51 | $H + \theta_{14} = \theta_{10}$ | 0.2438 | 0.20 | 0.25 |
| 52 | $H + \theta_{14} = H_2 + \theta_{15}$ | 0.2987 | 0.50 | 0.33 |
| 53 | $H_2 + \theta_{15} = H + \theta_{14}$ | 0.1729 | 0.00 | 0.30 |
| 54 | $CH_3 + \theta_{14} = CH_4 + \theta_{15}$ | 0.0022 | 1.50 | 0.41 |
| 55 | $CH_4 + \theta_{15} = CH_3 + \theta_{14}$ | 0.0309 | 2.00 | 0.44 |
| 56 | $H + \theta_{15} = \theta_{14}$ | 1 | 0 | 0 |
| 57 | $\theta_{14} = H + \theta_{15}$ | $8.59*10^{13}$, 1/s | 2.30 | 4.40 |
| 58 | $\theta_{16} = C_2H_2 + \theta_2$ | $1.90*10^{13}$, 1/s | 1.40 | 1.10 |
| 59 | $C_2H_2 + \theta_2 = \theta_{16}$ | 0.0036 | 1.60 | 0.53 |
| 60 | $H + \theta_{16} = \theta_{10}$ | 1 | 0 | 0 |
| 61 | $\theta_{10} = H + \theta_{16}$ | $8.59*10^{13}$, 1/s | 2.30 | 4.40 |
| 62 | $H_2 + \theta_{16} = H + \theta_{10}$ | 0.1729 | 0.00 | 0.30 |
| 63 | $H + \theta_{10} = H_2 + \theta_{16}$ | 0.2987 | 0.50 | 0.33 |
| 64 | $\theta_{16} = \theta_{14}$ | $3.75*10^{12}$, 1/s | 0.70 | 0.82 |
| 65 | $\theta_{14} = \theta_{16}$ | $6.28*10^{12}$, 1/s | 1.08 | 3.69 |
| 66 | $C_2H_2 + \theta_{15} = \theta_{17}$ | 0.076 | 1.60 | 0.07 |
| 67 | $\theta_{17} = C_2H_2 + \theta_{15}$ | $4.22*10^{13}$, 1/s | 1.60 | 1.66 |
| 68 | $\theta_{17} = H + \theta_{19}$ | $5.59*10^{12}$, 1/s | 1.50 | 1.87 |
| 69 | $H + \theta_{19} = \theta_{17}$ | 0.2141 | 0.20 | 0.26 |
| 70 | $H + \theta_{10} = \theta_{21}$ | 1.00 | 0 | 0 |
| 71 | $H_2 + \theta_{10} = H + \theta_{21}$ | 0.0759 | 1.10 | 0.29 |
| 72 | $H + \theta_{21} = H_2 + \theta_{10}$ | 1.00 | 0 | 0.62 |
| 73 | $H + \theta_{10} = \theta_{21}$ | 1.00 | 0 | 0 |
| 74 | $H_2 + \theta_{11} = H + \theta_{21}$ | 0.0377 | 1.10 | 0.33 |
| 75 | $H + \theta_{21} = H_2 + \theta_{11}$ | 1.00 | 0 | 0.45 |
| 76 | $H + \theta_{21} = \theta_9$ | 0.1409 | 0.70 | 0.22 |
| 77 | $\theta_9 = H + \theta_{21}$ | $4.74*10^{12}$, 1/s | 1.00 | 1.76 |
| 78 | $H + \theta_9 = \theta_8$ | 1.00 | 0 | 0 |
| $sp^2$-phase nucleation/decomposition | | | | |
| 79 | $\theta_{17} = \theta_{18}$ | $2.13*10^{12}$, 1/s | 0.06 | 0.36 |
| 80 | $\theta_{18} = \theta_{17}$ | $1.73*10^{13}$, 1/s | 1.60 | 2.17 |
| 81 | $H + \theta_{18} = \theta_{20}$ | 1 | 0 | 0 |
| 82 | $\theta_{20} = H + \theta_{18}$ | $9.40*10^{13}$, 1/s | 0 | 4.64 |
| 83 | $H_2 + \theta_{18} = H + \theta_{20}$ | 0.2392 | 0 | 0.25 |
| 84 | $H + \theta_{20} = H_2 + \theta_{18}$ | 0.3987 | 0.70 | 0.51 |
| "Dimer-opening-$CH_2C$-bridging" | | | | |
| 85 | $\theta_{10} = \theta_{11}$ | $1.52*10^{13}$, 1/s | 1.35 | 1.89 |
| 86 | $\theta_{11} = \theta_{10}$ | $1.21*10^{13}$, 1/s | 1.25 | 2.09 |
| 87 | $\theta_{11} = \theta_{12}$ | $7.01*10^{12}$, 1/s | 0.90 | 0.75 |
| 88 | $\theta_{12} = \theta_{11}$ | $1.61*10^{12}$, 1/s | -0.10 | 0.43 |
| 89 | $\theta_{12} = \theta_{13}$ | $8.83*10^{11}$, 1/s | -0.30 | 0.54 |
| 90 | $\theta_{13} = \theta_{12}$ | $1.02*10^{13}$, 1/s | 1.30 | 1.66 |
| 91 | $C_2H_2 + \theta_{13} = \theta_0$ | 0.0044 | 1.30 | 0.59 |

# Appendix A. DFT validation.

The WB97X-D functional was validated by comparison with available experimental data and previous simulations of the most important reactions for diamond growth, i.e., hydrogen abstraction in reactions with methane ($H + CH_4 = H_2 + CH_3$ [133–135]) and with isobutane (H + iso-$C_4H_{10}$ = $H_2$ + tret-$C_4H_9$ [103,136,137]). This approach of using an iso-$C_4H_{10}$ molecule as a model for diamond surfaces was pioneered by Frenklach and Wang [103] and then used in follow-up studies, which include modeling (Goodwin [138], Coltrin and Dandy [104], Mankelevich et al. [35,121]), and experiments (Krasnoperov [82] and Harris and Weiner [83]). Figure A1 shows that excellent agreement was achieved between the rate constants obtained using the DFT WB97X-D functional and the literature data for both reactions: $H + CH_4 \leftrightarrow H_2 + CH_3$ and H + iso-$C_4H_{10}$ $\leftrightarrow$ $H_2$ + tret-$C_4H_9$.

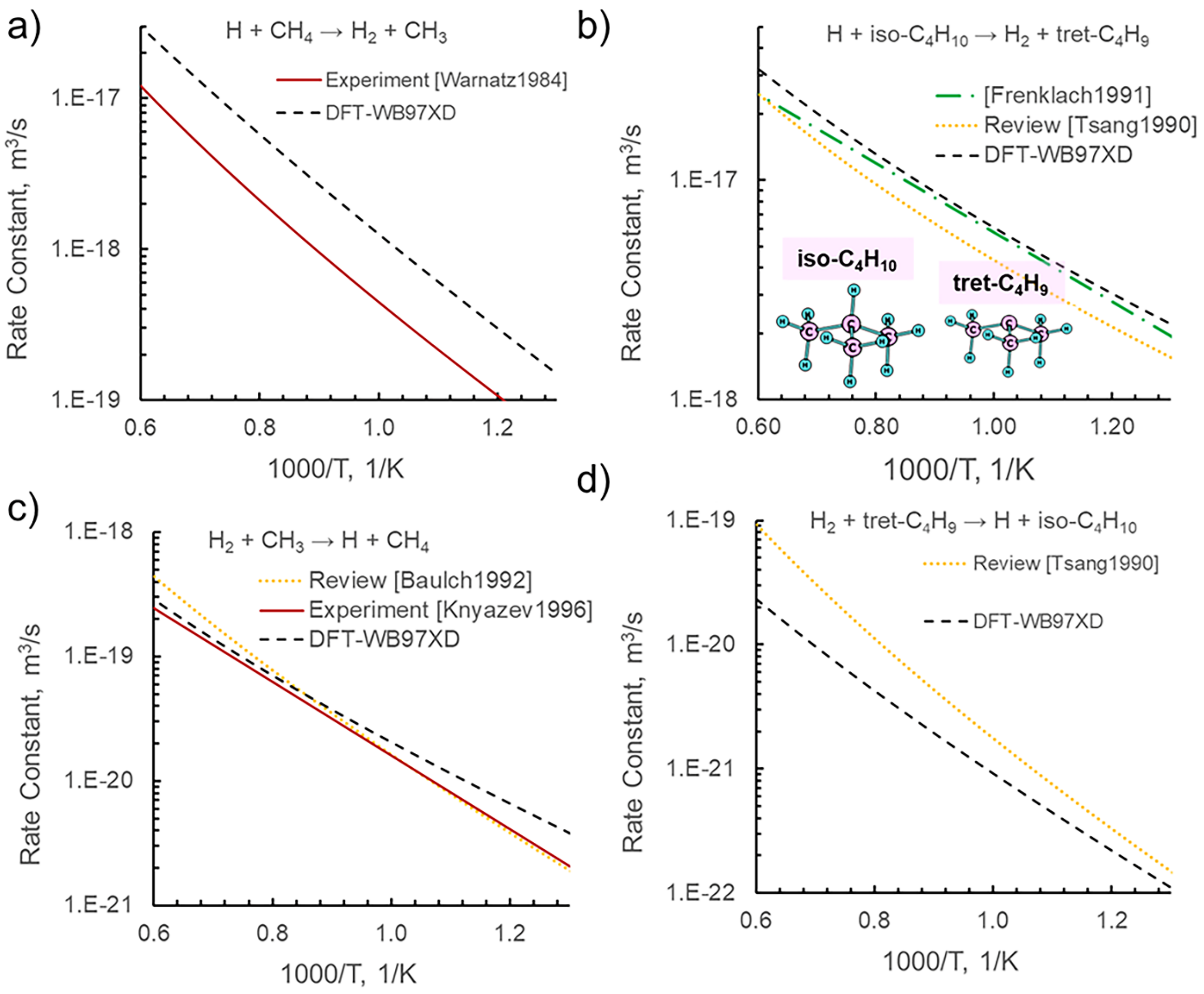

**Figure A1.** Rate constants from Refs. [103,133–136] and the rate constants calculated using the DFT WB97XD functional for the reactions: H + $CH_4$ = $H_2$ + $CH_3$ (a, c) and H + iso-$C_4H_{10}$ = $H_2$ + tret-$C_4H_9$ (b, d).

The validated WB97X-D functional was employed to simulate reactions with the $C_9H_{14}$ cluster as a substitute model for the iso-$C_4H_{10}$ molecule. This $C_9H_{14}$ cluster is designed to mimic the carbon dimer on the (100) reconstructed surface, serving as a model for the pristine surface site designated as $\theta_0$.

To accurately represent the (110) and (111) facets of diamond, we used $C_{32}H_{32}$ and $C_{43}H_{40}$ clusters, as shown in Fig. A2. These clusters were specifically chosen for the hydrogen abstraction reaction. We made sure that the size of the clusters (and the corresponding facet area) was sufficient to mitigate potential edge effects from the cluster edges being terminated by hydrogen atoms. We were increasing the facet area (and the cluster size) until the activation energy of reactions 1 and 3 became independent of the cluster size. For the (100), (110) and (111) facets, the smallest clusters for which this condition is achieved are $C_9H_{14}$, $C_{32}H_{32}$, and $C_{43}H_{40}$, respectively.

Our calculation can distinguish reactions on various diamond facets, determine the facet-selective probabilities of an H atom loss on diamond surfaces, similarly to what we did earlier for silicon in Ref. [139].

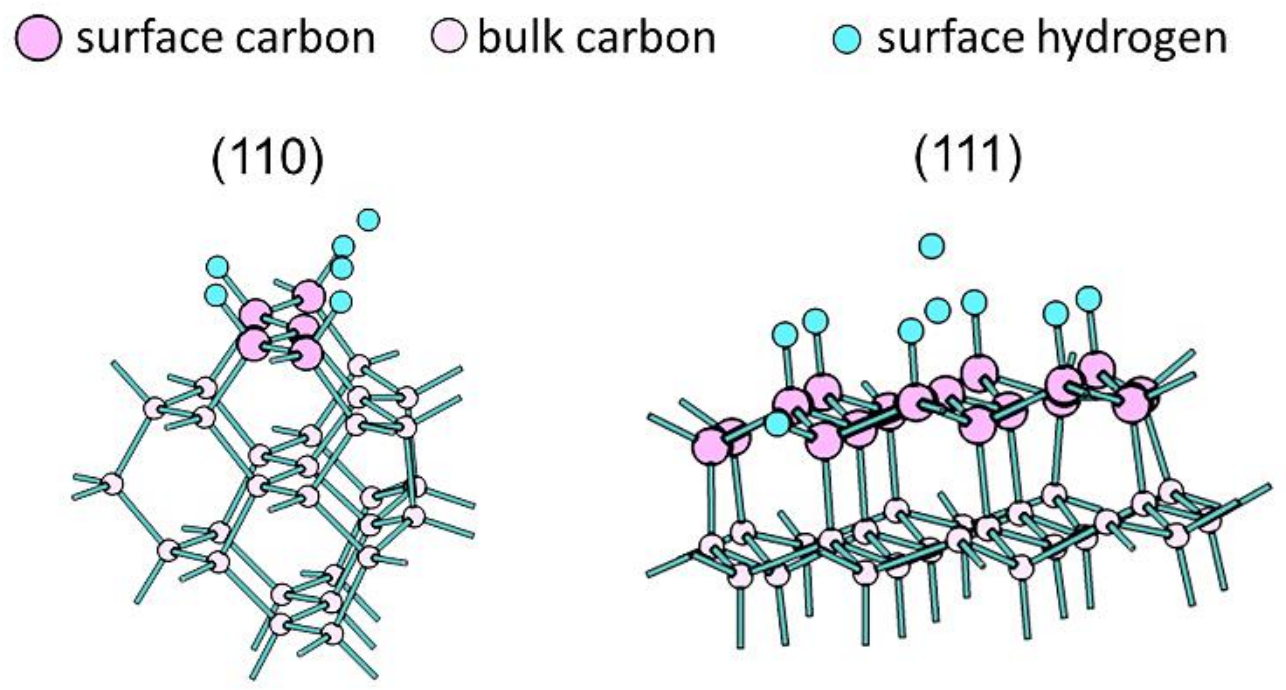

**Figure A2.** Transition states of hydrogen abstraction, $s_1$ reaction, on the (110) and (111) facets of the $C_{32}H_{32}$ and $C_{43}H_{40}$ clusters, respectively.

As shown in Fig. A3, the reaction $H_2 + \theta_1 \rightarrow H + \theta_0$, which leads to the loss of reactive sites by molecular hydrogen reaction (reaction $s_3$ in Table 1), has a high activation energy when $\theta_0$ and $\theta_1$ are modeled by the iso-$C_4H_{10}$ and tret-$C_4H_9$ molecules, respectively. This outcome differs from the results obtained using bigger cluster models ($C_9H_{14}$, $C_{32}H_{32}$, and $C_{43}H_{40}$), which predict substantially lower barriers for this reaction, making this reaction notably more significant. As a result, the fraction $\theta_1$ of the reactive sites is highly dependent on the gas-phase $H_2$ dissociation degree at the diamond surface. Further details are provided in Appendix B.

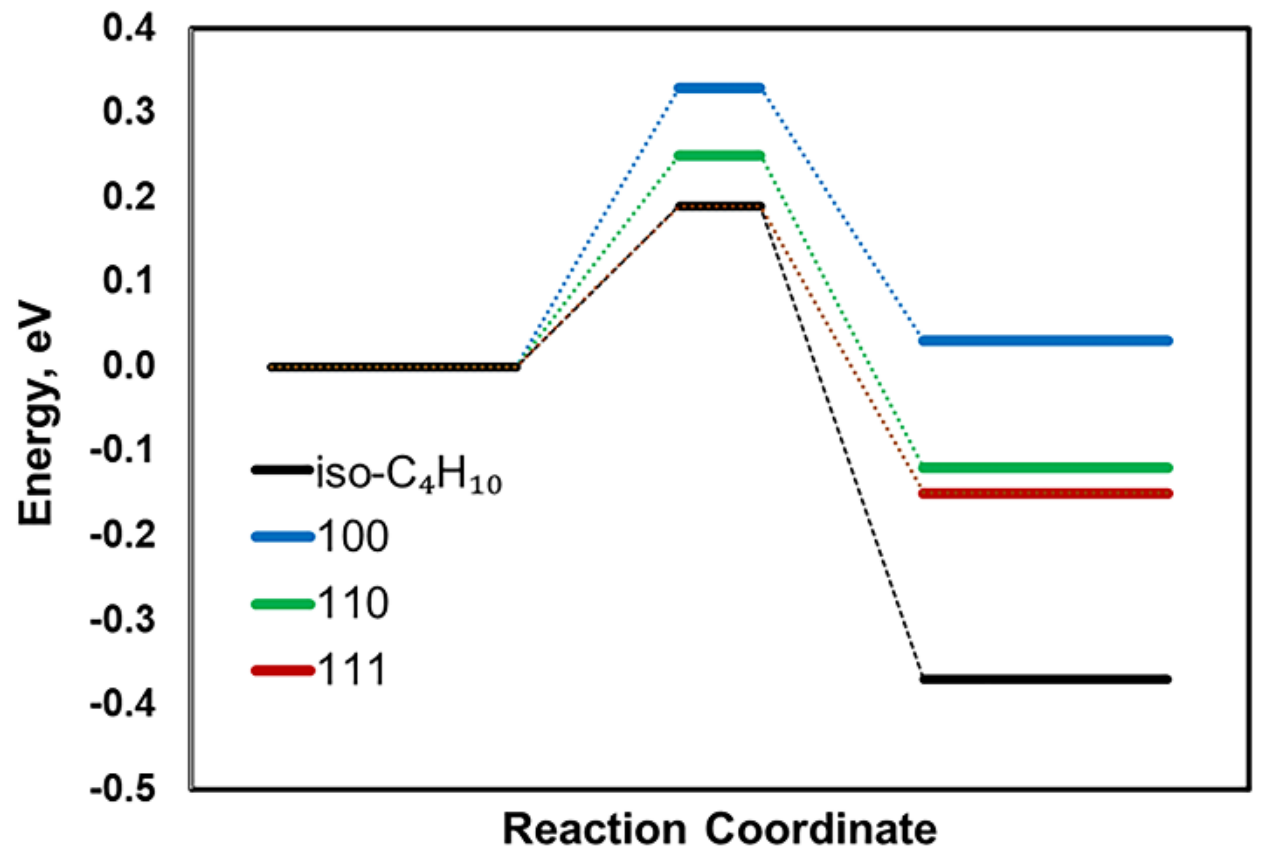

**Figure A3.** Reaction diagram for the hydrogen abstraction reaction. This diagram illustrates the reversible reaction $H + \theta_0 \leftrightarrow H_2 + \theta_1$ (reactions $s_1$ and $s_3$ in Table 1), where the surface is represented by either the iso-$C_4H_{10}$ molecule or by the bigger clusters ($C_9H_{14}$, $C_{32}H_{32}$, and $C_{43}H_{40}$) representing the (100), (110), and (111) surfaces.

Cheesman et al. [109] used the B3LYP function to calculate the potential energy surface for the insertion of the $CH_2$ adsorbate into the diamond network on the (100) surface. In our research, we employed the WB97XD functional because this functional is consistent with the result obtained by Cheesman et al. (see Fig. 4a) and accurately reproduces both the rate constant (see Fig. A1) and the activation energy (see Fig. A4b) for the hydrogen abstraction reaction.

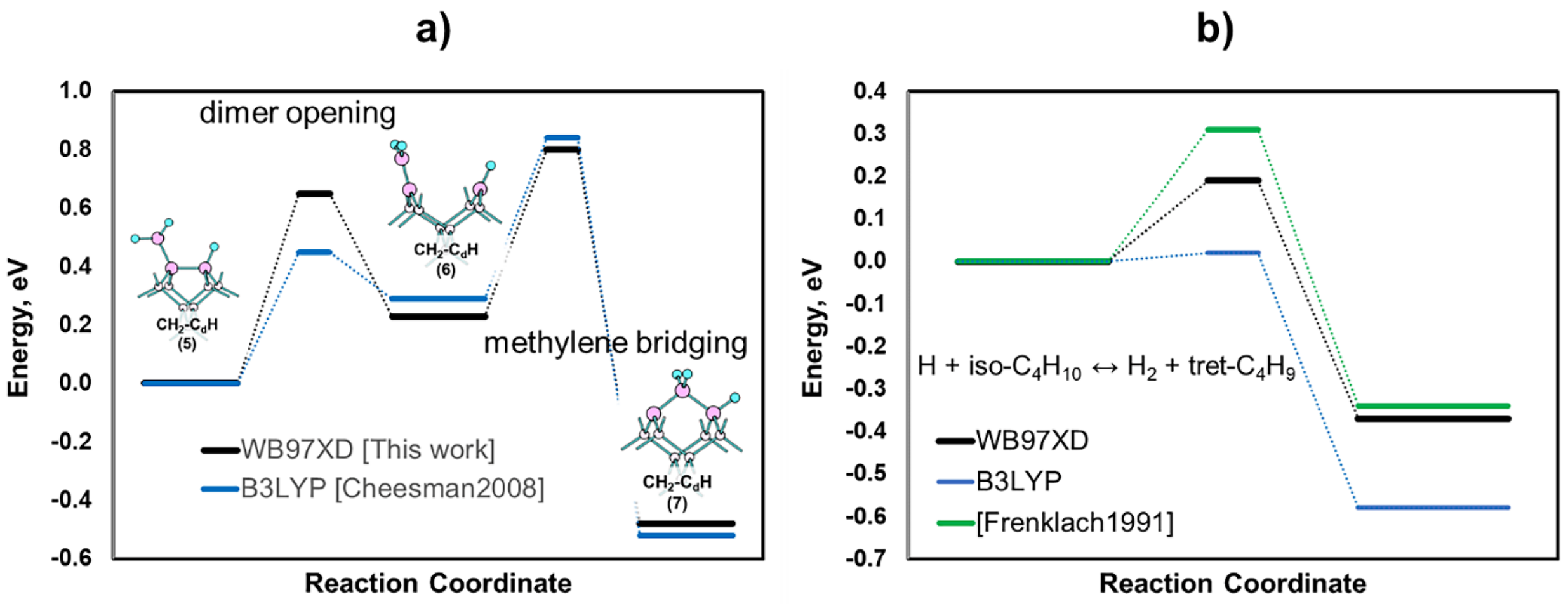

**Figure A4.** Reaction diagrams: (a) insertion of the $CH_2$ adsorbate into the diamond network through the "dimer-opening-methylene-bridging" mechanism, calculated using WB97XD and B3LYP [109] functionals (a); hydrogen abstraction reaction from the iso-$C_4H_{10}$ molecule as calculated by WB97XD and B3LYP functionals, in comparison with literature data from Ref. [103] (b).

# Appendix B. Models for the fractions of reactive sites.

## Original model

In the original model by Goodwin [138], the fraction of reactive sites $\theta_1$ was determined through a balance between the generation and deactivation of reactive sites by atomic hydrogen (reaction $s_1$ and $s_2$ in Table 1):

$$\gamma_1 \frac{v_{th}^H}{4} n_H \theta_0 - \gamma_2 \frac{v_{th}^H}{4} n_H \theta_1 = 0,$$
$$\theta_0 + \theta_1 = 1. \qquad (B1)$$

Here $p_s = (N_A \rho / M)^{2/3} = 3.1 \times 10^{15} \mathrm{cm}^{-2}$ is the surface site density, $N_A$ is the Avogadro number, $\rho$ is the bulk density of diamond, $M$ is the molar mass of diamond. According to the system of equations, it follows that the surface site density reaches equilibrium at 1200K in $4p_s/\gamma_1 v_{th}^H n_H = 2.15 \times 10^{-3} s$. Therefore, for standard CVD conditions, the surface site density can be obtained by assuming a detailed balance of creation and distraction of sites:

$$\theta_1 = \frac{\gamma_1}{\gamma_1 + \gamma_2} = \frac{\gamma_1}{\gamma_1 + 1}. \qquad (B2)$$

Here, $\gamma_1$ and $\gamma_2$ are, respectively, probabilities of the hydrogen abstraction by atomic hydrogen from the site $\theta_0$ and spontaneous adsorption of atomic hydrogen on a reactive site $\theta_1$. It is commonly assumed that $\gamma_2 \cong 1$ [76].

Goodwin calculated the probability $\gamma_1$ assuming that the rate of reaction $s_1$ is equal to the rate of the gas-phase reaction of hydrogen abstraction from an iso-$C_4H_{10}$ molecule, as discussed in Appendix A. The resulting formula for $\gamma_1(T)$ as a function of the diamond temperature is:

$$\gamma_1(T) = \frac{100}{\sqrt{T}} \exp\left(-\frac{0.32eV}{RT}\right).$$

However, this formula does not account for the dependence of $\gamma_1$ on the surface orientation which plays an important role. We calculated $\gamma_1(T)$ for three principal facets (100), (110), and (111). The results are presented in Table 2 and plotted in Fig. B1a. In this figure, we also show the results of the classical molecular dynamics simulation by Brenner et al. [140] for the (111) surface performed for the temperature range from 1200K to 1800K and approximated as $\gamma_1 = 1.162exp\,(-0.341/RT)$. From Fig. B1a, it is evident that the Goodwin's result accurately describes hydrogen abstraction probability from the (111) facet only, whereas $\gamma_1$ for the (100) and (110) surfaces are substantially lower.

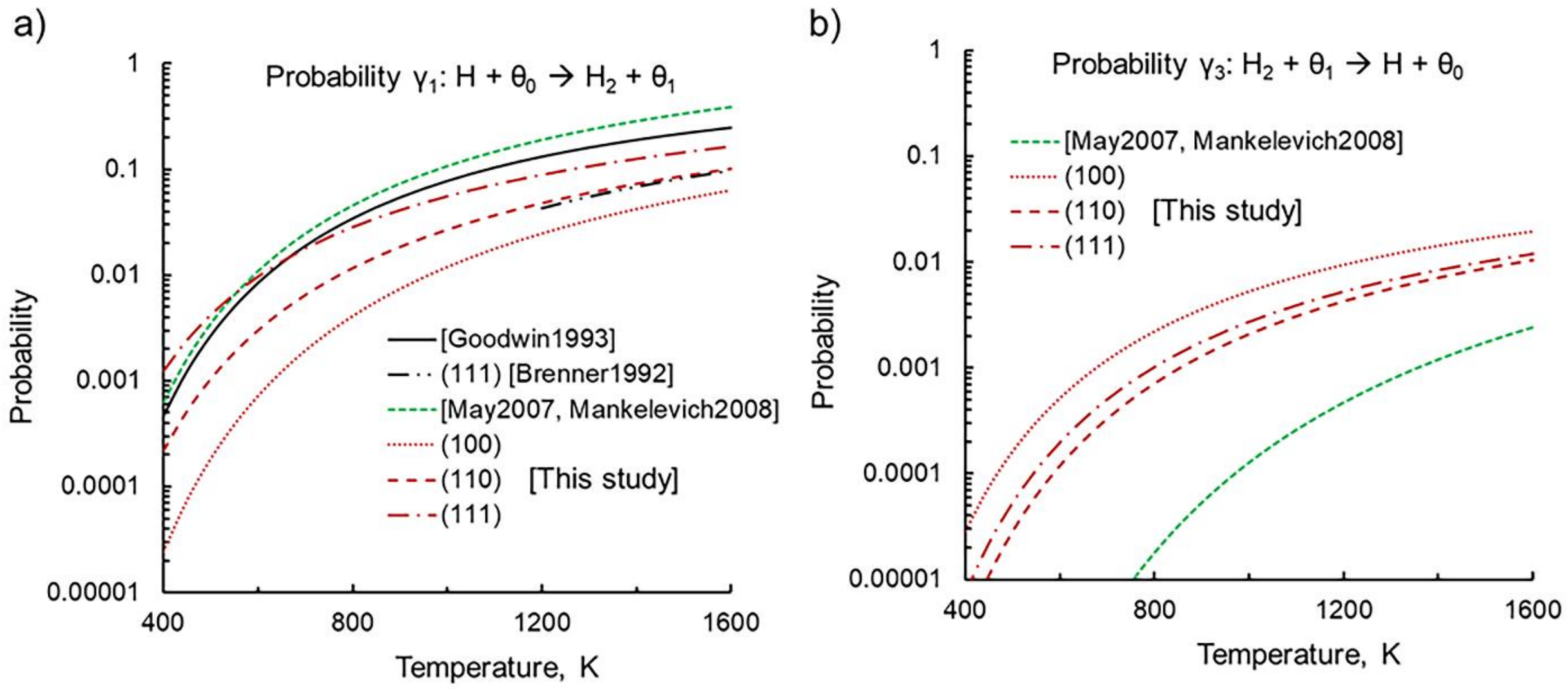

**Figure B1.** Probabilities $\gamma_1$ (a) and $\gamma_3$ (b) of the hydrogen abstraction (reaction $s_1$) and molecular dissociation (reactions $s_3$) reported by Goodwin [138], Mankelevich et al. [35,141], and calculated by Brenner et al. [140] and in this study (in red).

## Effect of reactive site deactivation by $H_2$

Equation (B2) does not account for the deactivation (loss) of reactive sites by molecular hydrogen (reaction $s_3$ in Table 1), which can be explicitly added to Eq.(B1), similarly to how it was done in Refs. [103,104,121]:

$$\frac{d\theta_1}{dt} p_s = \gamma_1 \frac{v_{th}^H}{4} n_H \theta_0 - \gamma_2 \frac{v_{th}^H}{4} n_H \theta_1 - \gamma_3 \frac{v_{th}^{H2}}{4} n_{H2} \theta_1 = 0,$$
$$\theta_0 + \theta_1 = 1. \qquad (B3)$$

Consequently, the fraction of the reactive sites is given by:

$$\theta_1 = \frac{\gamma_1}{\gamma_1 + \gamma_2 + \gamma_3 \frac{1-\alpha}{2\alpha}} \approx \frac{\gamma_1}{\gamma_1 + 1 + \gamma_3 \frac{1-\alpha}{2\sqrt{2}\alpha}}, \qquad (B4)$$

where $\alpha$ is the $H_2$ dissociation degree at the diamond surface and $\alpha \equiv \frac{n_H}{2n_{H_2}+n_H}$. Mankelevich et al. used the following equation for $\theta_1$ in Ref. [35,141]:

$$\theta_1 = \frac{1}{1 + 0.3\exp\left(\frac{3430}{T}\right) + 0.1\exp\left(-\frac{4420}{T}\right) * \frac{n_{H2}}{n_H}},$$

This equation can be transformed to the form identical to Eq.(B4) by dividing both the numerator and the denominator by $0.3\exp\left(\frac{3430}{T}\right)$, resulting in:

$$\theta_1 = \frac{3.33\exp\left(-\frac{3430}{T}\right)}{3.33\exp\left(-\frac{3430}{T}\right) + 1 + 0.33\exp\left(-\frac{7850}{T}\right)\frac{1-\alpha}{2\alpha}}. \qquad (B5)$$

This yields the following gamma coefficients: $\gamma_1 = 3.33\exp\left(-\frac{0.29eV}{RT}\right)$, $\gamma_2 = 1$, and $\gamma_3 = \sqrt{2} * 0.33\exp\left(-\frac{0.68eV}{RT}\right)$.

In the expression for $\gamma_3$ used by Mankelevich et al., the activation energy of 0.68 eV corresponds to the energy barrier of the gas-phase reaction $H_2$ + tret-$C_4H_9$ → H + iso-$C_9H_{10}$. Here, we calculated $\gamma_3$ for the (100), (110), and (111) diamond facets using larger clusters than the tret-$C_4H_9$ molecule, as discussed in Appendix A. As shown in Fig. A2, the activation energies of $\gamma_3$ are significantly lower for (100), (110), and (111) facets, being 0.30, 0.37, and 0.34 eV, respectively. As a result, the probability $\gamma_3$ of the molecular hydrogen dissociation on the diamond surface is underestimated (see Fig. B1b) and, consequently, $\theta_1$ is overestimated in Refs. [35,141], as shown in Fig. B2.

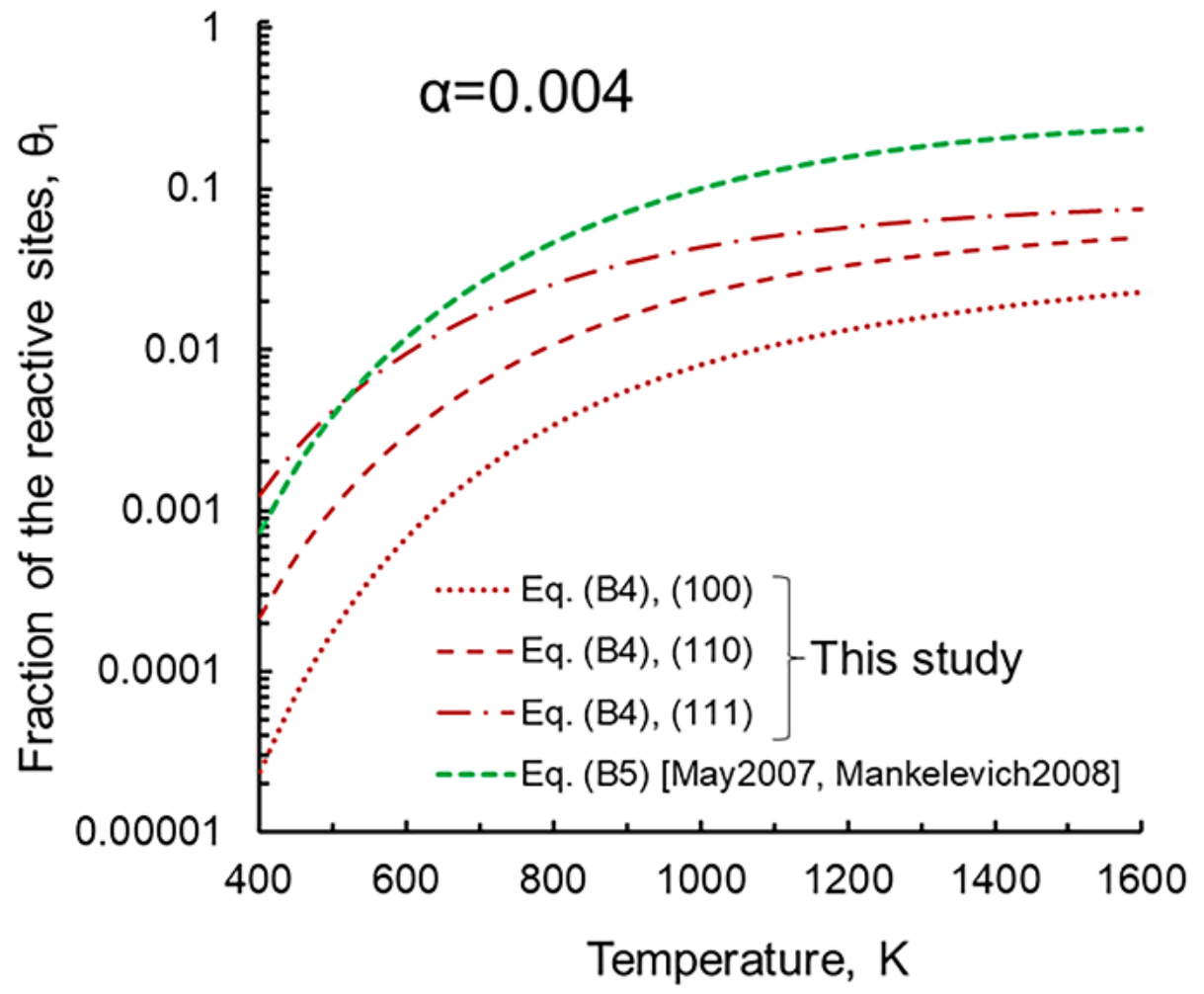

**Figure B2.** Fraction of the reactive site $\theta_1$ as a function of substrate, calculated using Eq. (B4) and Eq. (B5) for different diamond facets with probabilities $\gamma_1$ and $\gamma_3$ from Table 2.

Another important consequence of the inclusion of the reaction $s_3$ is that the fraction $\theta_1$ becomes a function of the $H_2$ dissociation degree for the standard conditions of diamond CVD growth. Indeed, when $\gamma_3$ is relatively high, i.e., $\alpha \ll 1$ and $\gamma_3 \frac{1-\alpha}{2\sqrt{2}\alpha} > 1$, the contribution of reaction $s_3$ in Eq. (B4) becomes significant, as it is for the standard conditions of HFCVD and MWCVD [84,85], where $\alpha\sim 0.004$ and $\gamma_3 \frac{1-\alpha}{2\sqrt{2}\alpha} \gg 1$. Therefore, under typical diamond CVD conditions, it is important to use Eq.(B4) rather than the original Goodwin model, Eq.(B2), as *the fraction of reactive sites linearly depends on the $H_2$ dissociation degree (see Fig. B3)*.

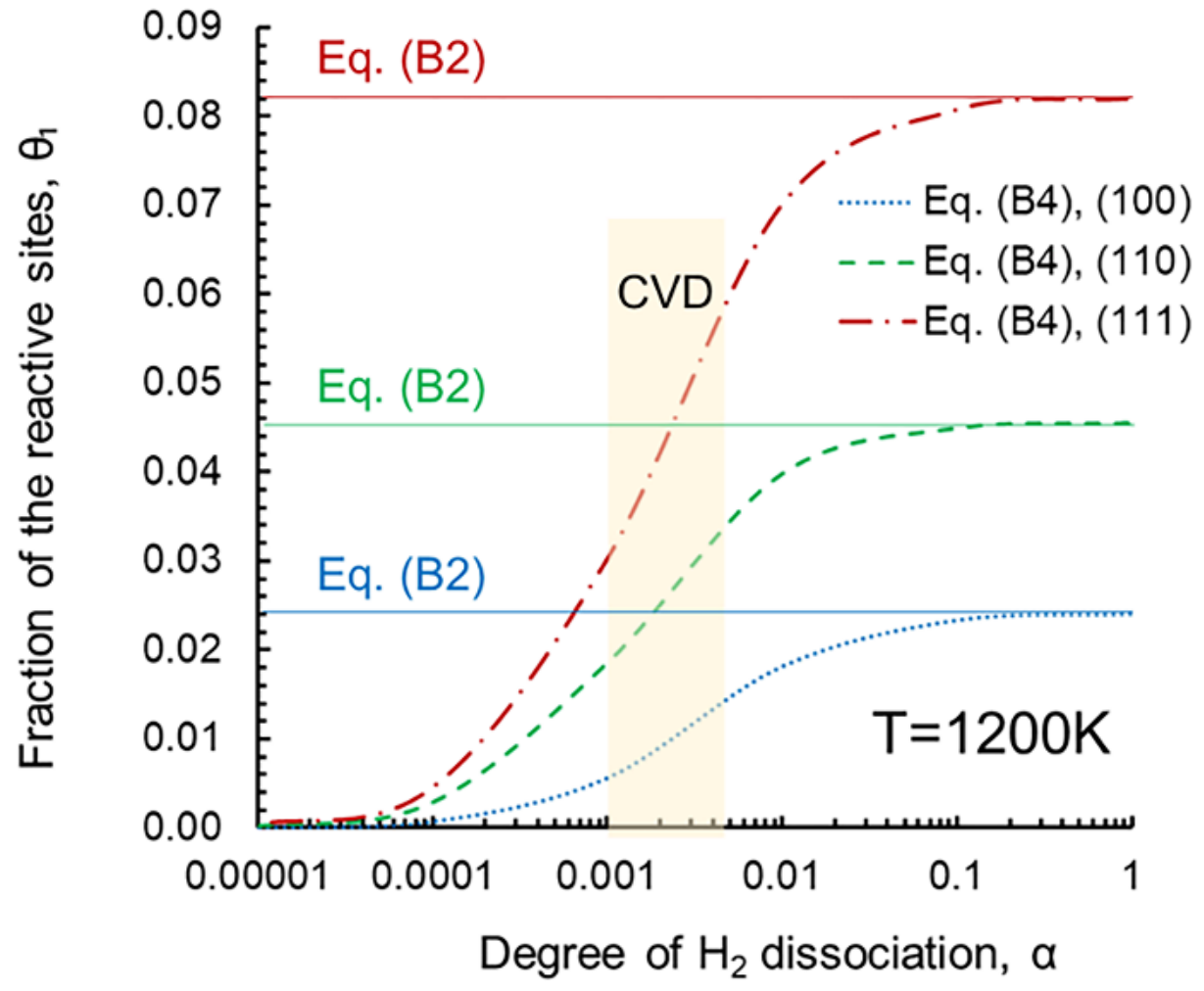

**Figure B3.** Fraction of the reactive site 1 as a function of $H_2$ dissociation degree, calculated using Eq. (B2) and Eq. (B4) for different diamond facets using probabilities $\gamma_1$ and $\gamma_3$ from Table 2.

## Effect of reactive site deactivation by $C_2H_2$

Equation (B4) does not account for the deactivation (loss) of reactive sites by acetylene (reaction $s_4'$ and $s_5'$ as shown in Fig. 8), which can be explicitly added to Eq.(B3). According to the reduced mechanism from Fig. 8, the five reactions that describe the composition of the diamond surface in the presence of H, $H_2$, and $C_2H_2$ reactants are:

$$H + \theta_0 \rightarrow H_2 + \theta_1 \qquad (s_1)$$
$$H + \theta_1 \rightarrow \theta_0 \qquad (s_2)$$
$$H_2 + \theta_1 \rightarrow H + \theta_0 \qquad (s_3)$$
$$2C_2H_2 + H + \theta_1 \rightarrow \theta_{20} \qquad (s_4')$$
$$\theta_{20} \rightarrow 2C_2H_2 + H + \theta_1 \qquad (s_5')$$

Here, $s_1$, $s_2$, and $s_3$ denote the elementary surface reactions, the same as in the full model and Eq. (B4). Reactions $s_4'$ and $s_5'$ are effective reactions combined from several elemental reactions of the full model. For this set of reactions, the system of balance equations for the fraction $\theta_1$ of reactive sites 1 is given by:

$$\gamma_1 \frac{v_{th}^H}{4} n_H \theta_0 - \gamma_2 \frac{v_{th}^H}{4} n_H \theta_1 - \gamma_3 \frac{v_{th}^{H2}}{4} n_{H2} \theta_1 - k`_4 n_{C2H2}^2 n_H p_s \theta_1 + k`_5 p_s \theta_{20} = 0,$$
$$\theta_0 + \theta_1 + \theta_{20} = 1. \qquad (B6)$$

Here, $\gamma_1$, $\gamma_2$, and $\gamma_3$ are probabilities of the elementary reactions $s_1$, $s_2$, and $s_3$. Rate constants $k_4'$ and $k_5'$ correspond to the effective reactions $s_4'$ and $s_5'$. The fraction $\theta_1$ of reactive sites 1 on the diamond surface, can be derived from this system of equations, assuming the detailed balance of creation and destruction of the reactive sites:

$$\theta_1 = \frac{\gamma_1}{\gamma_1 + 1 + \gamma_3 \frac{1-\alpha}{2\sqrt{2\alpha}} + \frac{2\alpha}{1+\alpha} \gamma_1 \frac{k'_4}{k'_5} n_{\mathrm{C2H2}}^2 n_0}. \qquad (B7)$$

Here, $\frac{2\alpha}{1+\alpha} n_0 = n_H$, where $n_0$ is the number density of $H_2$ in the feedstock gas. Figure B4 shows a comparison between the full reaction kinetic model and the reduced model represented by analytical expressions (B4) and (B7), with the following approximate expression for the ratio $k`_4/k`_5$:

$$\frac{k`_4}{k`_5} \approx 5.0 \times 10^{-67} \exp\left(\frac{4.5eV}{RT}\right), cm^9.$$

Eq. (B7) reduced to Eq.(B4), when $n_{C2H2}$ equals zero. However, when

$$\frac{2\alpha}{1+\alpha} \gamma_1(T) \frac{k'_4(T)}{k'_5(T)} n_{\mathrm{C2H2}}^2 n_0 \gg \gamma_1(\mathrm{T}) + 1 + \gamma_3(T) \frac{1-\alpha}{2\sqrt{2\alpha}}, \qquad (B8)$$

the effect of $C_2H_2$ becomes significant. As a result, Eq. (B7) deviates from Eq. (B4) when temperature drops below a critical value $T_{cr}$. We defined the critical temperature as a temperature at which the above condition (B8) is met:

$$T_{cr} \approx \frac{46418}{110 - \ln\left[P(Torr)n_{C2H2}^2(cm^{-6})\alpha\right]}. \tag{B9}$$

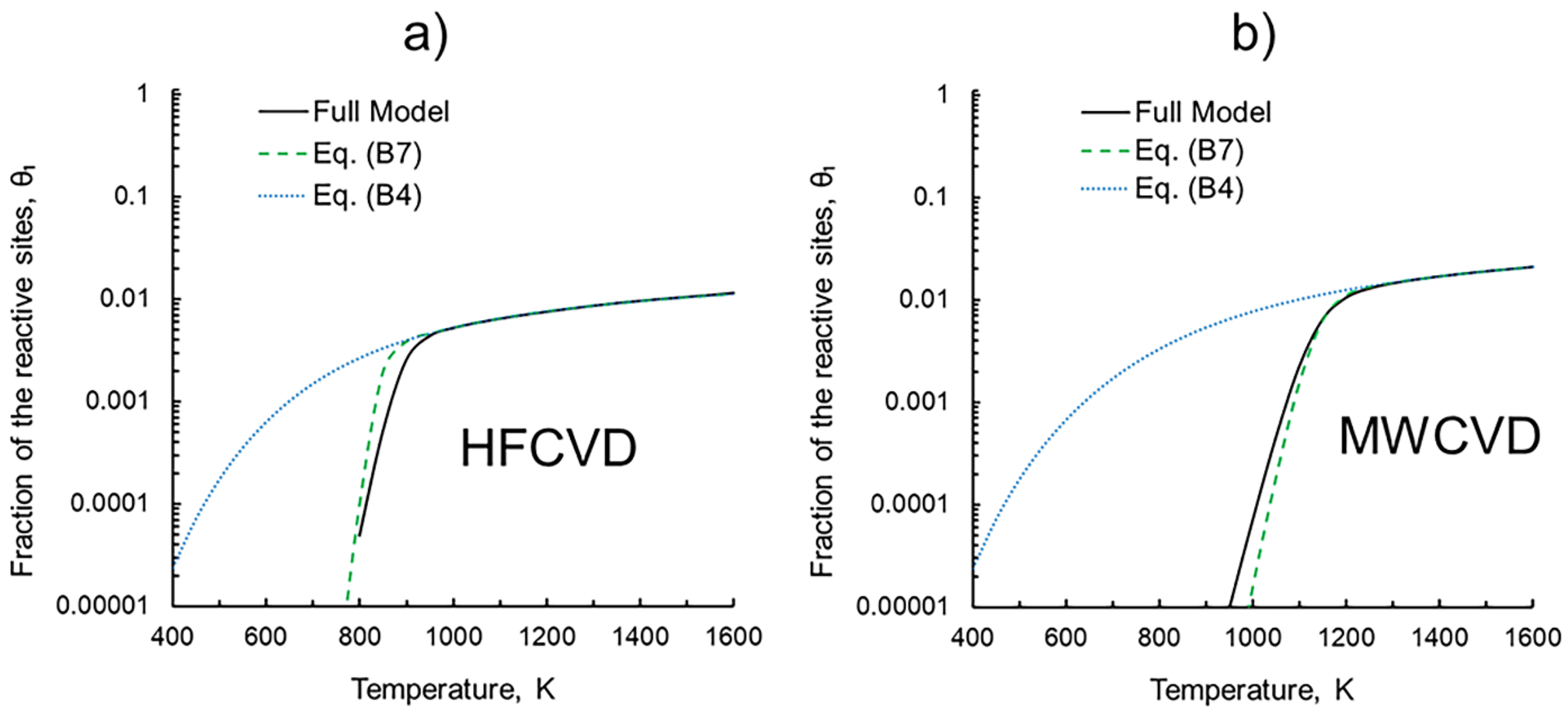

**Figure B4.** Fractions of the reactive sites $\theta_1$ calculated using the full reaction kinetic model and using Eq. (B4), and Eq. (B7) for the HFCVD (a) and MWCVD (b). The parameters for the models, including the densities of the gas-phase reactants and the $H_2$ dissociation degree, are taken from Refs. [84,85].

## Appendix C. Probability of H atom loss on a diamond surface.

For modeling volumetric physical and chemical processes, effective boundary conditions on the diamond surface are needed. The probability of an H atom loss is defined as

$$\gamma_H = 1 - \frac{outgoing\ flux\ of\ H}{incoming\ flux\ of\ H} \tag{C1}$$

In this section, we derive the expression for $\gamma_H$ on the diamond surfaces. The original expression for $\gamma_H$ was proposed by Goodwin [138]:

$$\gamma_H = \frac{2\gamma_1}{\gamma_1 + 1}. \tag{C2}$$

This expression did not take molecular hydrogen dissociation into account, which, as is shown in Appendix B, needs to be considered for the standard diamond CVD conditions. Fig. C1 illustrates that, considering reactions $s_1$, $s_2$, and $s_3$, the outgoing flux of H atoms becomes:

$$outgoing\ flux\ of\ H = (1-\gamma_1)\frac{v_{th}^{H}}{4}n_H\theta_0 + \gamma_3\frac{v_{th}^{H2}}{4}n_{H2}\theta_1 = (1-\gamma_1)\frac{v_{th}^{H}}{4}n_H(1-\theta_1) + \gamma_3\frac{v_{th}^{H2}}{4}n_{H2}\theta_1.$$

Substituting this expression for the outgoing flux into Eq. (C1) yields:

$$\gamma_H \approx 2\theta_1 \approx \frac{2\gamma_1}{\gamma_1 + 1 + \gamma_3\frac{1-\alpha}{2\sqrt{2}\alpha}}. \qquad (C3)$$

Here, we explicitly assumed that $\gamma_1, \gamma_3, \alpha < 1$. Mankelevich et al. in Refs. [35,141], essentially, used a similar expression for $\gamma_H$:

$$\gamma_H = \frac{0.83}{1 + 0.3\exp\left(\frac{3430}{T}\right) + 0.1\exp\left(-\frac{4420}{T}\right) * \frac{1-\alpha}{2\alpha}},$$

This expression can be transformed to the form similar to Eq. (C3) by dividing both the numerator and the denominator by $0.3\exp\left(\frac{3430}{T}\right)$, resulting in:

$$\gamma_H = \frac{2.77\exp\left(-\frac{3430}{T}\right)}{3.03\exp\left(-\frac{3430}{T}\right) + 1 + 0.33\exp\left(-\frac{7850}{T}\right)\frac{1-\alpha}{2\alpha}}, \qquad (C4)$$

where $\gamma_1$ is slightly modified compared to Eq. (B5).

Reactions $s_1$ and $s_3$ are dependent on the orientation of the diamond surface (facet-selective). The calculated probabilities $\gamma_1$ and $\gamma_3$ for the most common facets are listed in Table 2. As mentioned in the Introduction, the probability $\gamma_H$ was measured on polycrystalline diamond surfaces by Krasnoperov et al. [82], and by Harris and Weiner [83]. Figure C2 indicates that Eq. (C3) for the (111) facet agrees better with the measurements. Note that for an accurate interpretation of the experiments conducted by Krasnoperov et al. and by Harris and Weiner, it is necessary to know the $H_2$ dissociation degree at the diamond surface and the fractions of (111), (110), and (100) facets on the surfaces of a polycrystalline diamond.

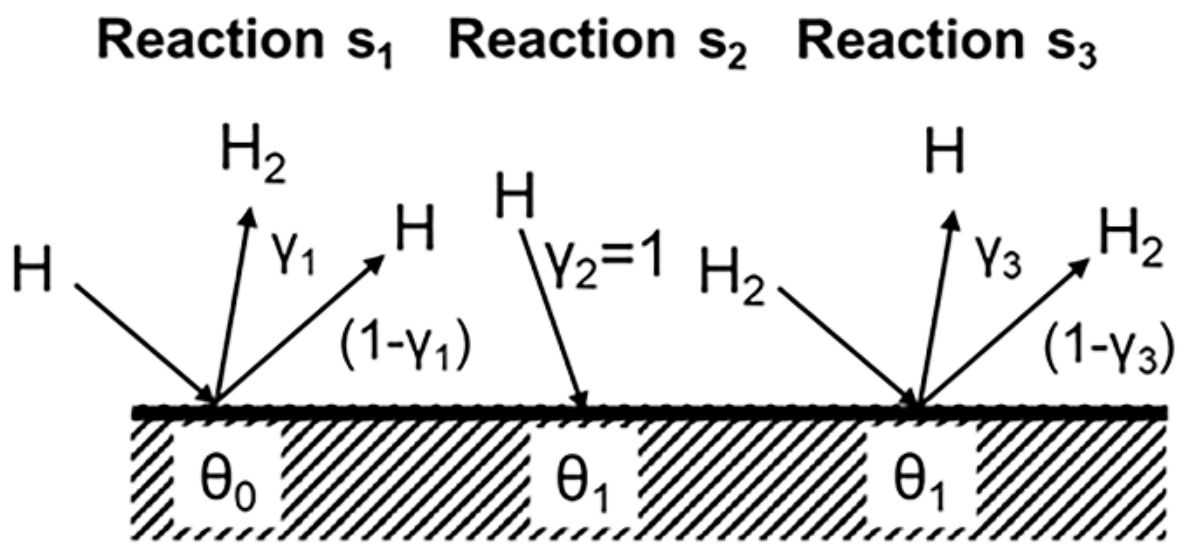

**Figure C1.** Incoming and outgoing fluxes of H and $H_2$ on the diamond surface caused by the $s_1$, $s_2$, and $s_3$ reactions.

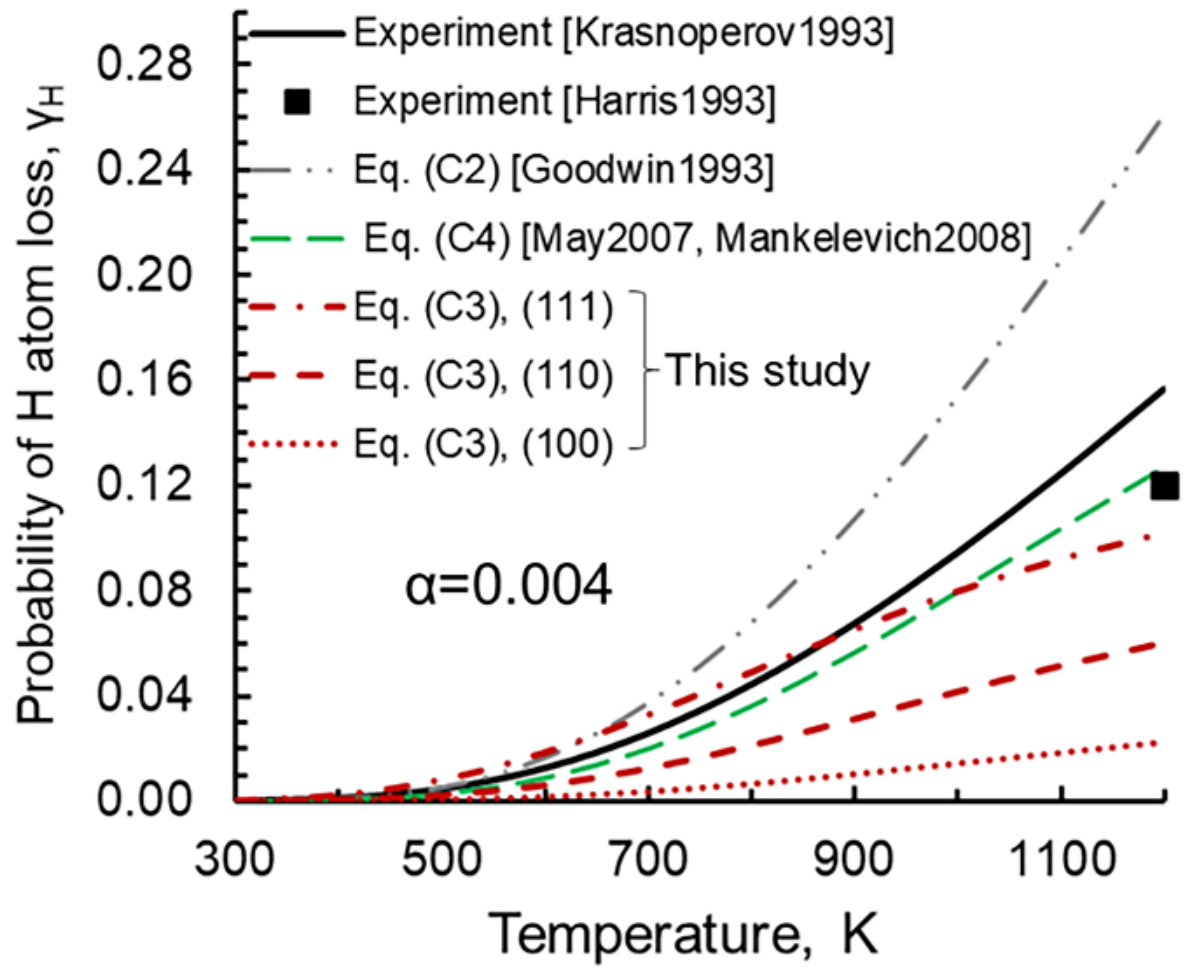

**Figure C2.** Measured and calculated probabilities of H atom loss $\gamma_H$ using Eqs. (C2), (C3), and (C4). The data for the (100), (110), and (111) facets presented here were calculated using Eq. (C3), along with the probabilities $\gamma_1$ and $\gamma_3$ listed in Table 2.

# Appendix D. Adsorbate desorption at high temperature.

In the previous sections, we explained how the reactive sites 1, which play a crucial role during diamond CVD, are formed. It is widely accepted that the diamond growth rate depends on the availability of these sites. According to the model by Mankelevich et al. [121,142], the growth rate ($GR$) correlates with the fraction $\theta_1$ of the reactive sites 1. However, the $CH_3$ radicals and $C_2H_2$ molecules contribute to the diamond growth rate through specific mechanisms: "dimer-opening-methylene-bridged" for $CH_3$ and "dimer-opening-vinylidene-bridged" for $C_2H_2$. Therefore, the contribution from $CH_3$ ($GR_{CH3}$) and $C_2H_2$ ($GR_{C2H2}$) should correlate with the fractions $\theta_7$ and $\theta_{13}$, respectively. (These fractions represent the surface sites with bridged $CH_2{}^{surf}$ and $C_2H_2{}^{surf}$ adsorbates, shown in Fig. 5b and c). Despite this, Fig. D1 shows that, at temperatures below 1200K, the fraction $\theta_7$ approximately equals to fraction $\theta_1$, making the equations where the $GR_{CH3}$ is proportional to $\theta_1$ and $\theta_7$ identical. The distinction between $GR_{CH3}\sim\theta_1$ and $GR_{CH3}\sim\theta_7$ equations appear above 1200K. As the fraction $\theta_1$ continues to increase, whereas $\theta_7$ peaks at 1200K. Consequently, to reproduce the growth rate peak, it is essential to use the approach that $GR_{CH3}{\sim}\theta_7$ rather than $GR_{CH3}{\sim}\theta_1$.

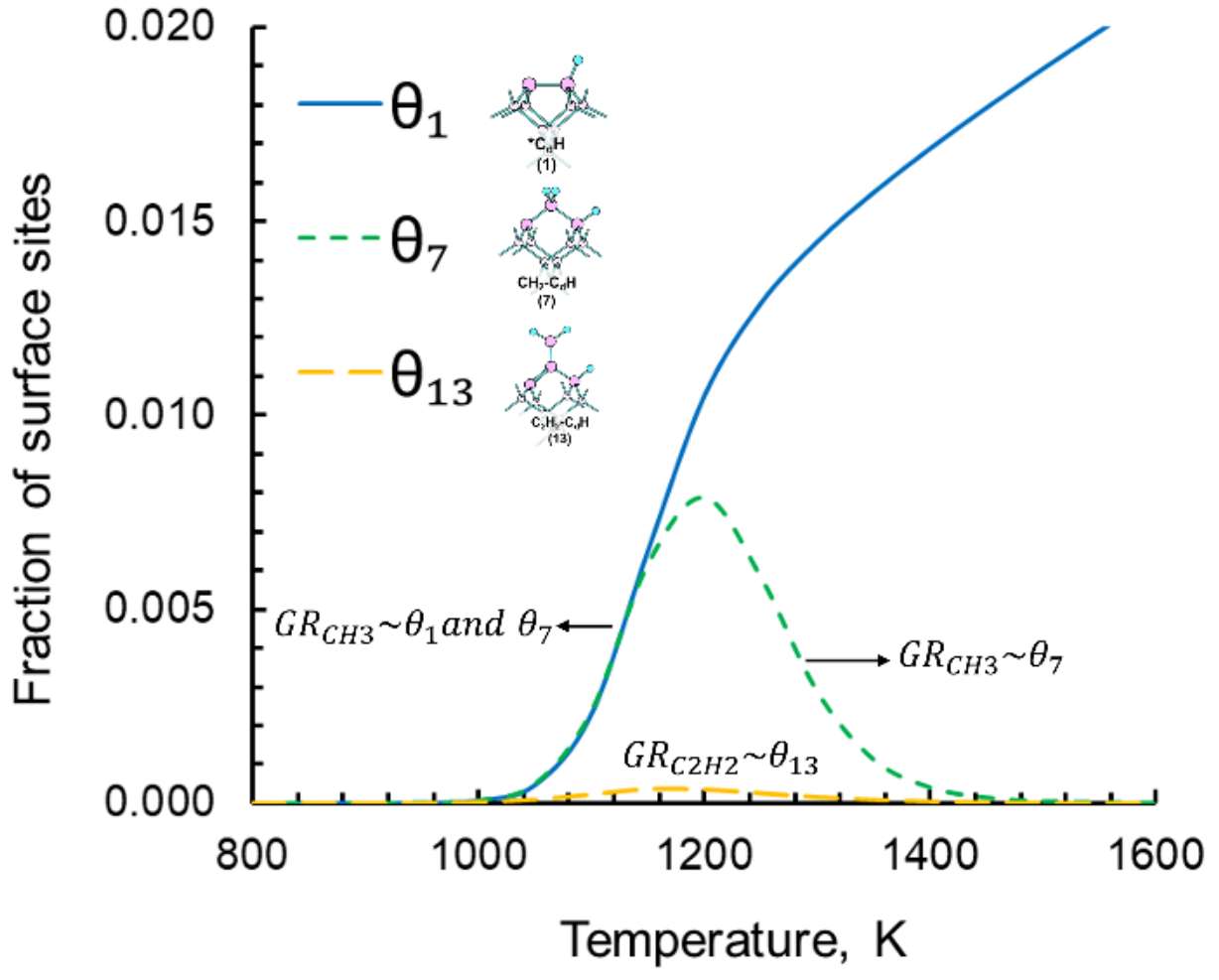

**Figure D1.** Fraction $\theta_1$, $\theta_7$ and $\theta_{13}$ of the reactive sites and sites with the bridged $CH_2{}^{surf}$ and $C_2H_2{}^{surf}$ adsorbates on (100) diamond surface during MWCVD, calculated by the full mode using densities of the gas-phase species from Refs. [84,85]. $GR_{CH3}$ and $GR_{C2H2}$ denote the contributions of $CH_3$ and $C_2H_2$ to diamond growth rates.

Fig. D1 demonstrates that the fraction $\theta_7$ is significantly higher than the fraction $\theta_{13}$. Consequently, $GR_{CH3}$ should be much higher than $GR_{C2H2}$, and $CH_3$ is the major precursor of diamond growth. Namely, the fraction of $C_2H_2{}^{surf}$ is 11 times lower than the fraction of $CH_2{}^{surf}$ at 1100K. It is consistent with Martin's measurements [97], where it was estimated that, at 1100K, the probability of $C_2H_2$ insertion into the diamond surface is roughly 10 times lower than that of the $CH_3$ radical.

The fraction $\theta_7$ deviates from the fraction $\theta_1$ above 1200K due to the reactions involving the removal of carbon adsorbates:

$$H + CH_2{}^{surf} \rightarrow CH_3{}^{surf} \qquad (s_{27})$$
$$H + CH_3{}^{surf} \rightarrow CH_4{}^{gas} + *^{surf} \qquad (s_{19}, s_{21})$$
$$CH_3{}^{surf} \rightarrow CH_3{}^{gas} + *^{surf} \qquad (s_{16}, s_{18})$$

These reactions are illustrated in Fig. 5e. To demonstrate the significance of these reactions, we performed a sensitivity test that excluded these four reactions. The result is shown in Fig. D2. Below 1200K, the fractions $\theta_7$ are identical in both the full model and the model, where these reactions were excluded. However, above 1200K, the model without these four reactions produce qualitatively different results: it incorrectly predicts that the growth rate continuously increases.

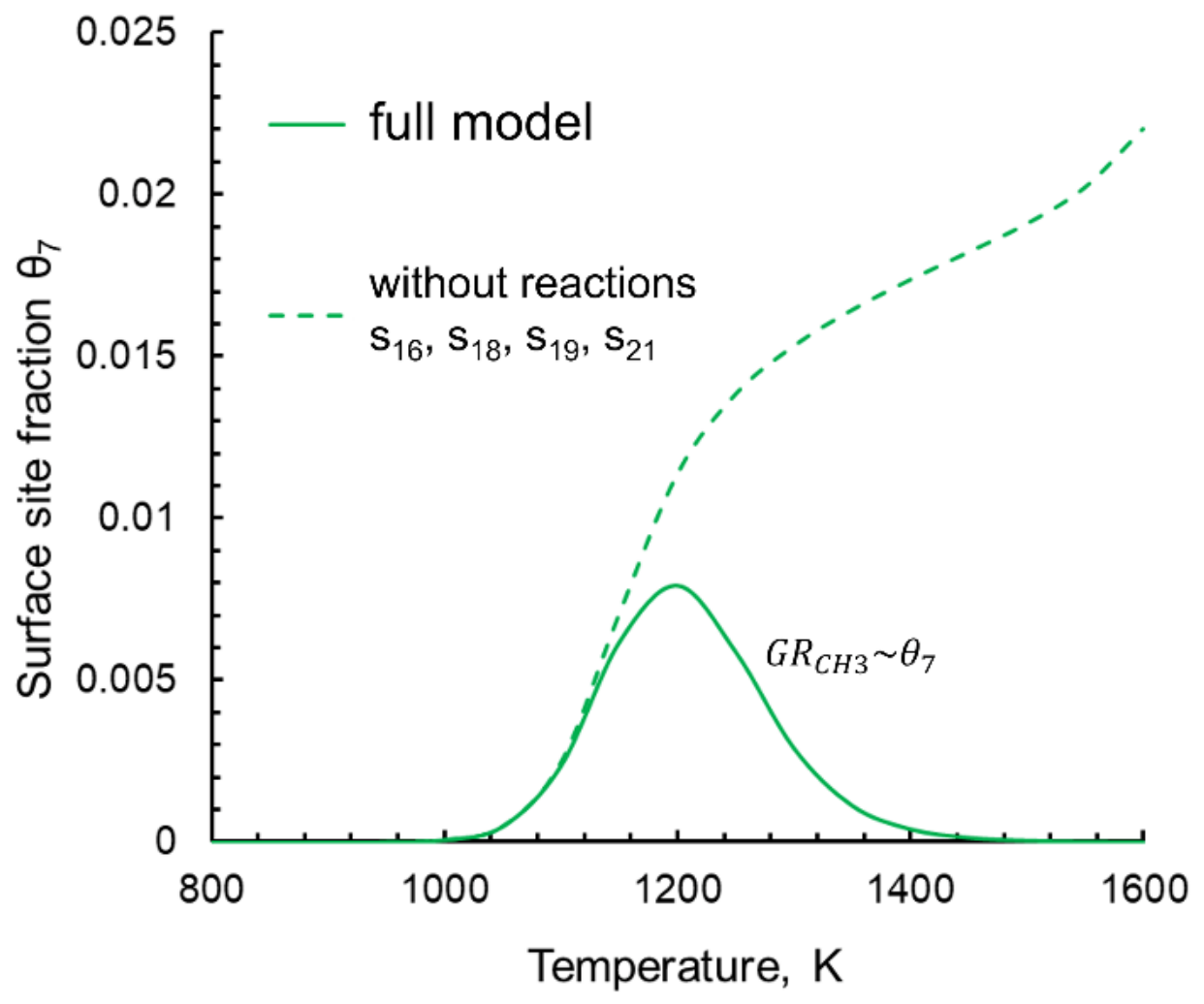

**Figure D2.** Sensitivity test for the fraction $\theta_7$. The solid line represents the result of the full model, and the dashed line represents the results of the model excluding reactions $s_{16}$, $s_{18}$, $s_{19}$, and $s_{21}$ of $CH_3$ adsorbate thermal desorption and $CH_3$ adsorbate abstractions by atomic hydrogen.

In addition, we modeled the reaction of the carbon adsorbate removal proposed in Ref. [114], involving adsorption of H atom on $CH_2$ adsorbate followed by $CH_3$ desorption:

$$H + {CH_2}^{surf} \rightarrow {CH_3}^{gas}.$$

This reaction should be described as a two-step process. The first step involves H atom adsorption and leads to vibrational excitation of the adsorbate:

$$H + {CH_2}^{surf} \rightarrow CH_3(v)^{surf}.$$

The vibrationally excited adsorbate can then follow the alternative reaction pathways:

$$CH_3(v)^{surf} \rightarrow {CH_3}^{gas} \quad (Path1)$$
$$CH_3(v)^{surf} \rightarrow H^{gas} + {CH_2}^{surf} \quad (Path2)$$
$$CH_3(v)^{surf} \rightarrow {CH_3}^{surf} + Heating \quad (Path3)$$

Ab initio molecular dynamics is a proper method to explore the evolution of the vibrationally excited adsorbate. The MD modeling indicates that the third reaction pathway, shown in Fig. D3a, is dominant. Indeed, the adsorption of H atom releases energy of 4.3 eV. Pathways 1 and 2 will be highly probable if most of the released energy is accumulated either in the C-H or C-C bonds. Due to energy dissipation, the pathways 1 and 2 represent rare events. The released energy is redistributed among all available vibrational modes, resulting in vibrational relaxation and surface heating. No $CH_3$ desorption was observed within $10^4$ fs. Instead, the surface reached thermal equilibrium after 350 fs. During this thermalization, the newly formed C-H bond oscillated with a large amplitude, exciting the neighboring bonds, including the $CH_3$-C bond between the adsorbate and diamond surface, as shown in Fig. D3b. We performed several ab initio MD simulations, with the initial substrate temperatures of 800, 950, and 1250 K. Based on these simulations, we concluded that the reaction pathway 3 is the most probable, as proposed by D'Evelyn et al. in Ref. [143]. Therefore, as mentioned above, we use reaction $s_{27}$ to describe the interaction of H atoms with ${CH_2}^{surf}$ adsorbates.

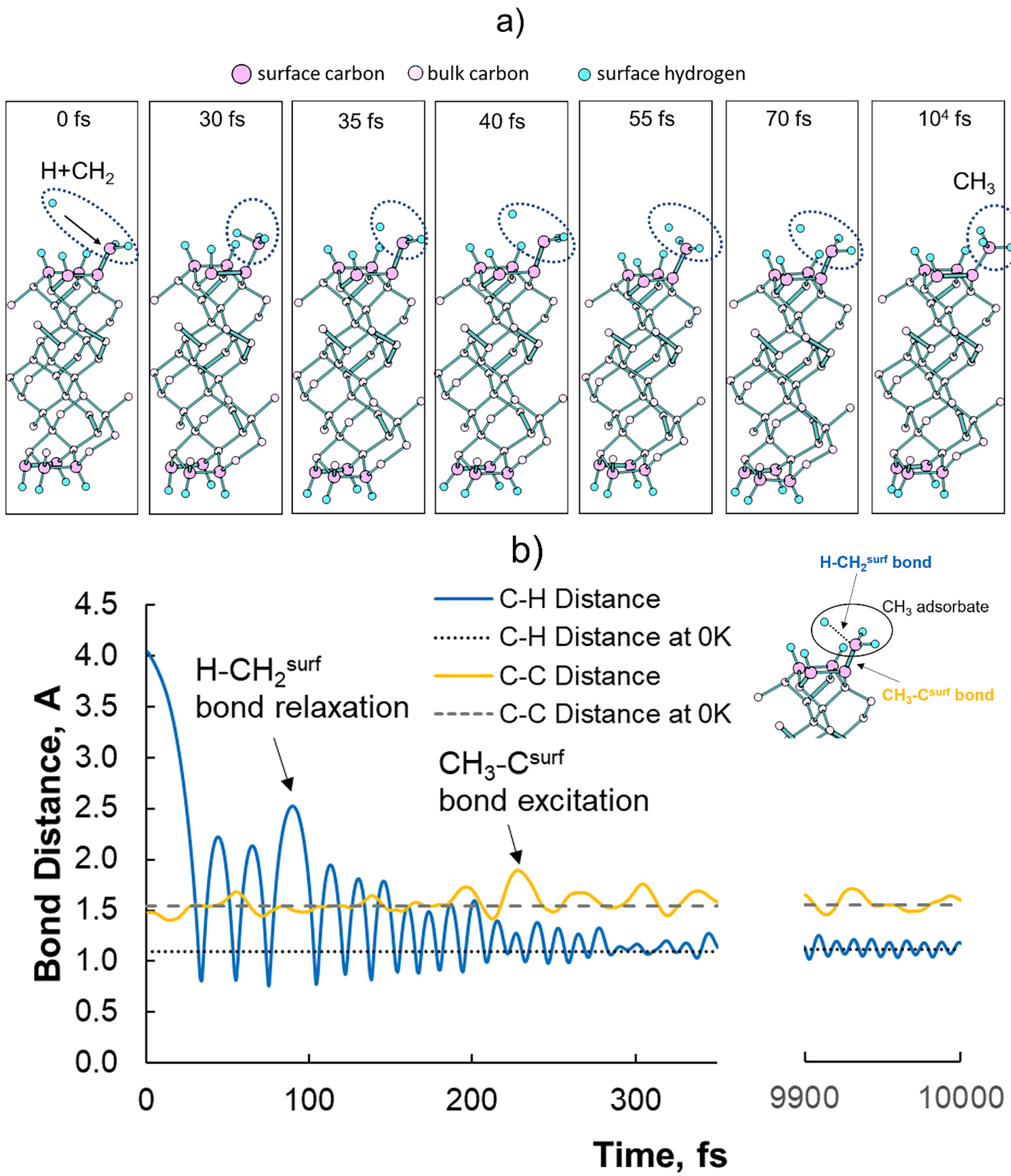

**Figure D3.** Results of the ab initio MD modeling of the reaction between H-atom and $CH_2$ adsorbate at 800 K during $10^4$ fs using an NVE ensemble: (a) structure evolution, where dashed circle indicates the reaction center; (b) H-$CH_2$ (blue solid line) and C-$CH_3$ (orange solid line) bond length oscillations with time; the dotted and dashed lines are the C-H and C-$CH_3$ bond distances at 0K.

# Appendix E. Diamond growth rate.

Appendix D shows that the diamond growth rate can be described through the fraction $\theta_7$. Here, we introduce an expression for the growth rate:

$$GR = \gamma_{CH3} \frac{v_{th}^{CH3}}{4} n_{CH3} \theta_7 \frac{M}{p * N_A}, \qquad (E1)$$

where $v_{th}^{CH3}$ is the thermal velocity of $CH_3$, *M* is the molar mass of carbon, *p* is diamond bulk density, $N_A$ is the Avogadro number. The parameter $\gamma_{CH3}$ represent the adsorption probabilities of $CH_3$ followed by surface reconstruction. Eq. (E1), with $\gamma_{CH3} = 0.85$ matches the experimentally measured growth rates reported in Ref. [38].

The fraction $\theta_7$ can be derived from $\theta_1$, considering the effective surface reactions $s_6'$, $s_7'$, and $s_8'$ from Fig. 8:

$$CH_3 + \theta_1 \to \theta_7 \qquad (s_6')$$
$$H + \theta_7 \to CH_3 + \theta_1 \qquad (s_7')$$
$$CH_3 + \theta_7 \to C_2H_4 + \theta_1, \qquad (s_8')$$

The balance of these reactions yields:

$$\theta_7 = \frac{k_6'/k_8'}{\frac{n_{\mathrm{H}}}{n_{\mathrm{CH3}}} k_7'/k_8' + 1} \theta_1. \qquad (E2)$$

Here $k_6'/k_8'$ and $k_7'/k_8'$ are dimensionless functions. Substituting from Eq. (E2) for $\theta_7$, Eq. (E1) becomes:

$$GR \approx \gamma_{CH3} \frac{v_{th}^{CH3}}{4} n_{CH3} \frac{k_6'/k_8'}{\frac{n_{\mathrm{H}}}{n_{\mathrm{CH3}}} k_7'/k_8' + 1} \frac{\gamma_1}{\gamma_1 + 1 + \gamma_3 \frac{1-\alpha}{2\sqrt{2}\alpha} + \frac{2\alpha}{1+\alpha} \gamma_1 \frac{k'_4}{k'_5} n_{\mathrm{C2H2}}^2 n_0} \frac{M}{p * N_A}. \qquad (E3)$$

Assuming $k_6'/k_8' = 1.1$ and $k_7'/k_8' = 7.0 \times 10^{18} \exp\left(-\frac{5.1eV}{RT}\right)$, Eq. (E3) closely approximates the results of the full model for microwave and hot-filament reactors (see Fig. E1).

Makelevich et. [142] proposed an equation where the growth rate is proportional to the fraction $\theta_1$:

$$GR \approx 2\sqrt{T}\left(\frac{n_{CH3}}{10^{13}}\right)\theta_1 f, \qquad (E4)$$

where $f \approx 0.03$ is an empirical factor, and $\theta_1$ was calculated using Eq. (B5). This equation was derived to predict the growth rate near its peak value, specifically ignoring the processes of surface passivation by $C_2H_2$ passivation and adsorbate desorption. Therefore, Eq. (E4) predicts a continuous increase in the growth rate with temperature. In contrast, both Eq. (E3) and the full reaction kinetic models indicate that the growth rate peaks at around 1200K.

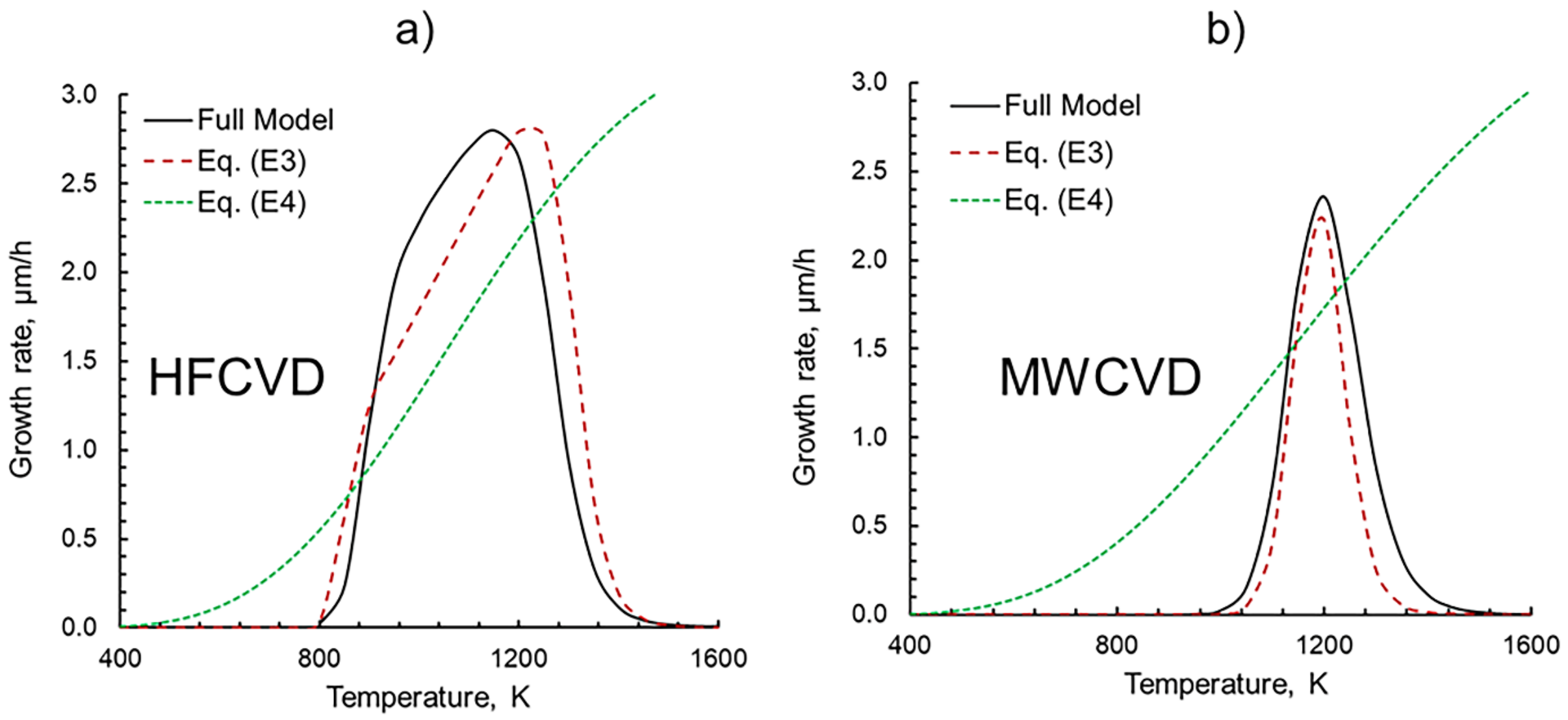

**Figure E1.** Growth rates calculated using the full reaction kinetic model, Eq. (E3), and Eq. (E4) for the HFCVD (a) and MWCVD (b). The parameters for the models, including the densities of the gas-phase reactants and the $H_2$ dissociation degree, are taken from Refs. [84,85].

# Appendix F. Desorption of the saturated and unsaturated chains.

The full reaction kinetic model includes the desorption of both saturated and unsaturated chains. In this context, “saturated chain” refers to a structure where each carbon atom is saturated with hydrogen atoms; therefore, all C-C bonds are single bonds ($sp^3$ hybridization). “Unsaturated chains” correspond to radical adsorbates such as $C_2H_4$* and $C_2H_2$*, with the asterisk indicating a dangling bond. Importantly, during desorption, saturated chains retain their $sp^3$ hybridization and single C-C bond structure, as shown in Fig. F1. Therefore, the desorption energy of the unsaturated chains is 4 eV, corresponding to the energy of the σ bond between the adsorbate C atom and a C atom of the diamond surface. During desorption, the unsaturated chains undergo a change in hybridization, leading to the formation of an additional π-bond between C-C atoms in the desorbed molecule. As a result, the desorption energy of the unsaturated chains is reduced to 1.5 eV for $C_2H_4$ and 1.7 eV for $C_2H_2$ (see Fig. F2). This mechanism is referred to as β-scission in organic chemistry.

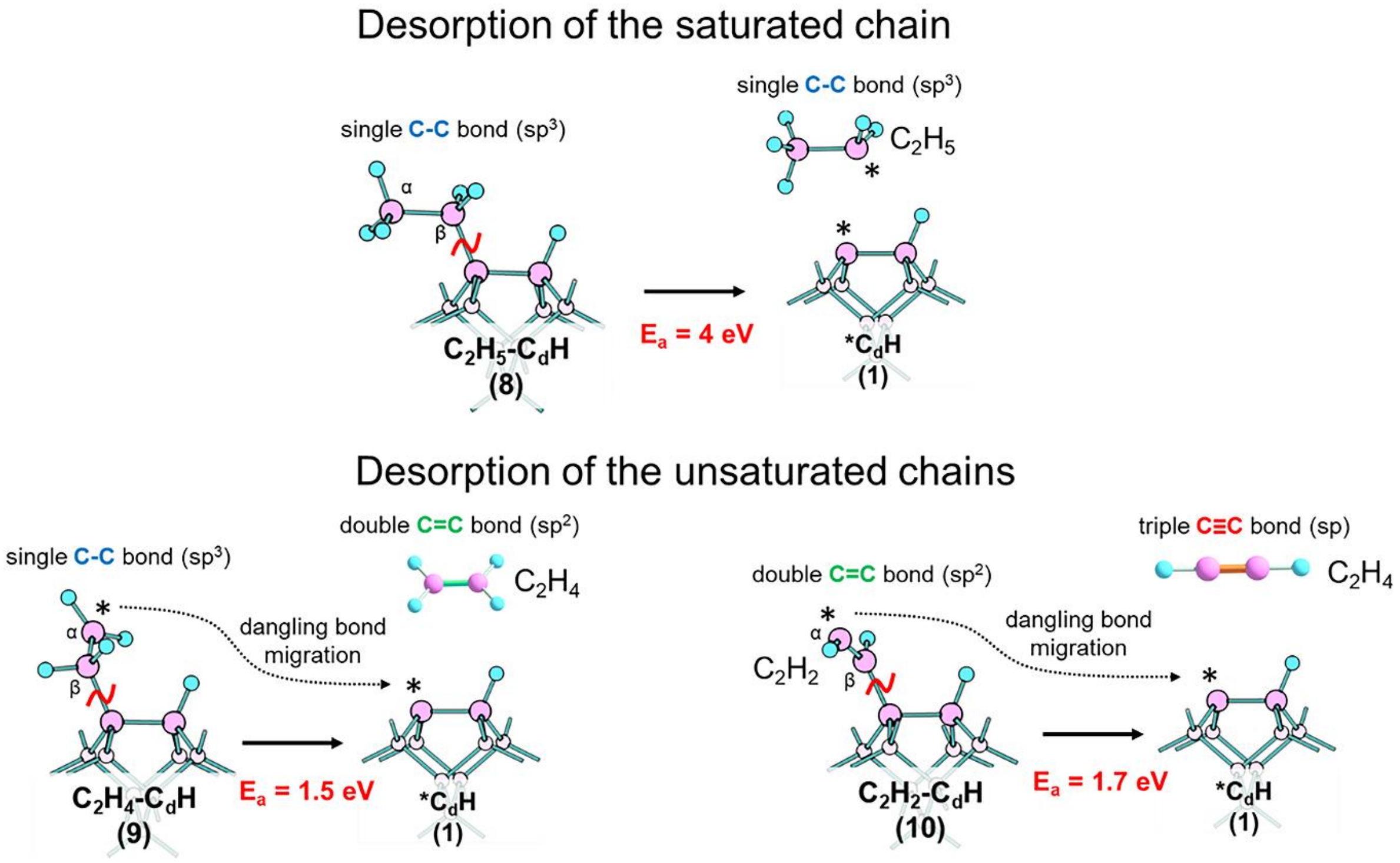

**Figure F1.** Desorption of the saturated and unsaturated chains from the (100) diamond surface via the β-scission mechanism. The $E_a$ indicates the desorption energy; asterisks represent dangling bonds. Carbon atoms of the adsorbates are labeled α and β, according to their position relative to the diamond surface.

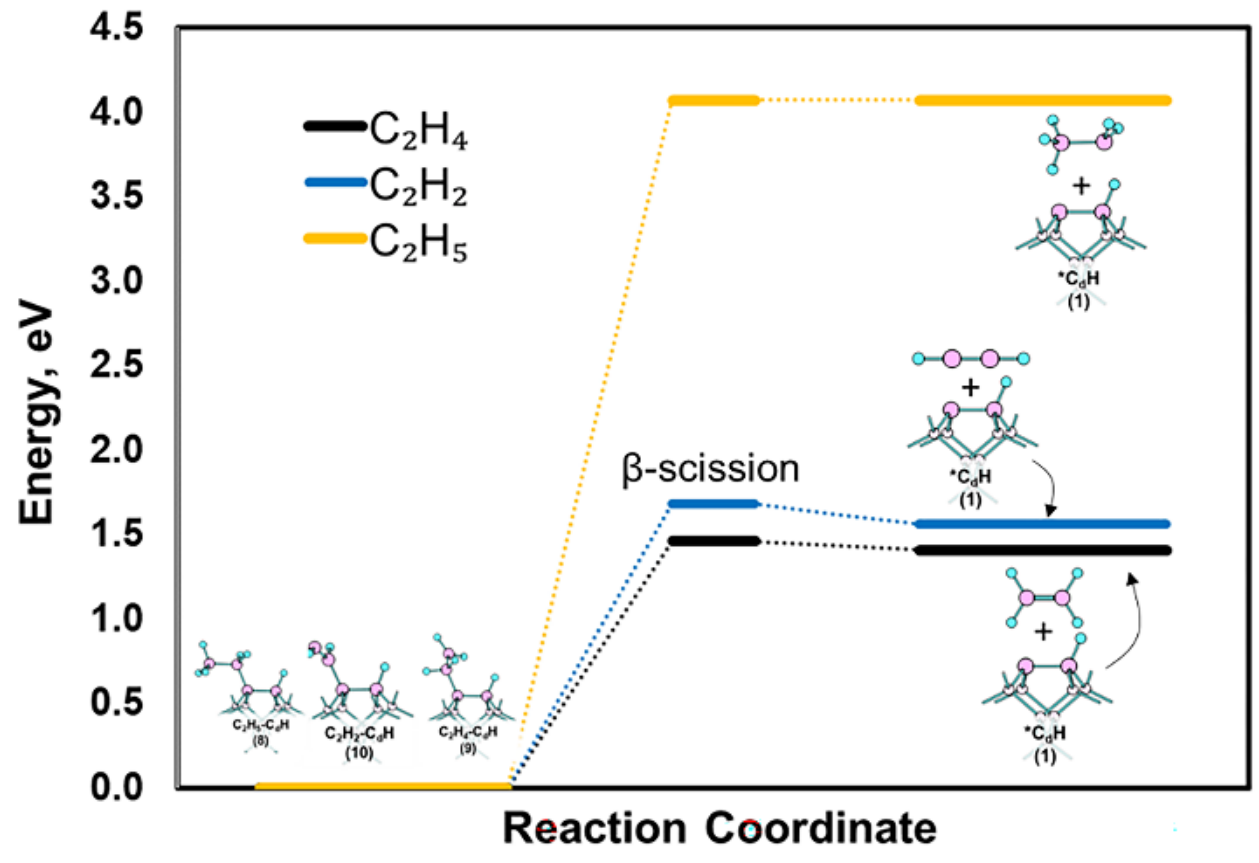

**Figure F2.** Reaction diagram of desorption of the saturated ($C_2H_5$) and unsaturated ($C_2H_4$ and $C_2H_2$) chains via the β-scission mechanism.

# Appendix G. New mechanism of sp$^2$-phase nucleation on the (100) diamond surface.

Here we introduce a new mechanism for sp$^2$-phase nucleation on the (100) diamond surface, as shown in Fig. G1. This mechanism follows the same steps as the well-known HACA mechanism [98–102,144–146], which describes soot growth. Both processes require two $C_2H_2$ molecules to create a six-membered ring on the surface. Initially, the first $C_2H_2$ adsorbs at the reactive sites (the radical site with a dangling bond). These dangling bonds are typically generated through a hydrogen abstraction reaction. Consequently, the presence of the H atoms is crucial, serving as a primary reactant that abstracts hydrogen from the surfaces. The final step leads to the ring closing. This step is an energetically favorable process, as shown in Fig. G2.

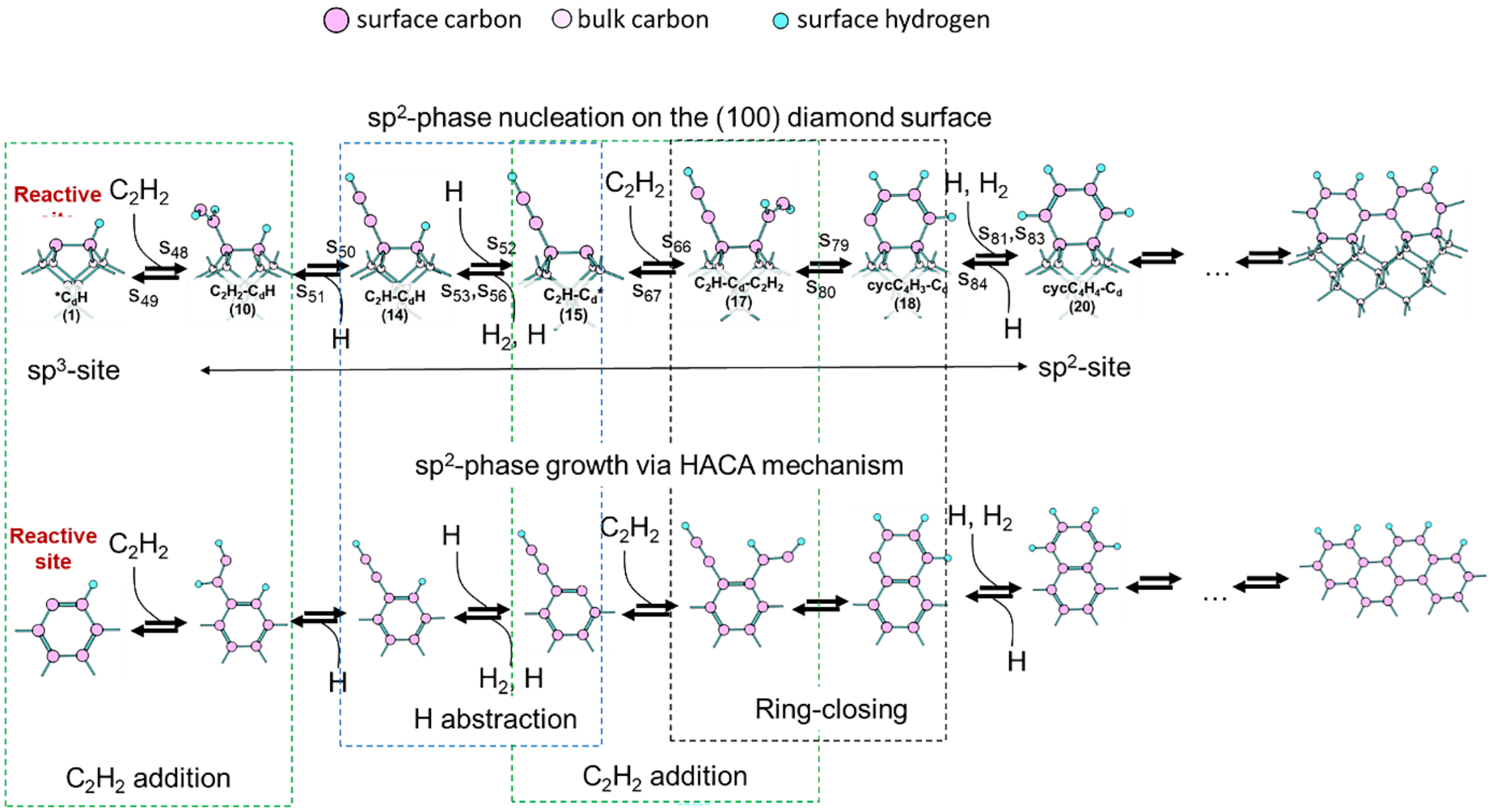

**Figure G1.** Mechanism of the sp$^2$-phase nucleation on the (100) diamond surface and HACA mechanism of soot growth.

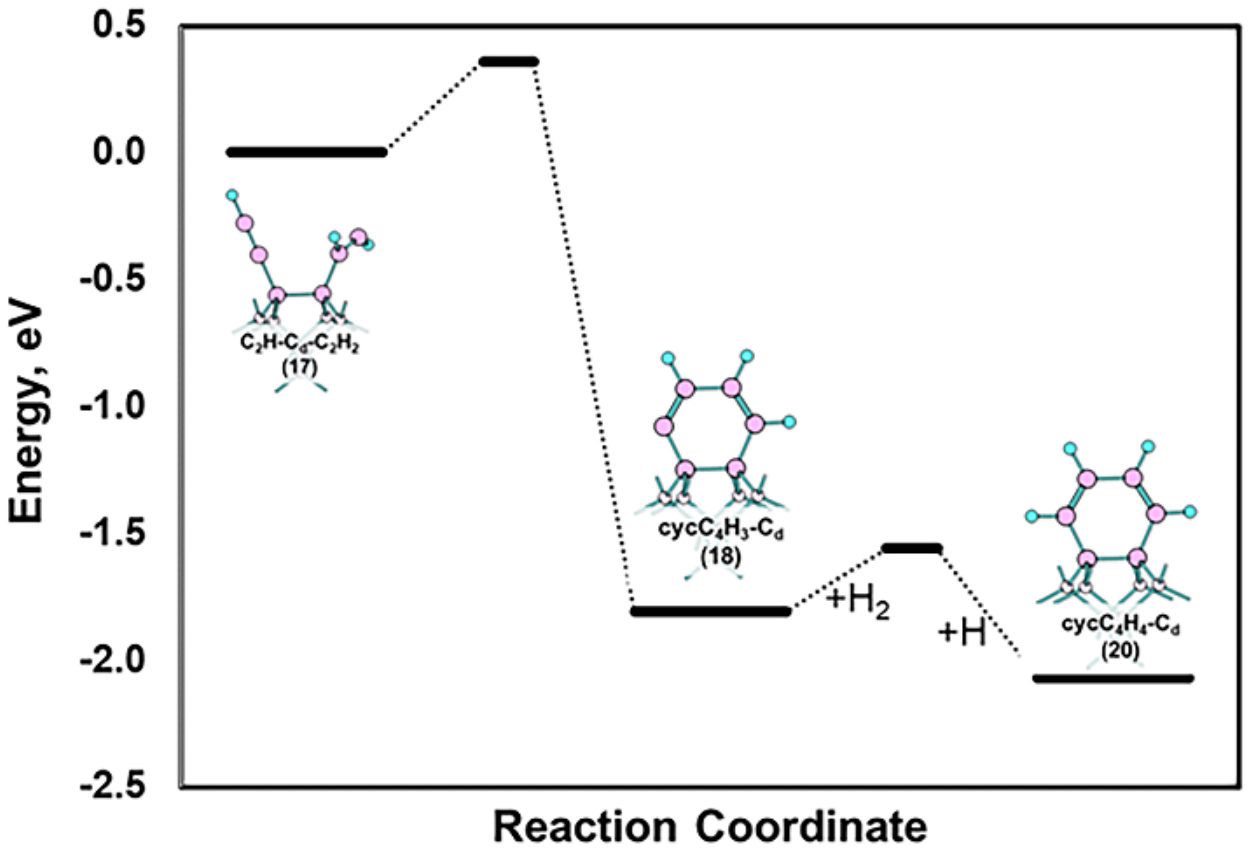

**Figure G2.** Reaction diagram for the ring-closing step during the $sp^2$-site nucleation on the (100) diamond surface.

# Appendix H. Simplified modelling of a hot-filament CVD reactor for fast estimates of concentrations of active chemical species in gas-phase.

One-dimensional [39,147] and comprehensive two [121] and three dimensional (2D and 3D) models [122,123] were used previously to simulate the physical and chemical processes in hot-filament reactors. The 2D and 3D models describe chemically non-equilibrium gas flow based on the rigorous conservation equations for mass, momentum, energy and species number densities. Nevertheless, there is a lack of data on the gas-phase concentrations of the chemically active species required for the surface kinetic models of the diamond growth. To fill this gap, we develop a one-dimensional model which considers radial diffusive transport of chemically reacting gas species at a given temperature distribution between filament and substrate. Although such modelling does not describe all the details of chemically reacting flow on the same level of accuracy as the full multi-dimensional models, the model is robust, and it can be used for the fast estimates of the active species concentrations in the reactors and used to reveal important chemical mechanisms at play.

We consider conditions typical for a standard hot filament reactor operating at pressures 20 - 100 Torr. The filament is made of refractory metal (tungsten or tantalum) of millimeter diameter and is typically positioned tens of millimeters away from the substrate surface. The reactor is filled up with $CH_4$ and $H_2$ as source gases. After pyrolysis of $CH_4/H_2$ mixture the gas becomes mixture consist of 12 main species: H, $CH_4$, $CH_3$, $CH_2$(S), $CH_2$, $C_2H_4$, $C_2H_3$, $C_2H_5$, $C_2H_2$, $C_2H_6$, $C_2H$, and $H_2$. The key reactions between these components are given in previous studies [39,121–123,147] and the list of the reactions used in our model is given in Table H1.

The reactor is represented in 1D cylindrical coordinates such that only the radial dependence of the concentrations of all species on the filament-to-substrate distance is considered in the model equations. When considering the diffusive transport of the species, we assume that the major component of the mixture is molecular hydrogen, while the concentrations of the other species are relatively small. Under these assumptions, the equations describing the diffusion of each component of the mixture in $H_2$ can be written as follows [148]:

$$\frac{\partial(cx_i)}{\partial t}-\frac{1}{r}\frac{\partial}{\partial r}\left(r\left[\frac{M_{H2}}{M_{i,H2}}cD_{i,H2}\frac{\partial x_i}{\partial r}+\frac{D_i^T}{M_i}\frac{1}{T}\frac{\partial T}{\partial r}\right]\right)=S_i \qquad (H1)$$

Here $x_i$ – is the molar fraction of a species *i*, *i*=1, *N*−1, where *N* is the total number of species in a mixture, c – is the gas concentration; $M_{H2}$ and $M_i$ are molar masses of $H_2$ and the *i*-th component of the mixture, $M_{i,H2}=M_i\cdot M_{H2}/(M_{H2}+M_i)$ is the reduced molar mass; $D_{i,H2}$ is the diffusion coefficient of the *i*th species in $H_2$ and $D_i^T$ is the thermal diffusion coefficient, $D_i^T=D_{i,H2}ck_T\frac{M_iM_{H2}}{M_{i,H2}}$, where $k_T$ is the thermal diffusion ratio; *T* is the gas temperature, and $S_i$ is the source term.

The source term $S_i$ in Eq.(H1) is written in the form:

$$S_i=R_i+\frac{c_0(x_{0i}-x_i)}{\tau}, \qquad (H2)$$

where $R_i$ is the total rate of transformation of the *i*th component due to chemical reactions, $\tau$ is the effective residence time of the chemical species due to species non-radial diffusion or gas flow, $x_{0i}$ are the reference molar fractions of the components in the initial mixture, $c_0$ is the total gas concentration corresponding to the temperature $T_0$ = 293 K and the gas pressure in the reactor $p=c_0RT_0$, *R* is the universal gas constant. The system (H1) consisting of *N*−1 equations for the molar concentrations $x_i$ of all species except for $H_2$ is solved together with the equation $\sum_{i=1}^{N}x_i=1$.

Instead of solving the gas energy balance in the reactor to obtain gas temperature profile, $T(r)$ we use the analytical expression from Ref. [121]:

$$T(r)=T_{\mathrm{nf}}\sqrt{1-a\cdot\ln\left(\frac{r}{R_{\mathrm{f}}}\right)},\ \ a=\frac{\left(1-\frac{T_{\mathrm{L}}{}^2}{T_{\mathrm{nf}}{}^2}\right)}{\ln\left(\frac{L}{R_{\mathrm{f}}}\right)} \qquad (H3)$$

Here $T_{\mathrm{nf}}$ and $T_L$ are the gas temperatures near the filament and at the substrate, $R_f$ is the filament radius, *L* is the filament-to-substrate distance. The expression (H3) gives the temperature profile in a gas between two coaxial cylinders sustained at $T_{\mathrm{nf}}$ and $T_L$, with the gas thermal conductivity being a linear function of the gas temperature $\lambda=aT$, where a=const. T The total concentration of gas at each spatial point c(r) is determined from the condition of constant pressure in reactor: *p*=*c*(r)*RT*(r)=*const*.

The boundary conditions to the equations (H1) are specified on the filament and substrate surfaces. It is well known that one of the main processes that activates the $CH_4/H_2$ mixture is $H_2$ dissociation on the filament [121]. Therefore, we assume that there is a flux of H atoms into the reactor coming from the wire surface due to this process, $Q_{cat}$ [$cm^{-2}s^{-1}$], with the value of $Q_{cat}$ taken from the previous studies [121–123]. At the substrate surface, the H atoms are lost with the probability of $\gamma_H$ [122,141]:

$$-D_{H,H2}\frac{\partial c_H}{\partial r}\bigg|_{r=L} = \gamma\frac{c_H v_T}{4},\ \ \gamma_H = \frac{0.83}{1 + 0.3\exp(3430/T) + 0.1\exp(-4420/T)c_{H_2}/c_H}$$

where $v_T$ is the mean thermal velocity of H atoms, and $c_H$ and $c_{H_2}$ are, respectively, H atoms and $H_2$ molecules concentrations near the substrate. For the other species, the zero diffusive fluxes on the filament and substrate are assumed. The binary diffusion coefficients $D_{i,H2}$ in Eq.(H1) were estimated as in Ref. [124] using the Chapman-Enskog kinetic theory. The data on the thermal diffusion ratio $k_T$ were taken from [124] as well, which assume a binary gas mixture for each species in Eq.(H1). The equations (H1) were discretized by the finite volume method on equidistant mesh and solved numerically up to the steady state.

Although the actual geometry of most hot-filament reactors is multi-dimensional, we tried to compare the results given by our model with the experimental data [149] to make sure that our simulations can reproduce the measured $CH_3$ concentrations within an order of magnitude. In these simulations, there is no substrate, and the gas flow velocity is very small such that we can neglect the last term in Eq. (H2). The properties of the gas mixture at the reactor periphery corresponding to r = 2 cm are assumed to coincide with the input mixture of $H_2$ with 0.5% $CH_4$ at p = 20 Torr and T = 300 K. The radius of the filament is 0.01 cm.

Because the actual gas temperature near the filament $T_{nf}$ is unknown, we put $T_{nf}$ to be equal to the filament temperature $T_f$ in the experiment $T_{nf} = T_f$ = 2400 K or assumed the difference $\Delta T$=400 K between $T_f$ and $T_{nf}$ such that $T_{nf}$ = 2000 K. The rate of the H atom production at the filament surface was taken to be $Q_{cat} = 6.5\times10^{19}$ $cm^{-2}s^{-1}$, as it follows from the previous detailed studies [150].

The comparison between our simulations and experiment is shown in Figure H1. As can be seen, the model can predict the concentration of $CH_3$ within a factor of 1.5. Given the complexity of the problem and many effects that require special treatment (for example, the calculation of the H atom production at the filament and the temperature distribution in the gas), we consider that the agreement between our simulations and the experiment is good. Thus we can use our gas phase model for calculations of the input data for the surface chemistry model.

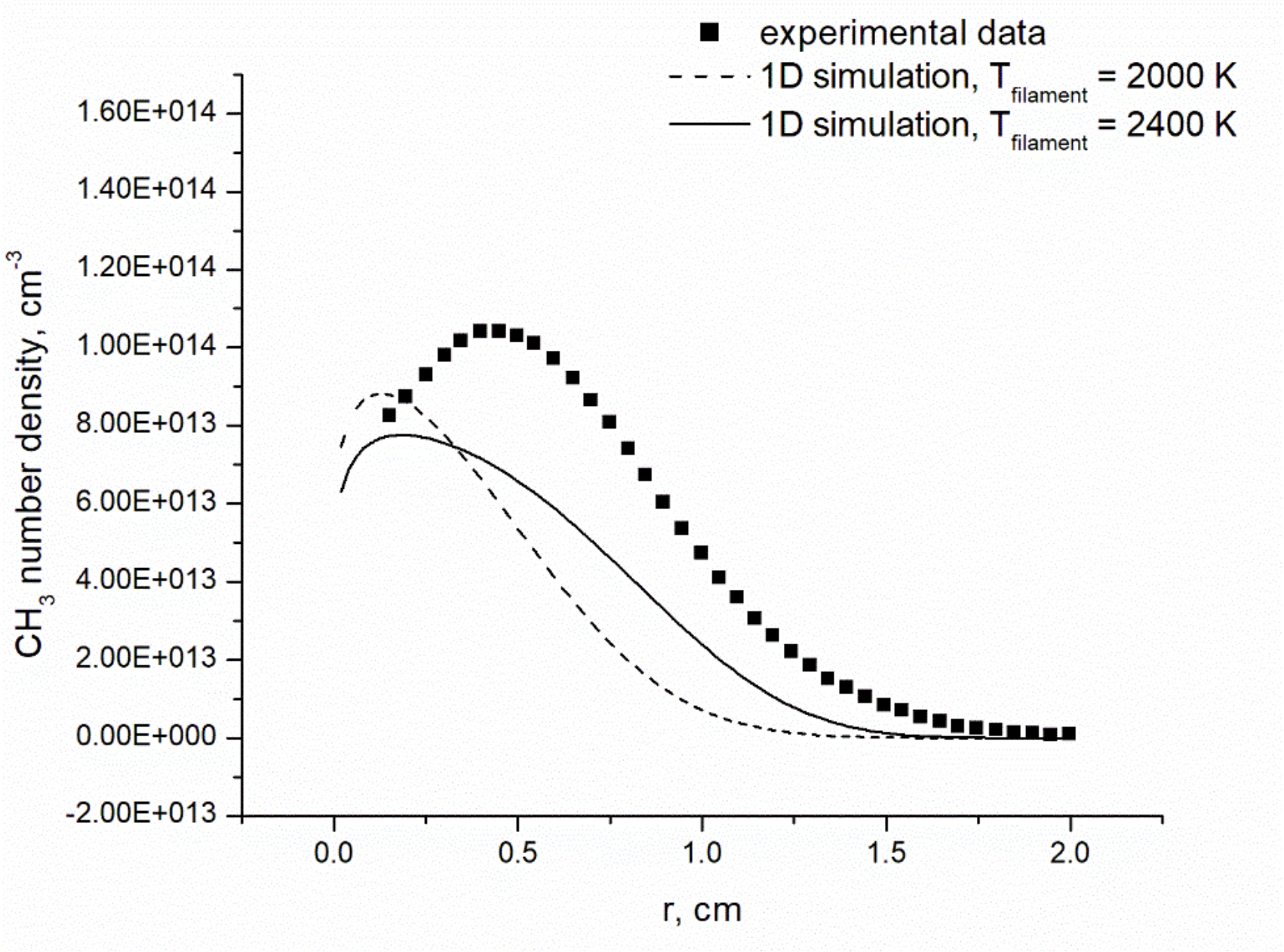

**Figure H1.** Radial distribution of the $CH_3$ concentration near the hot filament, as measured in [149] and calculated in this paper by simplified 1D model.

**Table H1.** The list of the reactions considered in the model with the references on data sources for the chemical reaction constants. With the equilibrium rate constant given, we calculate rate of the reverse reaction from the forward reaction and the equilibrium rate constant.

| No | Reaction | Data source for reaction constants | Data source for the equilibrium constant |
|---|---|---|---|
| 1 | $H + H + H_2 \rightleftharpoons H_2 + H_2$ | Ref. [151] | Ref. [152] |
| 2 | $H + CH_4 \rightleftharpoons CH_3 + H_2$ | Ref. [151] | Ref. [152] |
| 3 | $CH_4 + H_2 = H + CH_3 + H_2$ | Ref. [151] | Ref. [152] |
| 4 | $CH_2(S) + H_2 = CH_3 + H$ | Ref. [151] | Ref. [152] |
| 5 | $CH_2 + H_2 = CH_3 + H$ | Ref. [151] | Calculated from Thermochemical Tables |
| 6 | $CH_2(S) + H_2 = CH_2 + H_2$ | Ref. [151] | Ref. [152] |
| 7 | $CH_2 + CH_3 = C_2H_4 + H$ | Ref. [151] | Ref. [152] |
| 8 | $H + C_2H_4 = C_2H_3 + H_2$ | Ref. [151] | Ref. [152] |
| 9 | $H + C_2H_4 + H_2 = C_2H_5 + H_2$ | Ref. [151] | Ref. [152] |
| 10 | $CH_3 + CH_3 = H + C_2H_5$ | Ref. [151] | Ref. [152] |
| 11 | $C_2H_3 + H = C_2H_2 + H_2$ | Ref. [151] | Ref. [152] |
| 12 | $C_2H_2 + H + H_2 = C_2H_3 + H_2$ | Ref. [151] | Ref. [152] |
| 13 | $C_2H_4 + H_2 = C_2H_2 + H_2 + H_2$ | Ref. [151] | Ref. [152] |
| 14 | $C_2H_4 + CH_3 = C_2H_3 + CH_4$ | Ref. [151] | Ref. [152] |
| 15 | $H + C_2H_5 = H_2 + C_2H_4$ | Ref. [151] | Ref. [152] |
| 16 | $CH_3 + CH_3 + H_2 = C_2H_6 + H_2$ | Ref. [151] | Ref. [152] |
| 17 | $C_2H_6 + H = C_2H_5 + H_2$ | Ref. [151] | Ref. [152] |
| 18 | $C_2H + H_2 = H + C_2H_2$ | Ref. [151] | Ref. [152] |
| 19 | $H + C_2H + H_2 = C_2H_2 + H_2$ | Ref. [151] | Calculated from Thermochemical Tables |

The input data for the surface chemistry model were calculated for the conditions of experiment [38]. In our simulations we assume that the reactor is filled with 1%$CH_4$ in $H_2$ at the gas pressure 30 Torr, the gas temperature near the filament $T_{nf}$ = 2323 K (equal to filament temperature $T_f$ in the experiment), the substrate temperature $T_{sub}$ and filament-to-substrate distance are varied, respectively, in the ranges 700 - 1600 K and L = 0.6 – 1 cm. The rate of catalytic production of the H atoms at the filament was taken to be Q = 4.5×10$^{19}$ cm$^{-2}$s$^{-1}$. The radius of the filament $R_f$ and the geometric dimensions of the substrate are not specified in [38]. So we took $R_f$ = 0.15 cm, a typical value for the hot-filament reactors, and assumed that the substrate width is around 1 cm, which gives the characteristic diffusive time of $C_2H_2$ along the substrate surface $\tau = 10^{-3}$ s at T = 1000 K. Considering $\tau$ as the fitting parameter, we assume it the same for all species.

The calculated concentrations of the H, $CH_4$, $CH_3$, $C_2H_2$ and $H_2$ near the filament and substrate for the filament-substrate distance L = 0.6 cm are shown in Figure H2. These data were used in the surface chemistry model of the main text. For the range of the substrate temperatures $T_{sub}$ = 1000 – 1500 K at which a significant diamond growth rate was observed (see the data on Figure 12 of the main text), we see that variations in the active species concentrations near the substrate surface are not higher than by a factor of 1.5. For $CH_4$ , $CH_3$ and $H_2$, they correspond to decrease in the total gas density as the gas temperature increases at the constant pressure.

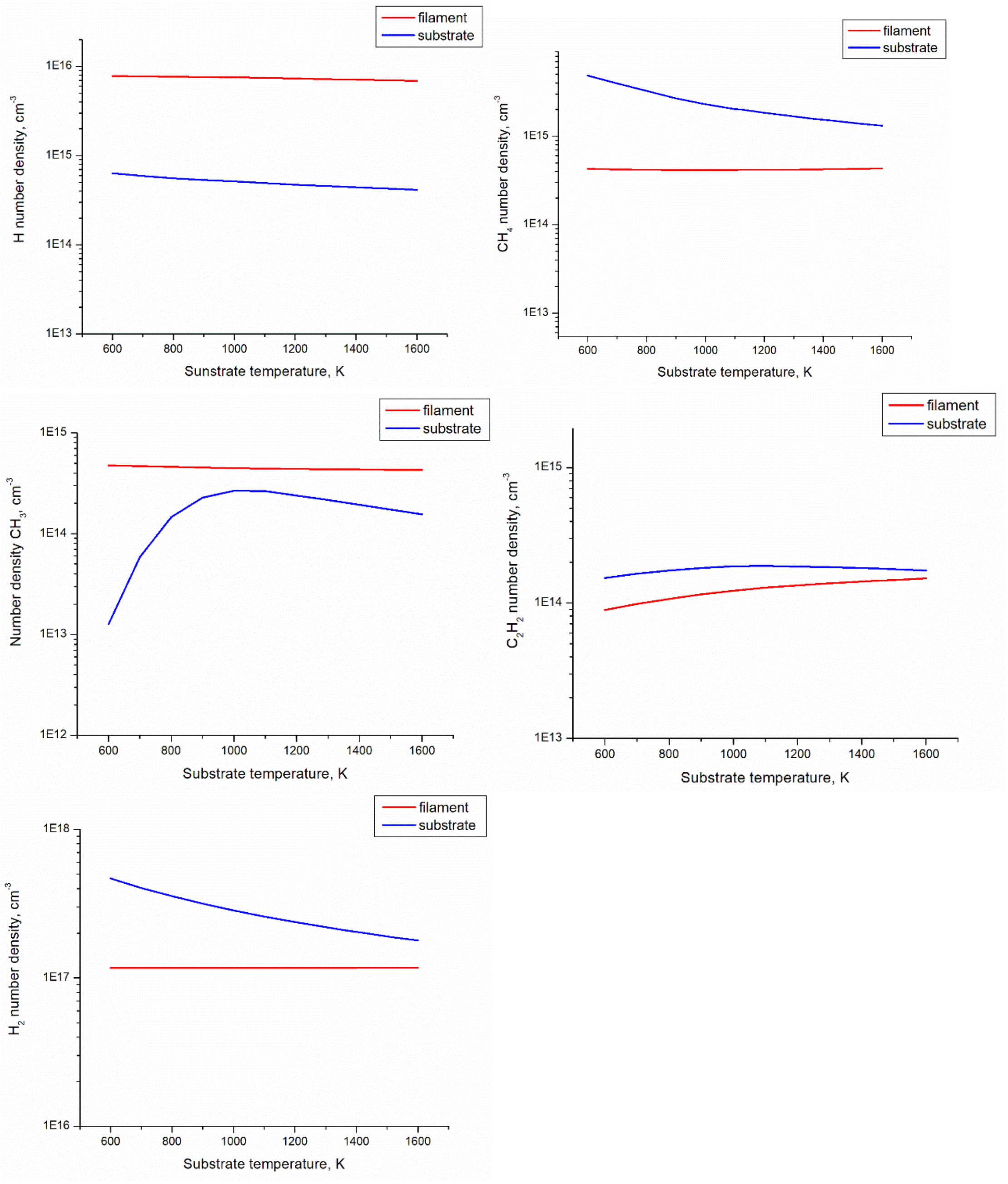

**Figure H2.** The number densities of the main species near the filament and substrate of a hot filament reactor for input gas mixture containing 1% $CH_4$ in $H_2$ under the conditions of experiment [38], as calculated by the developed 1D model. The filament and substrate are separated by L = 0.6 cm.

Let us discuss the physical and chemical processes that are important in the gas phase for the conditions corresponding to the maximum predicted growth rate, i.e. at the substrate temperature close to $T_{sub}$ = 1200 K. Figure H3 shows radial distributions of the H, $CH_4$, $CH_3$, $C_2H_2$ and $H_2$ concentrations at $T_{sub}$ = 1200 K and L = 0.6 cm.

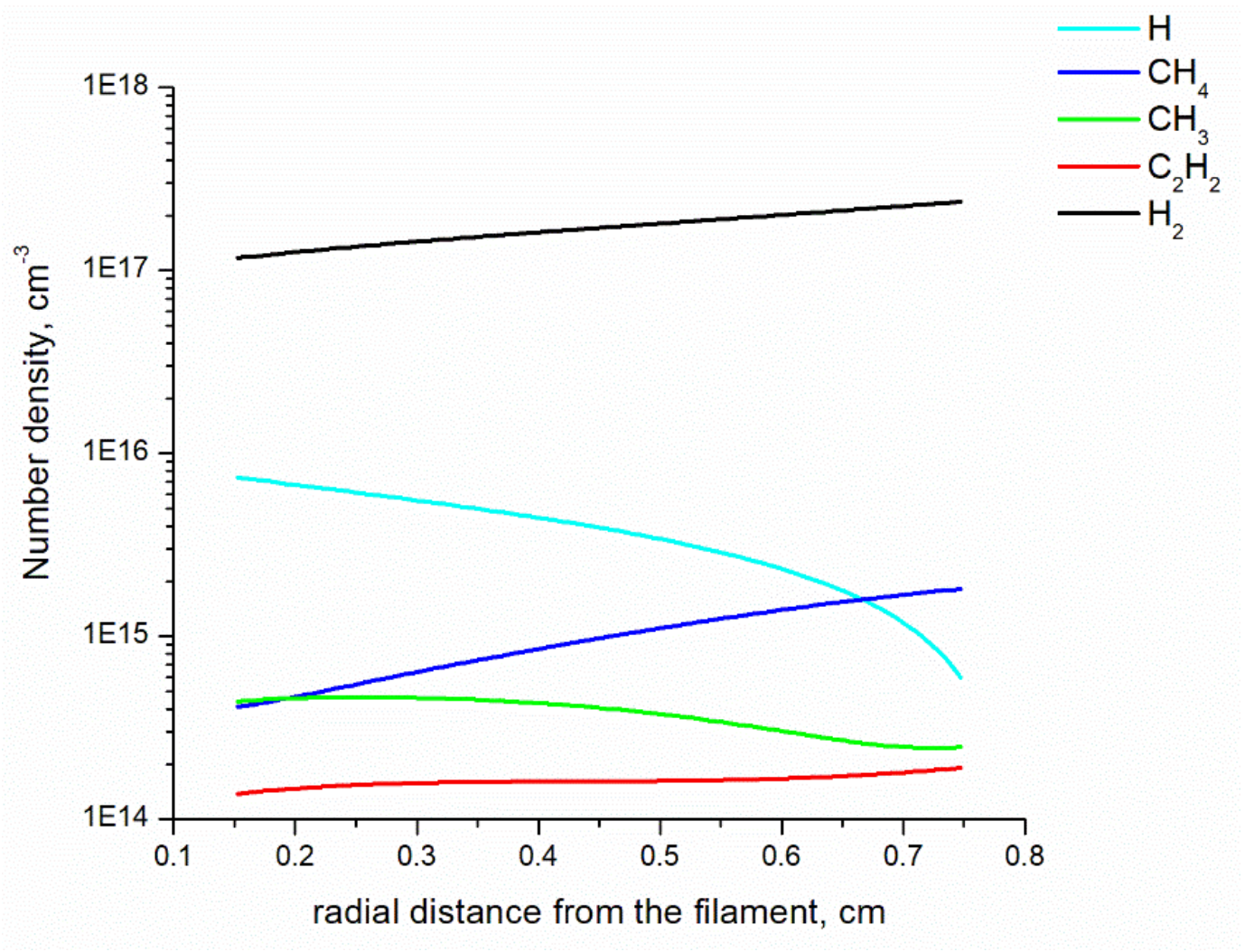

**Figure H3.** Radial dependences of H, $CH_4$, $CH_3$, $C_2H_2$ and $H_2$ number densities calculated for input gas mixture of 1% $CH_4$ in $H_2$ with $T_{sub}$ = 1200 K. The filament and substrate are separated by L = 0.6 cm.

The H atoms density profile of Figure H3 is explained by the H atoms catalytic production at the filament surface and their diffusion to the substrate. The volumetric processes of H atoms production and loss are negligible at the conditions considered. This can be seen from Figure H4, where the H atom distribution computed from the full model is compared with that obtained in the case when the source term in right-hand side of Eq.(H1) for H atoms is neglected.

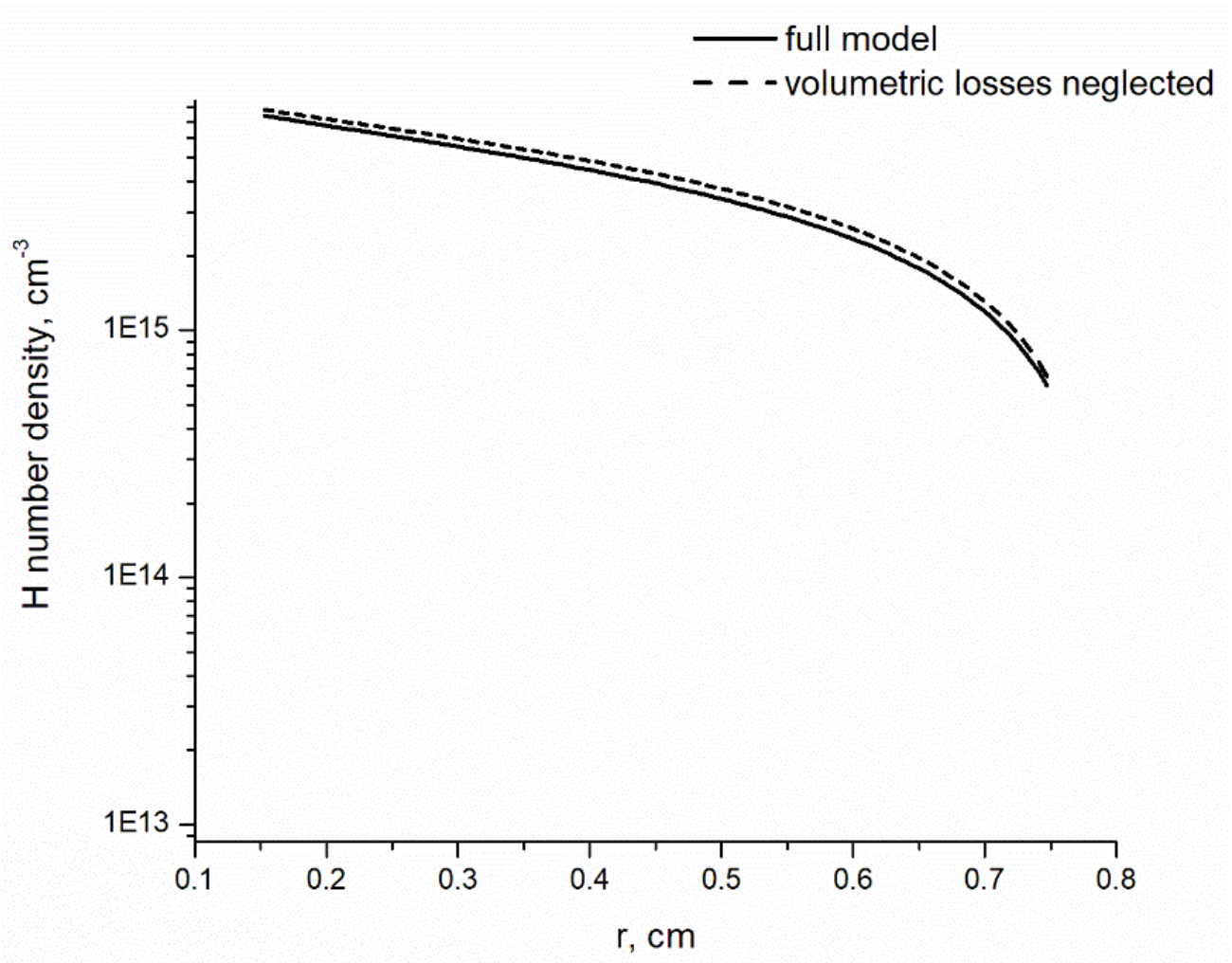

**Figure H4.** Radial dependencies of H number density calculated for input gas mixture of 1% $CH_4$ in $H_2$ using the full model and neglecting the source term $S_i$ in the right-hand-side of the Eq.(H1) for the H atoms.

In contrast to H, both gas-phase chemistry and diffusion transport are important to maintain steady-state concentrations of $CH_4$, $CH_3$ and $C_2H_2$ for the most parts of the reactor. Figures H5 and H6 show the rates of their total chemical transformations R (production minus loss, see Eq.(H2)) and the molar fractions of $CH_4$ , $CH_3$ and $C_2H_2$ as a function of radial coordinate r.

As Figure H5 shows, $CH_4$ molecules are decomposed near the filament, in the “hot” zone, and this depletion is compensated by the diffusion of $CH_4$ molecules from the “cold” zone of the reactor where the molar fraction of $CH_4$ is higher (see Figure H6). In the “cold” zone, the removal of $CH_4$ particles due to diffusion is balanced by their chemical production and by the diffusive income of $CH_4$ molecules of the initial mixture from lateral directions (the last term in Eq.(H2)).

As far as the radial distribution of $CH_3$ is concerned, $CH_3$ particles are produced by the chemical reactions in the “hot” zone, as shown by Figure H5, and they are lost by radial diffusion to the “cold” zone due to the gradient of the molar fraction of $CH_3$ (see Figure H6). In the cold zone the situation is reverse: the loss of the $CH_3$ due to chemical transformations is compensated by the diffusion of $CH_3$ from the “hot” zone. Some of these processes are important for $C_2H_2$ molecules as well. The $C_2H_2$ molecules are produced in the “hot” zone of the reactor (with the production rate much smaller than for $CH_3$) and they radially diffuse to the substrate due to the gradient of their molar fraction, as can be seen from Figures H5 and H6. But since the rate of chemical transformations of $C_2H_2$ in the cold zone is low, the influx of $C_2H_2$ molecules from the “hot” zone is balanced there by the $C_2H_2$ removal due to the transport in lateral directions, described by the second term in Eq.(H2). Indeed our simulations show significant sensitivity of the $C_2H_2$ number density at the substrate to the value of the residence time $\tau$.

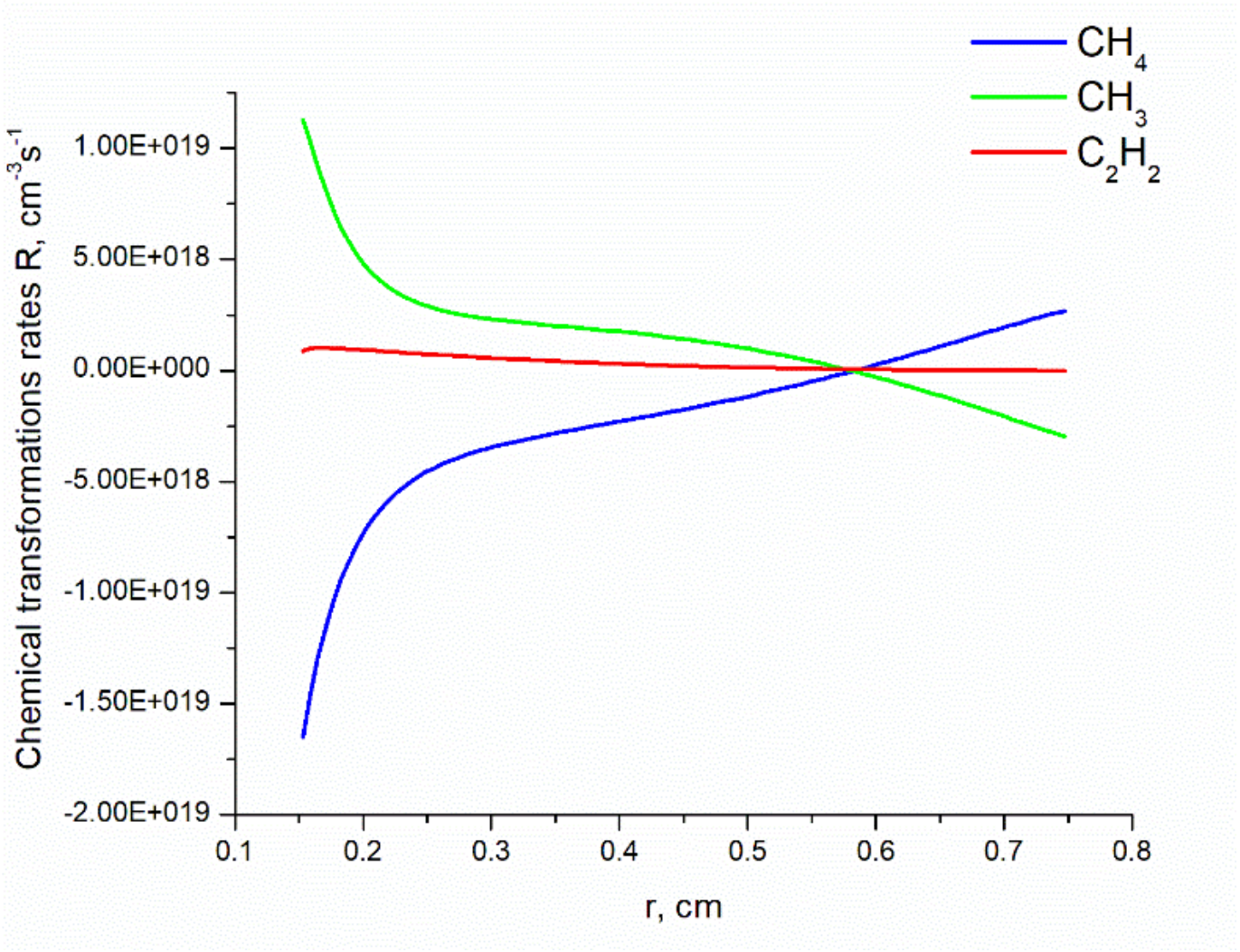

**Figure H5.** The radial dependence of the chemical transformation rates of $CH_4$, $CH_3$ and $C_2H_2$, as calculated by 1D radial model for the gas mixture of 1% $CH_4$ in $H_2$.

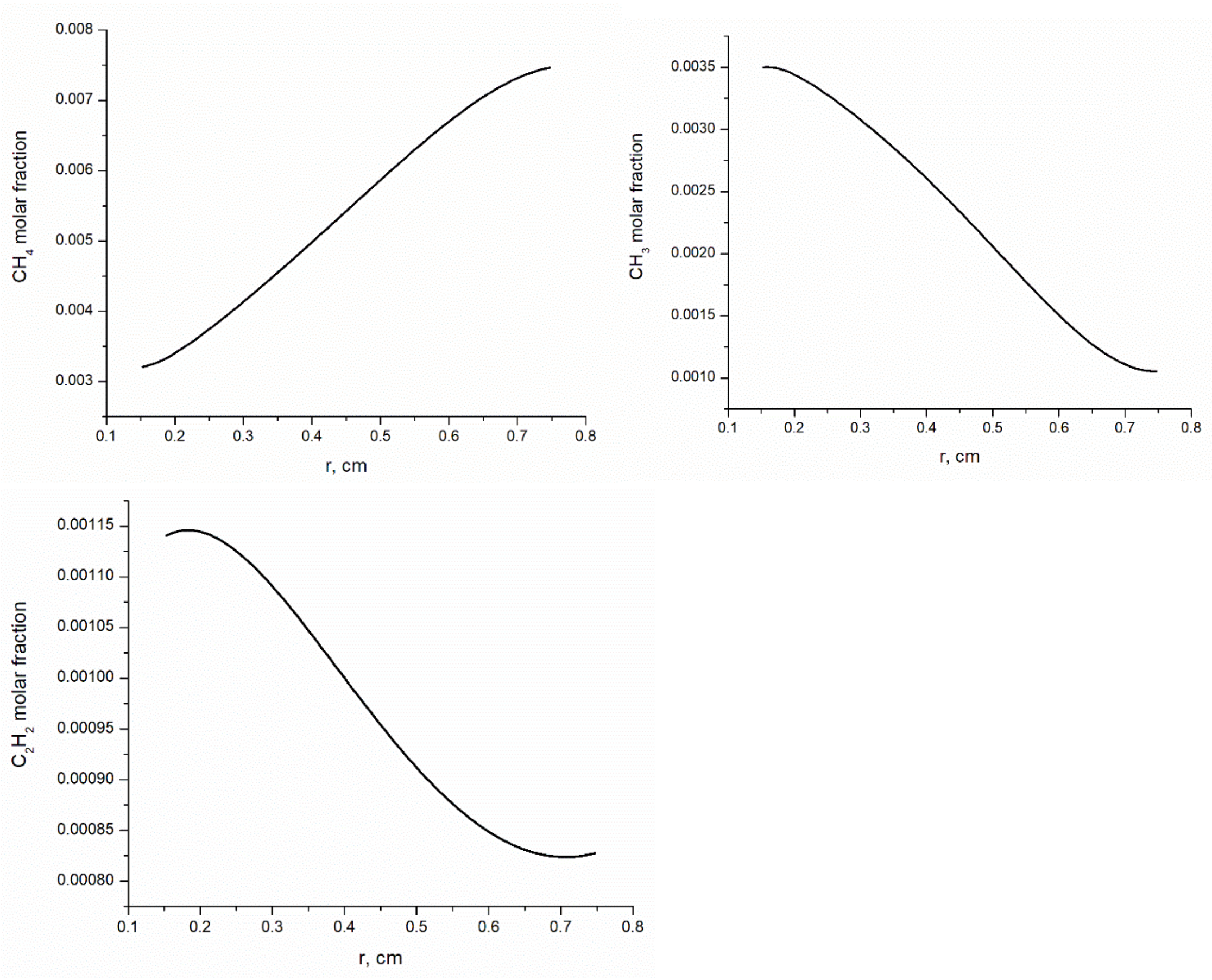

**Figure G6.** Radial dependences of molar fractions of H, $CH_4$, $CH_3$, $C_2H_2$ and $H_2$ calculated for input gas mixture of 1% $CH_4$ in $H_2$.

The following chemical reactions are most important. The $CH_4$ concentration in the “hot” and “cold” zones is determined by the forward and reverse reactions 2 of Table H1. In the “hot” zone the chemical production of $CH_3$ is governed mainly by the reactions 2 and 4. In the “cold” zone, the main chemical processes involved in $CH_3$ transformations are reverse and forward reactions 2 and reverse reaction 3 (see Table H1). In the “hot” zone the reverse reaction (4) produces $CH_2$(S) radicals from $CH_3$ molecules. The $CH_2$(S) radicals are intermediate product of the conversion $CH_4$ into $C_2H_2$ through the chain of reactions 6, 7, 8, and 11 of Table H1 (see diagram on figure 4 of the main text).

# References

[1] J. C. Angus and C. C. Hayman, *Low-Pressure, Metastable Growth of Diamond and “Diamondlike” Phases*, Science **241**, 913 (1988).

[2] W. A. Yarbrough and R. Messier, *Current Issues and Problems in the Chemical Vapor Deposition of Diamond*, Science **247**, 688 (1990).

[3] S. J. Harris and L. R. Martin, *Methyl versus Acetylene as Diamond Growth Species*, J. Mater. Res. **5**, 2313 (1990).

[4] M. Frenklach, *Theory and Models for Nucleation and Growth of Diamond Films*, in *Diamond and Diamond-like Films and Coatings*, edited by R. E. Clausing, L. L. Horton, J. C. Angus, and P. Koidl, Vol. 266 (Springer US, Boston, MA, 1991), pp. 499–523.
[5] Butler, J.E. and Woodin, R.L., *Thin Film Diamond Growth Mechanisms*, Phil. Trans. R. Soc. Lond. A **342**, 209 (1993).
[6] P. W. May, *Diamond Thin Films: A 21st-Century Material*, Philosophical Transactions: Mathematical, Physical and Engineering Sciences **358**, 473 (2000).
[7] H. Liu and D. S. Dandy, *Diamond Chemical Vapor Deposition*, (n.d.).
[8] Y. Zeng and Y. Sakamoto, *Effect of Frequency on Low Temperature Synthesis of Diamond by Pulsed Discharge MPCVD*, Results in Materials **19**, 100416 (2023).
[9] O. Auciello, *Science and Technology of a Transformational Multifunctional Ultrananocrystalline Diamond (UNCD ™ ) Coating*, Functional Diamond **2**, 1 (2022).
[10] J.-C. Arnault, S. Saada, and V. Ralchenko, *Chemical Vapor Deposition Single-Crystal Diamond: A Review*, Physica Rapid Research Ltrs **16**, 2100354 (2022).
[11] Th. D. Makris, R. Giorgi, N. Lisi, L. Pilloni, and E. Salernitano, *Bias Enhanced Nucleation of Diamond on Si(100) in a Vertical Straight Hot Filament CVD*, Diamond and Related Materials **14**, 318 (2005).
[12] X. T. Zhou, H. L. Lai, H. Y. Peng, C. Sun, W. J. Zhang, N. Wang, I. Bello, C. S. Lee, and S. T. Lee, *Heteroepitaxial Nucleation of Diamond on Si(100) via Double Bias-Assisted Hot Filament Chemical Vapor Deposition*, Diamond and Related Materials **9**, 134 (2000).
[13] K. Janischowsky, W. Ebert, and E. Kohn, *Bias Enhanced Nucleation of Diamond on Silicon (100) in a HFCVD System*, Diamond and Related Materials **12**, 336 (2003).
[14] S. Pecoraro, J. C. Arnault, and J. Werckmann, *BEN-HFCVD Diamond Nucleation on Si(111) Investigated by HRTEM and Nanodiffraction*, Diamond and Related Materials **14**, 137 (2005).
[15] Y. Li, J. Li, Q. Wang, Y. Yang, and C. Gu, *Controllable Growth of Nanocrystalline Diamond Films by Hot-Filament Chemical Vapor Deposition Method*, J. Nanosci. Nanotech. **9**, 1062 (2009).
[16] S. G. Ansari, T. L. Anh, H.-K. Seo, K.-G. Sung, D. Mushtaq, and H.-S. Shin, *Growth Kinetics of Diamond Film with Bias Enhanced Nucleation and H2/CH4/Ar Mixture in a Hot-Filament Chemical Vapor Deposition System*, Journal of Crystal Growth **265**, 563 (2004).
[17] E. V. Bushuev, V. Yu. Yurov, A. P. Bolshakov, V. G. Ralchenko, E. E. Ashkinazi, A. V. Ryabova, I. A. Antonova, P. V. Volkov, A. V. Goryunov, and A. Yu. Luk'yanov, *Synthesis of Single Crystal Diamond by Microwave Plasma Assisted Chemical Vapor Deposition with in Situ Low-Coherence Interferometric Control of Growth Rate*, Diamond and Related Materials **66**, 83 (2016).
[18] E. V. Bushuev, V. Yu. Yurov, A. P. Bolshakov, V. G. Ralchenko, A. A. Khomich, I. A. Antonova, E. E. Ashkinazi, V. A. Shershulin, V. P. Pashinin, and V. I. Konov, *Express in Situ Measurement of Epitaxial CVD Diamond Film Growth Kinetics*, Diamond and Related Materials **72**, 61 (2017).
[19] E. V. Grigoryev, V. N. Savenko, D. V. Sheglov, A. V. Matveev, V. A. Cherepanov, A. V. Zolkin, and B. A. Kolesov, *Synthesis of Diamond Crystals from Oxygen-Acetylene Flames on a Metal Substrate at Low Temperature*, Carbon **36**, 581 (1998).
[20] J. Lu, Y. Gu, T. A. Grotjohn, T. Schuelke, and J. Asmussen, *Experimentally Defining the Safe and Efficient, High Pressure Microwave Plasma Assisted CVD Operating Regime for Single Crystal Diamond Synthesis*, Diamond and Related Materials **37**, 17 (2013).
[21] W. G. S. Leigh, J. A. Cuenca, E. L. H. Thomas, S. Mandal, and O. A. Williams, *Mapping the Effect of Substrate Temperature Inhomogeneity during Microwave Plasma-Enhanced Chemical Vapour Deposition Nanocrystalline Diamond Growth*, Carbon **201**, 328 (2023).
[22] D. Das and R. N. Singh, *A Review of Nucleation, Growth and Low Temperature Synthesis of Diamond Thin Films*, International Materials Reviews **52**, 29 (2007).
[23] M. Malakoutian et al., *Cooling Future System-on-Chips with Diamond Inter-Tiers*, Cell Reports Physical Science 101686 (2023).

[24] J. Millán-Barba, A. Taylor, H. Bakkali, R. Alcantara, F. Lloret, R. G. De Villoria, M. Dominguez, V. Mortet, M. Gutiérrez, and D. Araújo, *Low Temperature Growth of Nanocrystalline Diamond: Insight Thermal Property*, Diamond and Related Materials **137**, 110070 (2023).
[25] J. Stiegler, T. Lang, M. Nyga˚rd-Ferguson, Y. Von Kaenel, and E. Blank, *Low Temperature Limits of Diamond Film Growth by Microwave Plasma-Assisted CVD*, Diamond and Related Materials **5**, 226 (1996).
[26] D. Dai, Y. Li, and J. Fan, *Room-Temperature Synthesis of Various Allotropes of Carbon Nanostructures (Graphene, Graphene Polyhedra, Carbon Nanotubes and Nano-Onions, n-Diamond Nanocrystals) with Aid of Ultrasonic Shock Using Ethanol and Potassium Hydroxide*, Carbon **179**, 133 (2021).
[27] H. Krzyżanowska et al., *Low Temperature Diamond Growth Arising from Ultrafast Pulsed-Laser Pretreatment*, Carbon **131**, 120 (2018).
[28] C. J. Chu, R. H. Hauge, J. L. Margrave, and M. P. D'Evelyn, *Growth Kinetics of (100), (110), and (111) Homoepitaxial Diamond Films*, Applied Physics Letters **61**, 1393 (1992).
[29] S.-F. Lee, C.-S. Yeh, Y.-P. Chang, D.-C. Wang, and B.-R. Huang, *Improved Quality of Polycrystalline Diamond Grown by Two-Stage Growth in Microwave Plasma Chemical Vapor Deposition*, Carbon **48**, 314 (2010).
[30] J.-W. Park, K.-S. Kim, and N.-M. Hwang, *Gas Phase Generation of Diamond Nanoparticles in the Hot Filament Chemical Vapor Deposition Reactor*, Carbon **106**, 289 (2016).
[31] L. Basso, M. Cazzanelli, M. Orlandi, and A. Miotello, *Nanodiamonds: Synthesis and Application in Sensing, Catalysis, and the Possible Connection with Some Processes Occurring in Space*, Applied Sciences **10**, 4094 (2020).
[32] X. Wang, X. Shen, J. Gao, and F. Sun, *Consecutive Deposition of Amorphous SiO2 Interlayer and Diamond Film on Graphite by Chemical Vapor Deposition*, Carbon **117**, 126 (2017).
[33] C.-H. Ku and J.-J. Wu, *Effects of CCl4 Concentration on Nanocrystalline Diamond Film Deposition in a Hot-Filament Chemical Vapor Deposition Reactor*, Carbon **42**, 2201 (2004).
[34] P. W. May, J. N. Harvey, J. A. Smith, and Yu. A. Mankelevich, *Reevaluation of the Mechanism for Ultrananocrystalline Diamond Deposition from Ar/CH4/H2 Gas Mixtures*, Journal of Applied Physics **99**, 104907 (2006).
[35] P. W. May, M. N. R. Ashfold, and Yu. A. Mankelevich, *Microcrystalline, Nanocrystalline, and Ultrananocrystalline Diamond Chemical Vapor Deposition: Experiment and Modeling of the Factors Controlling Growth Rate, Nucleation, and Crystal Size*, Journal of Applied Physics **101**, 053115 (2007).
[36] P. W. May and Y. A. Mankelevich, *From Ultrananocrystalline Diamond to Single Crystal Diamond Growth in Hot Filament and Microwave Plasma-Enhanced CVD Reactors: A Unified Model for Growth Rates and Grain Sizes*, J. Phys. Chem. C **112**, 12432 (2008).
[37] D.-W. Kweon, J.-Y. Lee, and D. Kim, *The Growth Kinetics of Diamond Films Deposited by Hot-Filament Chemical Vapor Deposition*, Journal of Applied Physics **69**, 8329 (1991).
[38] E. Kondoh, T. Ohta, T. Mitomo, and K. Ohtsuka, *Determination of Activation Energies for Diamond Growth by an Advanced Hot Filament Chemical Vapor Deposition Method*, Applied Physics Letters **59**, 488 (1991).
[39] E. Kondoh, T. Ohta, T. Mitomo, and K. Ohtsuka, *Experimental and Calculational Study on Diamond Growth by an Advanced Hot Filament Chemical Vapor Deposition Method*, Journal of Applied Physics **72**, 705 (1992).
[40] B. V. Spitsyn, *Chemical Crystallization of Diamond from the Activated Vapor Phase*, Journal of Crystal Growth **99**, 1162 (1990).
[41] A. Yamaguchi, M. Ihara, and H. Komiyama, *Temperature Dependence of Growth Rate for Diamonds Grown Using a Hot Filament Assisted Chemical Vapor Deposition Method at Low Substrate Temperatures*, Applied Physics Letters **64**, 1306 (1994).

[42] M. Yu. Plotnikov, Yu. E. Gorbachev, A. A. Emelyanov, D. V. Leshchev, A. K. Rebrov, N. I. Timoshenko, and I. B. Yudin, *Gas-Jet HFCVD Synthesis of Diamonds from Mixtures of Hydrogen with Ethylene and Methane*, Diamond and Related Materials **130**, 109505 (2022).
[43] G. R. Lai, E. N. Farabaugh, A. Feldman, and L. H. Robins, *Deposition of Diamond Films in A Closed Hot Filament Cvd System*, J. Res. Natl. Inst. Stand. Technol. **100**, 43 (1995).
[44] R. Haubner and B. Lux, *Diamond Growth by Hot-Filament Chemical Vapor Deposition: State of the Art*, Diamond and Related Materials **2**, 1277 (1993).
[45] S. Matsumoto, Y. Sato, M. Kamo, and N. Setaka, *Vapor Deposition of Diamond Particles from Methane*, Jpn. J. Appl. Phys. **21**, L183 (1982).
[46] S. Matsumoto, Y. Sato, M. Tsutsumi, and N. Setaka, *Growth of Diamond Particles from Methane-Hydrogen Gas*, J Mater Sci **17**, 3106 (1982).
[47] E. M. A. Fuentes-Fernandez et al., *Synthesis and Characterization of Microcrystalline Diamond to Ultrananocrystalline Diamond Films via Hot Filament Chemical Vapor Deposition for Scaling to Large Area Applications*, Thin Solid Films **603**, 62 (2016).
[48] S. Ohmagari, H. Yamada, H. Umezawa, N. Tsubouchi, A. Chayahara, and Y. Mokuno, *Growth and Characterization of Freestanding P+ Diamond (100) Substrates Prepared by Hot-Filament Chemical Vapor Deposition*, Diamond and Related Materials **81**, 33 (2018).
[49] S. Ohmagari, *Single-Crystal Diamond Growth by Hot-Filament CVD: A Recent Advances for Doping, Growth Rate and Defect Controls*, Functional Diamond **3**, 2259941 (2023).
[50] J. J. Alcantar-Peña, J. Montes, M. J. Arellano-Jimenez, J. E. O. Aguilar, D. Berman-Mendoza, R. García, M. J. Yacaman, and O. Auciello, *Low Temperature Hot Filament Chemical Vapor Deposition of Ultrananocrystalline Diamond Films with Tunable Sheet Resistance for Electronic Power Devices*, Diamond and Related Materials **69**, 207 (2016).
[51] T. Tabakoya et al., *High-Rate Growth of Single-Crystalline Diamond (100) Films by Hot-Filament Chemical Vapor Deposition with Tantalum Filaments at 3000 °C*, Physica Status Solidi (a) **216**, 1900244 (2019).
[52] Y. Takamori, M. Nagai, T. Tabakoya, Y. Nakamura, S. Yamasaki, C. E. Nebel, X. Zhang, T. Matsumoto, T. Inokuma, and N. Tokuda, *Insight into Temperature Impact of Ta Filaments on High-Growth-Rate Diamond (100) Films by Hot-Filament Chemical Vapor Deposition*, Diamond and Related Materials **118**, 108515 (2021).
[53] K. H. Lee, W. K. Seong, and R. S. Ruoff, *CVD Diamond Growth: Replacing the Hot Metallic Filament with a Hot Graphite Plate*, Carbon **187**, 396 (2022).
[54] Y. Mitsuda, T. Yoshida, and K. Akashi, *Development of a New Microwave Plasma Torch and Its Application to Diamond Synthesis*, Review of Scientific Instruments **60**, 249 (1989).
[55] H. Maeda, K. Ohtsubo, M. Irie, N. Ohya, K. Kusakabe, and S. Morooka, *Determination of Diamond [100] and [111] Growth Rate and Formation of Highly Oriented Diamond Film by Microwave Plasma-Assisted Chemical Vapor Deposition*, J. Mater. Res. **10**, 3115 (1995).
[56] A. Popovich, A. Martyanov, A. Khomich, P. Fedotov, S. Savin, V. Sedov, and V. Ralchenko, *CVD Diamond-SiC Composite Films: Structure and Electrical Properties*, Diamond and Related Materials **125**, 108975 (2022).
[57] Y. Muranaka, H. Yamashita, and H. Miyadera, *Low Temperature (∼400 °C) Growth of Polycrystalline Diamond Films in the Microwave Plasma of CO/H2 and CO/H2/Ar Systems*, Journal of Vacuum Science & Technology A: Vacuum, Surfaces, and Films **9**, 76 (1991).
[58] A. Martyanov, I. Tiazhelov, S. Savin, V. Voronov, V. Konov, and V. Sedov, *Synthesis of Polycrystalline Diamond Films in Microwave Plasma at Ultrahigh Concentrations of Methane*, Coatings **13**, 751 (2023).

[59] T. Nakai, K. Arima, O. Maida, and T. Ito, *High-Quality Diamond Films Grown at High Deposition Rates Using High-Power-Density MWPCVD Method with Conventional Quartz-Type Chamber*, Journal of Crystal Growth **309**, 134 (2007).
[60] H. Maeda, S. Masuda, K. Kusakabe, and S. Morooka, *Nucleation and Growth of Diamond in a Microwave Plasma on Substrate Pretreated with Non-Oxide Ceramic Particles*, Journal of Crystal Growth **121**, 507 (1992).
[61] W. G. S. Leigh, E. L. H. Thomas, J. A. Cuenca, S. Mandal, and O. A. Williams, *In-Situ Monitoring of Microwave Plasma-Enhanced Chemical Vapour Deposition Diamond Growth on Silicon Using Spectroscopic Ellipsometry*, Carbon **202**, 204 (2023).
[62] M. Kamo, Y. Sato, S. Matsumoto, and N. Setaka, *Diamond Synthesis from Gas Phase in Microwave Plasma*, Journal of Crystal Growth **62**, 642 (1983).
[63] Y. Saito, S. Matsuda, and S. Nogita, *Synthesis of Diamond by Decomposition of Methane in Microwave Plasma*, J Mater Sci Lett **5**, 565 (1986).
[64] K. Kitahama, K. Hirata, H. Nakamatsu, S. Kawai, N. Fujimori, T. Imai, H. Yoshino, and A. Doi, *Synthesis of Diamond by Laser-Induced Chemical Vapor Deposition*, Applied Physics Letters **49**, 634 (1986).
[65] P. A. Molian, B. Janvrin, and A. M. Molian, *Laser Chemical Vapor Deposition of Fluorinated Diamond Thin Films for Solid Lubrication*, Wear **165**, 133 (1993).
[66] J. H. D. Rebello, D. L. Straub, and V. V. Subramaniam, *Diamond Growth from a CO/CH4 Mixture by Laser Excitation of CO: Laser Excited Chemical Vapor Deposition*, Journal of Applied Physics **72**, 1133 (1992).
[67] K. A. Snail and C. M. Marks, In Situ *Diamond Growth Rate Measurement Using Emission Interferometry*, Applied Physics Letters **60**, 3135 (1992).
[68] B. Zhang and S. Chen, *Morphological Evolution of Diamonds in Combustion Synthesis*, Journal of Applied Physics **79**, 7241 (1996).
[69] R. J. H. Klein-Douwel, J. J. L. Spaanjaars, and J. J. Ter Meulen, *Two-Dimensional Distributions of C2, CH, and OH in a Diamond Depositing Oxyacetylene Flame Measured by Laser Induced Fluorescence*, Journal of Applied Physics **78**, 2086 (1995).
[70] W. Zhu, B. H. Tan, J. Ahn, and H. S. Tan, *Crystal Growth and Gas Ratio Effect of Diamond Films Synthesized by Oxyacetylene Flames*, Diamond and Related Materials **2**, 491 (1993).
[71] D. Y. Wang, Y. H. Song, J. J. Wang, and R. Y. Cheng, *Implementation of Large-Scale Deposition of Diamond Films by Combustion Synthesis*, Diamond and Related Materials **2**, 304 (1993).
[72] J. A. Von Windheim, F. Sivazlian, M. T. McClure, J. T. Glass, and J. T. Prater, *Nucleation and Growth of Diamond Using a Computer-Controlled Oxy-Acetylene Torch*, Diamond and Related Materials **2**, 438 (1993).
[73] P. Alers, W. Hänni, and H. E. Hintermann, *A Comparative Study of Laminar and Turbulent Oxygen-Acetylene Flames for Diamond Deposition*, Diamond and Related Materials **2**, 393 (1993).
[74] L. M. Hanssen, W. A. Carrington, J. E. Butler, and K. A. Snail, *Diamond Synthesis Using an Oxygen-Acetylene Torch*, Materials Letters **7**, 289 (1988).
[75] M. Murayama, S. Kojima, and K. Uchida, *Uniform Deposition of Diamond Films Using a Flat Flame Stabilized in the Stagnation-Point Flow*, Journal of Applied Physics **69**, 7924 (1991).
[76] Goodwin, D.G. and Butler, J.E., *Theory of Diamond Chemical Vapor Deposition*, in *Handbook of Industrial Diamonds and Diamond Films*, edited by M. A. Prelas, G. Popovici, and L. K. Bigelow (Marcel Dekker, New York, 1998).
[77] Y. Liou, R. Weimer, D. Knight, and R. Messier, *Effect of Oxygen in Diamond Deposition at Low Substrate Temperatures*, Applied Physics Letters **56**, 437 (1990).
[78] Y. Liou, A. Inspektor, R. Weimer, and R. Messier, *Low-Temperature Diamond Deposition by Microwave Plasma-Enhanced Chemical Vapor Deposition*, Applied Physics Letters **55**, 631 (1989).

[79] T. Kawato and K. Kondo, *Effects of Oxygen on CVD Diamond Synthesis*, Jpn. J. Appl. Phys. **26**, 1429 (1987).
[80] H. Matsuyama, N. Sato, and H. Kawakami, *Effect of Oxygen on Diamond Deposition in $CH_4/O_2/H_2$ Gas Mixtures*, MRS Proc. **334**, 135 (1993).
[81] M. Malakoutian, R. Soman, K. Woo, and S. Chowdhury, *Development of 300–400 °C Grown Diamond for Semiconductor Devices Thermal Management*, MRS Advances **9**, 7 (2023).
[82] L. N. Krasnoperov, I. J. Kalinovski, H. N. Chu, and D. Gutman, *Heterogeneous Reactions of Hydrogen Atoms and Methyl Radicals with a Diamond Surface in the 300-1133 K Temperature Range*, J. Phys. Chem. **97**, 11787 (1993).
[83] S. J. Harris and A. M. Weiner, *Reaction Kinetics on Diamond: Measurement of H Atom Destruction Rates*, Journal of Applied Physics **74**, 1022 (1993).
[84] B. S. Truscott, M. W. Kelly, K. J. Potter, M. N. R. Ashfold, and Y. A. Mankelevich, *Microwave Plasma-Activated Chemical Vapor Deposition of Nitrogen-Doped Diamond. II: CH4/N2/H2 Plasmas*, J. Phys. Chem. A **120**, 8537 (2016).
[85] P. W. May and Yu. A. Mankelevich, *Experiment and Modeling of the Deposition of Ultrananocrystalline Diamond Films Using Hot Filament Chemical Vapor Deposition and Ar/CH4/H2 Gas Mixtures: A Generalized Mechanism for Ultrananocrystalline Diamond Growth*, Journal of Applied Physics **100**, 024301 (2006).
[86] C. A. Rego, P. W. May, C. R. Henderson, M. N. R. Ashfold, K. N. Rosser, and N. M. Everitt, *In-Situ Mass Spectrometric Study of the Gas-Phase Species Involved in CVD of Diamond as a Function of Filament Temperature*, Diamond and Related Materials **4**, 770 (1995).
[87] M. Frenklach and K. E. Spear, *Growth Mechanism of Vapor-Deposited Diamond*, J. Mater. Res. **3**, 133 (1988).
[88] S. Skokov, B. Weiner, and M. Frenklach, *Elementary Reaction Mechanism for Growth of Diamond (100) Surfaces from Methyl Radicals*, J. Phys. Chem. **98**, 7073 (1994).
[89] D. Huang, M. Frenklach, and M. Maroncelli, *Energetics of Acetylene-Addition Mechanism of Diamond Growth*, J. Phys. Chem. **92**, 6379 (1988).
[90] C. J. Chu, M. P. D'Evelyn, R. H. Hauge, and J. L. Margrave, *Mechanism of Diamond Film Growth by Hot-Filament CVD: Carbon-13 Studies*, J. Mater. Res. **5**, 2405 (1990).
[91] C. J. Chu, M. P. D'Evelyn, R. H. Hauge, and J. L. Margrave, *Mechanism of Diamond Growth by Chemical Vapor Deposition on Diamond (100), (111), and (110) Surfaces: Carbon-13 Studies*, Journal of Applied Physics **70**, 1695 (1991).
[92] M. P. D'Evelyn, C. J. Chu, R. H. Hange, and J. L. Margrave, *Mechanism of Diamond Growth by Chemical Vapor Deposition: Carbon-13 Studies*, Journal of Applied Physics **71**, 1528 (1992).
[93] C. E. Johnson, W. A. Weimer, and F. M. Cerio, *Efficiency of Methane and Acetylene in Forming Diamond by Microwave Plasma Assisted Chemical Vapor Deposition*, J. Mater. Res. **7**, 1427 (1992).
[94] W. A. Yarbrough, K. Tankala, and T. DebRoy, *Diamond Growth with Locally Supplied Methane and Acetylene*, J. Mater. Res. **7**, 379 (1992).
[95] L. R. Martin and M. W. Hill, *A Flow-Tube Study of Diamond Film Growth: Methane versus Acetylene*, J Mater Sci Lett **9**, 621 (1990).
[96] L. R. Martin, *Diamond Film Growth in a Flowtube: A Temperature Dependence Study*, Journal of Applied Physics **70**, 5667 (1991).
[97] L. R. Martin, *High-Quality Diamonds from an Acetylene Mechanism*, J Mater Sci Lett **12**, 246 (1993).
[98] M. Frenklach, D. W. Clary, W. C. Gardiner, and S. E. Stein, *Detailed Kinetic Modeling of Soot Formation in Shock-Tube Pyrolysis of Acetylene*, Symposium (International) on Combustion **20**, 887 (1985).
[99] H. Wang and M. Frenklach, *Calculations of Rate Coefficients for the Chemically Activated Reactions of Acetylene with Vinylic and Aromatic Radicals*, J. Phys. Chem. **98**, 11465 (1994).

[100] H. Wang and M. Frenklach, *A Detailed Kinetic Modeling Study of Aromatics Formation in Laminar Premixed Acetylene and Ethylene Flames*, Combustion and Flame **110**, 173 (1997).

[101] A. M. Mebel, Y. Georgievskii, A. W. Jasper, and S. J. Klippenstein, *Temperature- and Pressure-Dependent Rate Coefficients for the HACA Pathways from Benzene to Naphthalene*, Proceedings of the Combustion Institute **36**, 919 (2017).

[102] E. Reizer, B. Viskolcz, and B. Fiser, *Formation and Growth Mechanisms of Polycyclic Aromatic Hydrocarbons: A Mini-Review*, Chemosphere **291**, 132793 (2022).

[103] M. Frenklach and H. Wang, *Detailed Surface and Gas-Phase Chemical Kinetics of Diamond Deposition*, Phys. Rev. B **43**, 1520 (1991).

[104] M. E. Coltrin and D. S. Dandy, *Analysis of Diamond Growth in Subatmospheric Dc Plasma-Gun Reactors*, Journal of Applied Physics **74**, 5803 (1993).

[105] S. Skokov, B. Weiner, and M. Frenklach, *Elementary Reaction Mechanism of Diamond Growth from Acetylene*, J. Phys. Chem. **98**, 8 (1994).

[106] S. Skokov, C. S. Carmer, B. Weiner, and M. Frenklach, *Reconstruction of (100) Diamond Surfaces Using Molecular Dynamics with Combined Quantum and Empirical Forces*, Phys. Rev. B **49**, 5662 (1994).

[107] B. Weiner, S. Skokov, and M. Frenklach, *A Theoretical Analysis of a Diamond (100)-(2×1) Dimer Bond*, The Journal of Chemical Physics **102**, 5486 (1995).

[108] B. J. Garrison, E. J. Dawnkaski, D. Srivastava, and D. W. Brenner, *Molecular Dynamics Simulations of Dimer Opening on a Diamond {001}(2 x 1) Surface*, Science **255**, 835 (1992).

[109] A. Cheesman, J. N. Harvey, and M. N. R. Ashfold, *Studies of Carbon Incorporation on the Diamond {100} Surface during Chemical Vapor Deposition Using Density Functional Theory*, J. Phys. Chem. A **112**, 11436 (2008).

[110] L. M. Oberg, M. Batzer, A. Stacey, and M. W. Doherty, *Nitrogen Overgrowth as a Catalytic Mechanism during Diamond Chemical Vapour Deposition*, Carbon **178**, 606 (2021).

[111] I. I. Oleinik, D. G. Pettifor, A. P. Sutton, C. C. Battaile, D. J. Srolovitz, J. E. Butler, D. S. Dandy, S. J. Harris, and M. P. D'evelyn, *Diamond CVD Growth Mechanisms and Reaction Rates From First-Principles*, MRS Proc. **616**, 123 (2000).

[112] D. Huang and M. Frenklach, *Energetics of Surface Reactions on (100) Diamond Plane*, J. Phys. Chem. **96**, 1868 (1992).

[113] C. C. Battaile, D. J. Srolovitz, and J. E. Butler, *A Kinetic Monte Carlo Method for the Atomic-Scale Simulation of Chemical Vapor Deposition: Application to Diamond*, Journal of Applied Physics **82**, 6293 (1997).

[114] C. C. Battaile, D. J. Srolovitz, I. I. Oleinik, D. G. Pettifor, A. P. Sutton, S. J. Harris, and J. E. Butler, *Etching Effects during the Chemical Vapor Deposition of (100) Diamond*, The Journal of Chemical Physics **111**, 4291 (1999).

[115] M. J. Frisch et al., *Gaussian 16 Rev. C.01*, (2016).

[116] H. Eyring, *The Activated Complex in Chemical Reactions*, J. Chem. Phys. **3**, 107 (1935).

[117] G. Kresse and J. Hafner, Ab Initio *Molecular Dynamics for Liquid Metals*, Phys. Rev. B **47**, 558 (1993).

[118] G. Kresse and J. Furthmüller, *Efficiency of Ab-Initio Total Energy Calculations for Metals and Semiconductors Using a Plane-Wave Basis Set*, Computational Materials Science **6**, 15 (1996).

[119] G. Kresse and J. Furthmüller, *Efficient Iterative Schemes for* Ab Initio *Total-Energy Calculations Using a Plane-Wave Basis Set*, Phys. Rev. B **54**, 11169 (1996).

[120] G. Kresse and D. Joubert, *From Ultrasoft Pseudopotentials to the Projector Augmented-Wave Method*, Phys. Rev. B **59**, 1758 (1999).

[121] Y. A. Mankelevich, A. T. Rakhimov, and N. V. Suetin, *Two-Dimensional Simulation of a Hot-Filament Chemical Vapor Deposition Reactor*, Diamond and Related Materials **5**, 888 (1996).

[122] Y. A. Mankelevich, A. T. Rakhimov, and N. V. Suetin, *Three-Dimensional Simulation of a HFCVD Reactor*, Diamond and Related Materials **7**, 1133 (1998).

[123] M. N. R. Ashfold, P. W. May, J. R. Petherbridge, K. N. Rosser, J. A. Smith, Y. A. Mankelevich, and N. V. Suetin, *Unravelling Aspects of the Gas Phase Chemistry Involved in Diamond Chemical Vapour Deposition*, Phys. Chem. Chem. Phys. **3**, 3471 (2001).

[124] D. S. Dandy and M. E. Coltrin, *A Simplified Analytical Model of Diamond Growth in Direct Current Arcjet Reactors*, J. Mater. Res. **10**, 1993 (1995).

[125] M. Frenklach, *The Role of Hydrogen in Vapor Deposition of Diamond*, Journal of Applied Physics **65**, 5142 (1989).

[126] J. E. Butler and I. Oleynik, *A Mechanism for Crystal Twinning in the Growth of Diamond by Chemical Vapour Deposition*, Phil. Trans. R. Soc. A. **366**, 295 (2008).

[127] M. A. Lieberman and A. J. Lichtenberg, *Principles of Plasma Discharges and Materials Processing*, 1st ed. (Wiley, 2005).

[128] T. W. Mercer, J. N. Russell, and P. E. Pehrsson, *The Effect of a Hydrogen Plasma on the Diamond (110) Surface*, Surface Science **392**, L21 (1997).

[129] E. Y. Guillaume, D. E. P. Vanpoucke, R. Rouzbahani, L. Pratali Maffei, M. Pelucchi, Y. Olivier, L. Henrard, and K. Haenen, *First-Principles Investigation of Hydrogen-Related Reactions on (100)–(2 × 1) : H Diamond Surfaces*, Carbon **222**, 118949 (2024).

[130] S. Bühlmann, E. Blank, R. Haubner, and B. Lux, *Characterization of Ballas Diamond Depositions*, Diamond and Related Materials **8**, 194 (1999).

[131] R. E. Rawles and M. P. D'Evelyn, *Diamond Homoepitaxy Kinetics: Growth, Etching, and the Role of Oxygen*, MRS Proc. **339**, 279 (1994).

[132] S. Osswald, G. Yushin, V. Mochalin, S. O. Kucheyev, and Y. Gogotsi, *Control of $Sp^2/Sp^3$ Carbon Ratio and Surface Chemistry of Nanodiamond Powders by Selective Oxidation in Air*, J. Am. Chem. Soc. **128**, 11635 (2006).

[133] J. Warnatz, *Rate Coefficients in the C/H/O System*, in *Combustion Chemistry*, edited by W. C. Gardiner (Springer New York, New York, NY, 1984), pp. 197–360.

[134] D. L. Baulch et al., *Evaluated Kinetic Data for Combustion Modelling*, Journal of Physical and Chemical Reference Data **21**, 411 (1992).

[135] V. D. Knyazev, Á. Bencsura, S. I. Stoliarov, and I. R. Slagle, *Kinetics of the $C_2H_3 + H_2 \rightleftarrows H + C_2H_4$ and $CH_3 + H_2 \rightleftarrows H + CH_4$ Reactions*, J. Phys. Chem. **100**, 11346 (1996).

[136] W. Tsang, *Chemical Kinetic Data Base for Combustion Chemistry Part 4. Isobutane*, Journal of Physical and Chemical Reference Data **19**, 1 (1990).

[137] W. C. Gardiner, editor , *Combustion Chemistry* (Springer, New York Berlin Heidelberg, 1984).

[138] D. G. Goodwin, *Scaling Laws for Diamond Chemical-Vapor Deposition. I. Diamond Surface Chemistry*, Journal of Applied Physics **74**, 6888 (1993).

[139] O. D. Dwivedi, Y. Barsukov, S. Jubin, J. R. Vella, and I. Kaganovich, *Orientation-Dependent Etching of Silicon by Fluorine Molecules: A Quantum Chemistry Computational Study*, Journal of Vacuum Science & Technology A **41**, 052602 (2023).

[140] D. W. Brenner, D. H. Robertson, R. J. Carty, D. Srivastava, and B. J. Garrison, *Combining Molecular Dynamics and Monte Carlo Simulations to Model Chemical Vapor Deposition: Application to Diamond*, MRS Proc. **278**, 255 (1992).

[141] Y. A. Mankelevich, M. N. R. Ashfold, and J. Ma, *Plasma-Chemical Processes in Microwave Plasma-Enhanced Chemical Vapor Deposition Reactors Operating with C/H/Ar Gas Mixtures*, Journal of Applied Physics **104**, 113304 (2008).

[142] M. N. R. Ashfold and Y. A. Mankelevich, *Two-Dimensional Modeling of Diamond Growth by Microwave Plasma Activated Chemical Vapor Deposition: Effects of Pressure, Absorbed Power and*

*the Beneficial Role of Nitrogen on Diamond Growth*, Diamond and Related Materials **137**, 110097 (2023).
[143] M. P. D'Evelyn, J. D. Graham, and L. R. Martin, *[100] versus [111] Diamond Growth from Methyl Radicals and/or Acetylene*, Journal of Crystal Growth **231**, 506 (2001).
[144] M. Frenklach and H. Wang, *Detailed Mechanism and Modeling of Soot Particle Formation*, in *Soot Formation in Combustion*, edited by H. Bockhorn, Vol. 59 (Springer Berlin Heidelberg, Berlin, Heidelberg, 1994), pp. 165–192.
[145] M. Frenklach, *On the Driving Force of PAH Production*, Symposium (International) on Combustion **22**, 1075 (1989).
[146] M. Frenklach, R. I. Singh, and A. M. Mebel, *On the Low-Temperature Limit of HACA*, Proceedings of the Combustion Institute **37**, 969 (2019).
[147] M. C. McMaster, W. L. Hsu, M. E. Coltrin, and D. S. Dandy, *Experimental Measurements and Numerical Simulations of the Gas Composition in a Hot-Filament-Assisted Diamond Chemical-Vapor-Deposition Reactor*, Journal of Applied Physics **76**, 7567 (1994).
[148] R. B. Bird, W. E. Stewart, and E. N. Lightfoot, *Transport Phenomena*, Revised ed (Wiley, New York, 2007).
[149] P. Zalicki, Y. Ma, R. N. Zare, E. H. Wahl, T. G. Owano, and C. H. Kruger, *Measurement of the Methyl Radical Concentration Profile in a Hot-Filament Reactor*, Applied Physics Letters **67**, 144 (1995).
[150] Mankelevich, Yuri, Doctoral Thesis, 2013.
[151] *GRI-Mech 3.0*, http://combustion.berkeley.edu/gri-mech/data/frames.html.
[152] D. L. Baulch et al., *Evaluated Kinetic Data for Combustion Modeling: Supplement II*, Journal of Physical and Chemical Reference Data **34**, 757 (2005).